\documentclass[journal=jpcafh,manuscript=article]{achemso}

\usepackage[version=3]{mhchem} % Formula subscripts using \ce{}
\usepackage{xcolor}
\usepackage{multirow}
\usepackage{subcaption}
\usepackage[colorlinks]{hyperref}
\usepackage{xcolor}
\newcommand{\red}[1]{\textcolor{black}{#1}}
\usepackage{ulem}
\usepackage{amssymb}
\usepackage{textcomp}
\usepackage{booktabs}
\usepackage{amsmath}

\newcommand*{\rttensortwo}[1]{\bar{\bar{#1}}}

\author{Riccardo Dettori}
\affiliation{Physical and Life Sciences Directorate, Lawrence Livermore National Laboratory, 7000 East Avenue, Livermore, California 94550, United States}
\alsoaffiliation{Department of Physics, University of Cagliari, Monserrato, CA, 09042 Italy}
\email{riccardo.dettori@dsf.unica.it}
\author{Nir Goldman}
\email{ngoldman@llnl.gov}
\affiliation{Physical and Life Sciences Directorate, Lawrence Livermore National Laboratory, 7000 East Avenue, Livermore, California 94550, United States}
\alsoaffiliation{Department of Chemical Engineering, University of California, Davis, California 95616, United States}

\title{Creation of an Fe$_3$P Schreibersite Density Functional Tight Binding Model for Astrobiological Simulations}

\abbreviations{IR,NMR,UV}

\begin{document}

\begin{tocentry}

  \includegraphics[width=9.0cm,height=3.5cm]{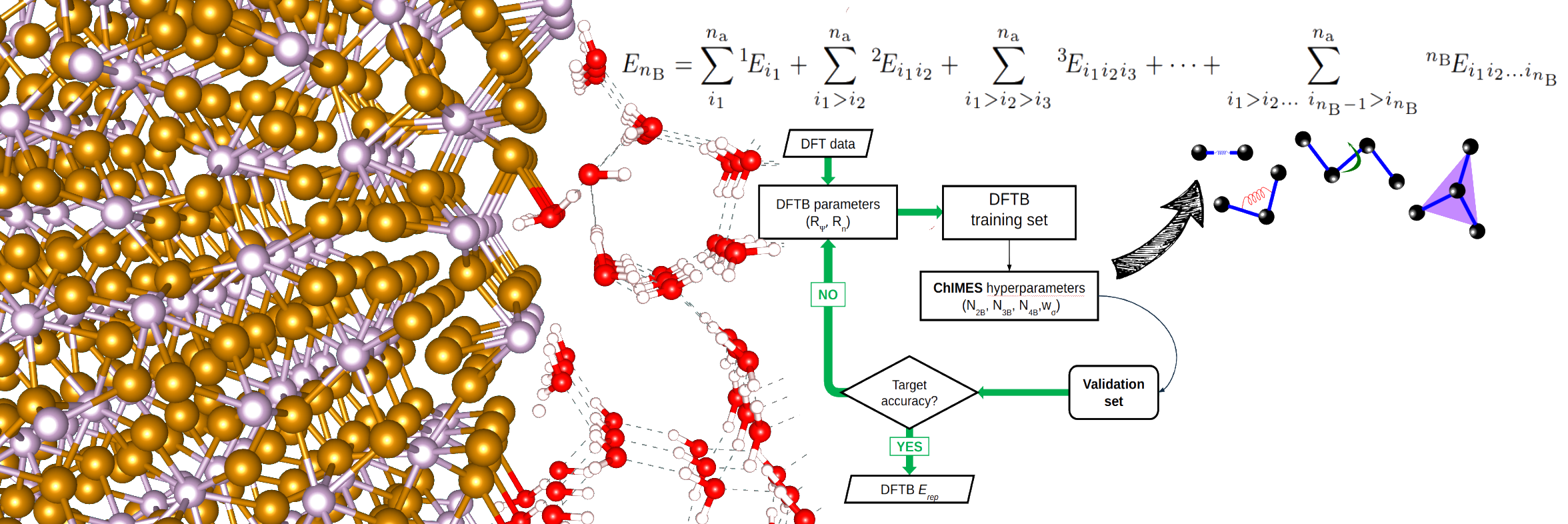}

\end{tocentry}

\begin{abstract}
The mineral schreibersite, e.g., Fe$_3$P, is commonly found in iron-rich meteorites and could have served as an abiotic phosphorus source for prebiotic chemistry. However, atomistic calculations of its degradation chemistry generally require quantum simulation approaches, which can be too computationally cumbersome to study sufficient time and length scales for this process. In this regard, we have created a computationally efficient semi-empirical quantum Density Functional Tight Binding (DFTB) model for iron and phosphorus-containing materials by adopting an existing semi-automated workflow that represents many-body interactions by linear combinations of Chebyshev polynomials. We have utilized a relatively small training set to optimize a DFTB model that is accurate for schreibersite physical and chemical properties, including its bulk properties, surface energies, and water absorption. We then show that our model shows strong transferability to several iron phosphide solids as well as multiple allotropes of iron metal. Our resulting DFTB parameterization will allow us to interrogate schreibersite aqueous decomposition at longer time and length scales than standard quantum approaches, providing for more detailed investigations of its role in prebiotic chemistry on early Earth.
\end{abstract}

\section{Introduction}
Phosphorous plays a central role in the formation of many life-building compounds, including nucleotides, phospholipids, adenosine triphosphate (ATP) for metabolic energy storage, and calcium phosphate (apatite) found in bone.\cite{Pasek2017,Gulick,Pasek2011,Schwartz2006,Pasek2017b} Extraterrestrial sources such as meteoritic impacts could have served as one possible source of prebiotic phosphorous on early Earth,\cite{Schwartz2006} including those with the mineral schreibersite, (Fe,Ni)$_3$-P.\cite{Pasek2005,Pasek2008} Schreibersite is a ferromagnetic material with tetrahedral symmetry (orthogonal lattice constants with $a = b > c$). It has been suggested to play a role in the formation of prebiotic organics,\cite{Schwartz2006,Gull2014,Gull2015, Pasek2005} where experimental studies have shown that it can catalyze the formation of amino acids, sugars, and nucleotides.\cite{Pasek2017,Gull2015,Cruz2016} However, early Earth likely experienced a wide range of conditions, including electrical storms\cite{Miller53} and extreme thermodynamic conditions due to impacts\cite{Manaa09,Goldman10,Goldman13}, that are difficult to probe with experimental trial and error alone. Schreibersite degradation studies would thus benefit from atomistic simulations, which can identify specific reaction mechanisms and kinetics, and can help narrow the relevant thermodynamic conditions for prebiotic synthesis.  

To date, computational studies on schreibersite have largely focused on quantum simulations of bulk properties and water surface adsorption using Kohn-Sham Density Functional Theory (DFT).\cite{Dettori_Fe3p,Pantaleone2021,Pantaleone2022} DFT is known to yield accurate descriptions of condensed phase physical and chemical data, as well as the energetics of chemical bond breaking and forming under reactive conditions (e.g., Refs.~\citenum{Manaa09, Goldman13}). However, DFT simulations generally require heavy computational effort, which limits calculations to systems containing few hundreds of atoms and to molecular dynamics (MD) timescales of tens of picoseconds. These scales are generally orders of magnitude smaller than those probed by experiments, where for example, chemistry at elevated pressure and temperature can \red{require} nanoseconds or longer to equilibrate,\cite{Cannella15,Lindsey_DNTF_2020} and extended structures can form that are nanometer-scale in diameter.\cite{Armstrong2020,Lindsey_CO_2022} Reactive MD models can help alleviate this time and length scale gap by optimizing empirical\cite{vanduin07} or machine-learning functional forms\cite{Drautz_ACE1,ANI-1ccx,Csanyi_ACE_organic,Pham_HN3_2021} to DFT or other quantum mechanical data. These methods are generally fit to the properties of specific chemical reactants over a narrow range of thermodynamic conditions. As a result, they can yield high computational efficiency, but tend to yield completely unphysical results outside of their fitting regime. 

The Density Functional Tight Binding method (DFTB) represents a promising alternative for simulations of reactive materials. DFTB is an approximate quantum simulation technique whereby the majority of the electronic interactions are represented by a minimal atomic basis set and are precomputed and mapped to a radial grid.\cite{Elstner, Gaus_mioPP,DFTB+_current}.
The DFTB total energy expression contains an empirical repulsive energy $E_{\rm rep}$ that can be fit to experimental or high-level computational data in order to extend the technique to a wide variety of materials and conditions.\cite{Goldman15-fm} There is considerable flexibility in the $E_\mathrm{Rep}$ functional form, including cubic splines\cite{Kullgren_CCS_2021}, Gaussian Process Regression\cite{Margraf_GPR_2021}, many-body polynomial based representations\cite{Goldman_SEQM_2023}, and neural network potentials.\cite{ISO34_DFTB_DTNN} The DFTB methodology allows for several orders of magnitude increase in computational efficiency while retaining most of the accuracy of standard DFT simulations. Consequently, DFTB simulations can achieve close to chemical equilibrium timescales for reactivity under extreme conditions ($\sim1 $ ns) using relatively low computational effort. One of the main limitations of DFTB, though, is the availability of existing repulsive energies in literature, as one repository of interaction parameters\cite{Yang2008,Elstner,Gaus_mioPP,Niehaus,Guimares2007,Luschtinetz2008,Sieck2003,Khler2005,Gaus2012,Kubillus2014,Gaus2014,Zheng2007} does not contain specific interactions between iron and phosphorus. In addition, the DFTB method itself has several hyperparameters, mainly the wavefunction and density confining radii (discussed below), that require challenging optimizations as well. 

In this work, we overcome these issues for Fe$_3$P schreibersite by determining the Fe-Fe and Fe-P DFTB interaction potentials through a semi-automated workflow that uses the Chebyshev Interaction Model for Efficient Simulation (ChIMES) method for determination of the repulsive energies.\cite{Lindsey2017,Lindsey2020,Goldman2023} ChIMES is a many-body reactive force field model for MD simulation based on linear combinations of Chebyshev polynomials. This linear parameterization yields rapid MD or $E_{\rm rep}$ model optimization, allowing us to efficiently cycle through large numbers of DFTB and ChIMES hyperparameters. DFTB/ChIMES models have been created for a wide variety of systems, including TiH$_2$,\cite{Goldman_TiH2} energetic materials\cite{Lindsey_DNTF_2020}, actinides\cite{Goldman_DFTB_H_Pu}, and as a general purpose model for organic materials.\cite{Pham_DFTB_JPCL} Here, we first briefly review the DFTB and ChIMES approaches and explain our workflow for optimization of both DFTB and ChIMES hyperparameters. We then downselect our model through validation against a wide range of DFT data that was not included in our training set, such as the Fe$_3$P lattice constants, bulk moduli, surface facet energies, and bulk water absorption. We then examine our optimal DFTB/ChIMES model further by computing water surface adsorption energies, \red{the phonon density of states of the perfect crystal}, and MD simulations under ambient and extreme conditions. Finally, we explore the transferability of our model through the prediction of properties for different materials, including FeP, Fe$_2$P, and three allotropes of iron. Our results indicate that our DFTB/ChIMES model yields comparable accuracy to DFT for all properties investigated here, including the intense pressures and temperatures that can be accessed during planetary impact.

\section{Methods}
\subsection{DFT Calculations and Training Set Construction}
All DFT simulations were performed using the Vienna Ab initio Simulation Package (VASP)\cite{Kresse1,Kresse2,Kresse3}, with projector augmented wave (PAW) pseudopotentials\cite{Kresse4,Blochl} and the Perdew, Burke, and Ernzerhof (PBE) exchange-correlation functional\cite{Perdew}. Partial occupancies of the electronic states were set with fourth-order Methfessel-Paxton smearing\cite{Methfessel}, using a width of 0.2 eV. We observed converged energies for the bulk system with a plane-wave energy cut-off of $500$ eV, self-consistent field (SCF) convergence criteria of $10^{-6}$~eV, and $6\times 6\times 6$ $\mathbf{k}$-point Monkhorst-Pack mesh sampling of Brillouin zone\cite{Monkhorst76}. The force convergence tolerance for geometry optimizations was set to 0.01 eV/\AA{} for each atom in each direction. Our calculations were performed with collinear spin polarization initialized in a ferromagnetic configuration, which was maintained during all optimizations. As mentioned in Ref.~\citenum{Dettori_Fe3p}, the inclusion of van der Waals (vdW) correction in our calculations resulted in a constant offset of the water absorption energies, with minimal change to optimized configurations. Furthermore, the observed effect is strongly dependent upon the choice of the vdW functional\cite{Bedolla}. Hence, for the model development in this study, we have chosen to adopt the standard PBE functional without vdW, though we note that those effects can be included at a later date.

Our training set was built by performing DFT-MD simulations of a 32 atom Fe$_3$P schreibersite supercell, initially optimized to dimensions of $9.045$\AA\ $\times9.045$\AA\ $\times4.380$\AA, yielding a density of 7.36 g/cm$^3$. We then used the relaxed structure to initiate 20~ps long DFT-MD trajectories with a timestep of 1.0~fs at temperatures of $300, 1500,$ and $2000$ K and pressures of $0, 10, 20, 30,$ and $40$ GPa. From these simulations, we then aggregated the following 310 training configurations in total:
\begin{itemize}
    \item 120 \red{configurations equally spaced in time (e.g., sampled every $\sim$167 fs from the MD trajectory)} at $P=0$ GPa and $T=300, 1500,$ and $2000$ K.
    \item 120 equally spaced configurations at $P=10,20,30,$ and $40$ GPa and $T=1500$ K.
    \item 70 configurations obtained by randomly displacing by $0.1$ \AA{} the atoms of the optimized Fe$_3$P structure, similar to previous reactive MD model development.\cite{Pham_HN3_2021}
\end{itemize}

\subsection{DFTB calculations} 
DFTB calculations were performed with the DFTB$+$ code\cite{DFTB+_current} using orbital resolved self-consistent charge calculations (SCC)\cite{Elstner1998} with a convergence criteria of $2.72\times10^{-5}$ eV ($10^{-6}$ au). We adopted the same electron thermal smearing and $\mathbf{k}$-point mesh of our DFT calculations. Similarly, all DFTB calculations were performed with collinear spin polarization initialized in a ferromagnetic spin geometry. The DFTB formalism has been extensively discussed in literature\cite{DFTB+,Elstner,Koskinen09,Gaus_mioPP,DFTB+_current}, with a brief outline provided here. The total energy expression with SCC is derived by assuming spherically symmetric charge densities and then expanding the Kohn-Sham total DFT total energy expression to either second\cite{Elstner} or third order\cite{Gaus11} about small charge fluctuations. We note that in this work we rely on the second-order charge fluctuation model only. 

The resulting total energy expression reads as:
\begin{equation}
E_{\rm DFTB} = E_{\rm BS}+E_{\rm Coul}+E_{\rm rep}.
\end{equation}
\noindent Here, $E_{\rm BS}$ is the band structure energy, which is computed from the Kohn-Sham eigenstates determined from diagonalization of the approximate Hamiltonian with a minimal, Slater-type basis set. The term $E_{\rm Coul}$ corresponds to the charge-transfer energy from second-order charge fluctuations and is computed self-consistently. Finally, $E_{\rm rep}$ is the repulsive energy, which contains the ion-ion repulsions, as well as Hartree and exchange-correlation double counting terms. $E_{\rm rep}$ is generally represented as a short-ranged empirical function that includes first coordination shell or bonded interactions only. Its parameters are usually fit to reproduce DFT or experimental data, and it can describe pairwise\cite{Goldman15-fm,Vuong2020} or many-body interactions.\cite{Goldman_DFTB_H_Pu,Dantanarayana2020,Lindsey_DNTF_2020,Pham_DFTB_JPCL,Goldman_SEQM_2023} The Hamiltonian and overlap matrices used in $E_{\rm BS}$ are determined from pretabulated Slater-Koster (SK) tables that assume two-center interactions, only. However, prior to SK tabulation, both the electronic wave functions and electron density are subjected to separate confining or compression potentials, which has the effect of improving transferability of the calculations. For our work, we use the following simple power function for confinement:

\begin{equation}
V_{conf}(r) = \left( \frac{r}{R_\chi} \right)^2,
\label{eq:vconf}
\end{equation} 

\noindent where $\chi = \Psi$ or $n$, $R_\Psi$ corresponds to wavefunction compression, and $R_n$ to density compression. Previous efforts have used larger exponents in the power function or alternate functional forms (e.g., Ref.~\citenum{DFTB_transferability_1}), which can be investigated in future work.

In general, optimal choice of confining radii is system-dependent and cannot be determined \red{\textit{a priori}}, frequently requiring a semi-exhaustive search.\cite{Goldman2021,Goldman_SEQM_2023} In our study, we choose to focus on optimizing the Fe density and wavefunction confining radii, only, in order to make our search more tractable. Similarly, we have only created $E_\mathrm{Rep}$ models for Fe-Fe and Fe-P interactions. The remaining DFTB$+$ interaction potentials (including tabulated SK Hamiltonian and overlap matrices, and/or $E_\mathrm{Rep}$ functions) were taken from known parameter sets. All non-iron interactions, such as those involving P, O, and H were taken from the mio-1-1 parameter set.\cite{Elstner,Gaus11} The Fe-O and Fe-H repulsive energies were determined by first computing new SK tables using our choice of confining radii for Fe and those from mio-1-1 for O and H. This was followed by matching energetics for dimer calculations from the trans3d parameter set\cite{Zheng2007} to approximate the repulsive energies. \red{A recent publication has made available DFTB parameterizations for a large portion of the periodic table\cite{Margraf-DFTB-2024}, though these were not available at the time of this study and have not been tested for schreibersite/water systems, specifically.}

\subsection{ChIMES Methodology and Many-body $E_\mathrm{Rep}$ determination} 
The ChIMES method has been discussed in detail elsewhere,\cite{Lindsey2017,Lindsey2020,Goldman2021,Goldman2023} and hence will only be given a short introduction here. The starting point in ChIMES is a many-body expansion of the total energy of the system:
\begin{equation}
E_\mathrm{DFT} =
\sum^{n_\mathrm{a}}_{i_1}                      {}^1             \!E_{i_1} +
\sum^{n_\mathrm{a}}_{i_1>i_2}                  {}^2             \!E_{i_1 i_2} +
\sum^{n_\mathrm{a}}_{i_1>i_2>i_3}              {}^3             \!E_{i_1 i_2 i_3} +
\sum^{n_\mathrm{a}}_{i_1>i_2>i_3>i_4}           {}^4             \!E_{i_1 i_2 i_3 i_4}  \\+ \dots +
\sum^{n_\mathrm{a}}_{i_1>i_2\dots\ i_{n{_\mathrm{B}-1}}>i_{n{_\mathrm{B}}}}{}^{n_\mathrm{B}}\!E_{i_1 i_2 \dots i_n{_\mathrm{B}}}.
\label{eq:genchimes}
\end{equation}

\noindent Here, the index $n_\mathrm{a}$ allows for a sum over the total number of atoms. The one-body contributions (${}^1\!E_{i_1}$) are element-specific constants, the two-body \red{(2B)} or dimer contributions (${}^2\!E_{i_1 i_2}$) correspond to pair-wise interactions, the three-body \red{(3B)} or trimer contributions (${}^3\!E_{i_1 i_2 i_3}$) to triplet energies, etc., until a user-defined maximum of $n_\mathrm{B}$ is reached. 

We now express each of the greater than one-body terms as a sum of linear combinations of Chebyshev polynomials of the first kind. In this case, $T_m(x)$ corresponds to a Chebyshev polynomial of order $m$, and $x = \mathrm{cos}\, \theta$, which limits the Chebyshev polynomial range to $\left[-1:1\right]$. For the 2B energy ${}^2\!E_{i_1 i_2}$, we use the following expression:

\begin{equation}
        {}^2\!E_{i_1 i_2} = f_{\mathrm{p}}\left(r_{i_1 i_2}\right) + f^{e_{i_1} e_{i_2}}_{\mathrm{c}}\left(r_{i_1 i_2}\right)
             \sum^{N_{2B}}_{m=1} C_m^{e_{i_1} e_{i_2}} T_m (s^{e_{i_1} e_{i_2}}_{i_1 i_2})
        \label{eqn:FF2B}
\end{equation}

\noindent Starting with the terms in the summation, $C_{m}^{e_{i_1} e_{i_2}}$ is the permutationally invariant coefficient for $m$th order polynomial (taken to maximum order of $N_{2B}$) for pairwise interactions between atoms $i$ and $j$, which are represented by element types $e_{i_1}$ and $e_{i_2}$. The term $s^{e_{i_1} e_{i_2}}_{i_1 i_2}$ is the transformed pairwise distance ($r_{i_1 i_2}$) such that the user-defined radial range $\left[r^{e_{i_1} e_{i_2}}_\mathrm{min}:r^{e_{i_1} e_{i_2}}_\mathrm{max}\right]$ is scaled to $\left[-1:1\right]$, using an element-pair specific scaling function. Similar to previous work\cite{Lindsey2017}, we rely on the Morse-like exponential function, where $r_{i_1 i_2}$ is scaled by a user chosen constant $\lambda_{e_1 e_2}$, i.e.,  $s^{e_{i_1} e_{i_2}}_{i_1 i_2} \propto \mathrm{exp}\left(-r_{i_1 i_2}/\lambda_{e_1 e_2}\right)$. The value of $\lambda_{e_1 e_2}$ is usually taken to be the peak position of the first coordination shell from the radial distribution function for that element pair. We ensure smooth behavior and zero energy at the radial boundary ($r^{e_{i_1} e_{i_2}}_{\mathrm{max}}$) through the term $f^{e_{i_1} e_{i_2}}_\mathrm{c}(r_{i_1 i_2})$, which we take to be is a Tersoff smoothing function\cite{Tersoff88}. Lastly, we use a smoothly varying penalty function $f_\mathrm{p} (r_{i_1 i_2})$ in order to prevent sampling of distances below the minimum value sampled in our training set ($r^{e_{i_1} e_{i_2}}_{\mathrm{min}}$), thus preventing calculation of the Chebyshev polynomials outside of their natural range.

Greater than two-body terms can be determined through creation of many-body orthogonal polynomials. These can be obtained by defining a cluster of size $n$ and taking the tensorial product of the Chebyshev polynomials derived from the constituent ${n \choose 2}$ unique pairs. In this case, the three-body polynomials will be products of ${3 \choose 2} = 3$ two-body polynomials, i.e.:
\begin{equation} %%% 3B FF EQUATION %%%
{}^3\!E_{i_1 i_2 i_3} =       f^{e_{i_1} e_{i_2}}_{\mathrm{c}}\left(r_{i_1 i_2}\right)
                                f^{e_{i_1} e_{i_3}}_{\mathrm{c}}\left(r_{i_1 i_3}\right)
                                f^{e_{i_2} e_{i_3}}_{\mathrm{c}}\left(r_{i_2 i_3}\right)
                \sum^{N_{3B}}_{m=0}
                \sum^{N_{3B}}_{p=0}
                {\sum^{N_{3B}}_{q=0}}^\prime
                C^{e_{i_1} e_{i_2} e_{i_3}}_{mpq}
                                T_{m}\left(s^{e_{i_1} e_{i_2}}_{i_1 i_2}\right)
                                T_{p}\left(s^{e_{i_1} e_{i_3}}_{i_1 i_3}\right)
                                T_{q}\left(s^{e_{i_2} e_{i_3}}_{i_2 i_3}\right)
        \label{eqn:FF3B}
\end{equation}
\noindent The triple sum over the three-body polynomials is taken over the hypercube up to a predefined order $N_{3B}$, and the primed sum denotes that we include only terms for which two or more of the polynomial powers are greater than zero (to guarantee that three distinct triplets are evaluated).  Four-body (4B) terms are similarly included in ChIMES optimizations\cite{Pham_HN3_2021}, where ${}^4\!E_{i_1 i_2 i_3 i_4}$ is now determined from the sum over the product of the ${4 \choose 2} = 6$ constituent pair-wise polynomials multiplied by a single permutationally invariant coefficient.

The training set for the $E_\mathrm{Rep}$ was then created by computing the DFTB quantum mechanical terms for each configuration (e.g., $E_\mathrm{BS}$ and $E_\mathrm{Coul}$) and subtracting these quantities from the corresponding DFT data. This generally includes the cartesian forces, $\vec{F}$, stress tensor components, $\rttensortwo{\sigma}$, and the total system energy. Optimal ChIMES parameters (the coefficients of linear combination) can then readily be determined through linear least squares fitting of the overdetermined matrix equation $\boldsymbol{w} \boldsymbol{A} \boldsymbol{C} = \boldsymbol{w} \boldsymbol{B}_{\mathrm {rep}}$, The matrix $\boldsymbol{A}$ corresponds to the derivatives of the ChIMES energy or force expression with respect to the fitting coefficients (i.e., the unweighted polynomial values, including the smoothing function). The column vectors $\boldsymbol{C}$ and $\boldsymbol{B}_{\mathrm {rep}}$ correspond to the linear ChIMES coefficients for which we are solving and the numerical values for the training data, respectively. The symbol $\boldsymbol{w}$ corresponds to a diagonal matrix of weights to be applied to the elements of $\boldsymbol{B}_{\mathrm {rep}}$ and rows of $\boldsymbol{A}$. 

The Chebyshev polynomial coefficients can then be solved for by minimizing the objective function:

\begin{footnotesize}
%%% RMSE EQUATION %%%
\begin{equation}
        F_{\mathrm{obj}}  = \sqrt{
        \frac{1}{N_d}
        \times
\left(
\sum_{\tau=1}^M
\sum_{i=1}^{n_a}
\sum_{\alpha=1}^{3} w_{\mathrm{F}_{\alpha_i}}^2\left(\Delta \mathrm{F}_{\alpha_i} \right)^{2}
+
\sum_{\tau=1}^{M}
\sum_{\alpha=1}^3 w_{\sigma_{\alpha\alpha}}^2 \left(\Delta\sigma_{\alpha\alpha}\right)^{2}
+
\sum_{\tau=1}^{M} w_{\mathrm{E}_{\tau}}^2 \left(\Delta E_{\tau}\right)^{2}
\right)}.
\label{eqn:rmse}
\end{equation}
\end{footnotesize}

\noindent Here, $M$ is the total number of configurations in the training set, $N_d$ is the total number of data entries ($3Mn_a$ force
components plus $3M$ stress tensor components plus $M$ energy components), the $w$ values correspond to the weights, and $\Delta F_{\alpha_i} = F^{\tau}_{\mathrm{ChIMES}_{\alpha_i}} - F^{\tau*}_{\mathrm{Rep}_{\alpha_i}}$, $\Delta\sigma_{\alpha\beta} = \sigma^{\tau}_{\mathrm{ChIMES}_{\alpha\beta}} - \sigma^{\tau*}_{\mathrm{Rep}_{\alpha\beta}}$, and $\Delta E_{i} = E^{\tau}_{\mathrm{ChIMES}} - E^{\tau*}_{\mathrm{Rep}}$. In this case, $F^{\tau*}_{\mathrm{Rep}_{\alpha_i}}$, $\sigma^{\tau*}_{\mathrm{Rep}_{\alpha\beta}}$, and $E^{\tau*}_{\mathrm{Rep}}$ correspond to the residuals determined from the difference between the DFT computed quantities and those computed from DFTB without the repulsive energy for $i$th atom in the $\tau$th configuration, and $\alpha/\beta$ correspond to the Cartesian direction. Previous work has indicated that forces and the diagonal stress tensor components can be the most significant data for $E_\mathrm{Rep}$ determination.\cite{Goldman15-fm,Lindsey_DNTF_2020,Goldman2021} Consequently, for Fe$_3$P DFTB/ChIMES model development, we have chosen to optimize to these quantities, only. 

\begin{figure}
  \centering
\includegraphics[width=.8\linewidth]{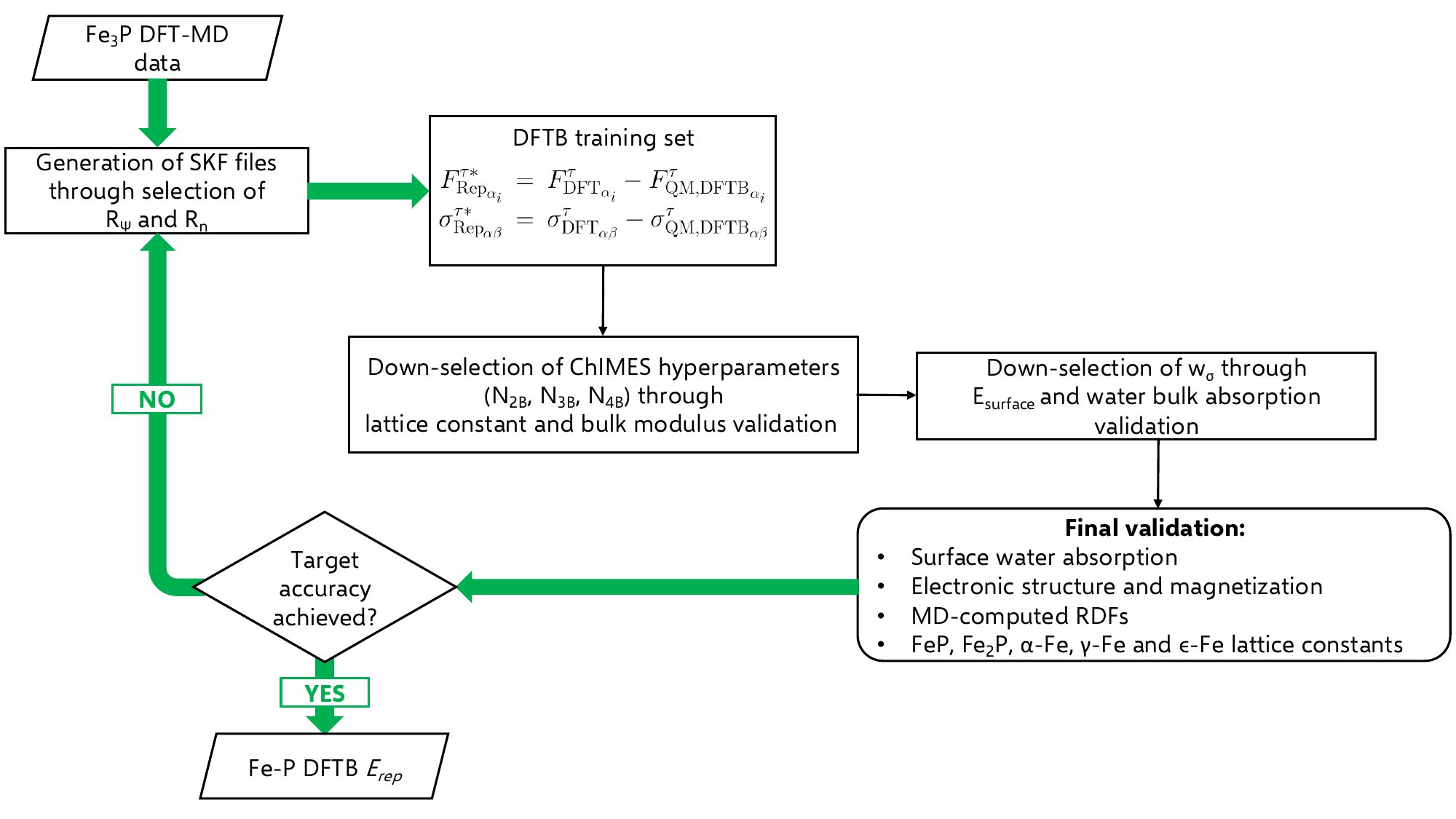}
\caption{Workflow for DFTB/ChIMES optimization.}
\label{fig:workflow}
\end{figure}

We then optimized the DFTB$+$ confining radii and ChIMES hyperparameters through a workflow similar to our previous efforts\cite{Goldman2021} (Figure~\ref{fig:workflow}). For ChIMES $E_{\rm Rep}$ parametrization, we set the minimum and maximum element-pair radii to span the first coordination shell of the configurations in our training set: $1.6\le r_{FeFe}\le 3.5\,$\AA{} and $1.5\le r_{\rm FeP}\le 2.9\,$\AA. We use Morse-like scaling constants of $\lambda_{\rm FeFe}=2.6$ and $\lambda_{\rm FeP}=2.25$, which correspond to the average nearest neighbor distances in our training set. We have limited our search to the following sets of $\{n_{\rm 2B}, n_{\rm 3B}, n_{\rm 4B}\}$ ChIMES polynomials orders: $\{8,4,0\}$, $\{12,8,0\}$, $\{12,8,4\}$, and $\{16,12,0\}$. We have also applied different weights to stress tensor components,  with values of $w_{\rm \sigma}=1, 5, 10, 25, 50$, in order to improve agreement with the validation data. 

\section{Results and discussion}
\subsection{$R_\Psi$ and $R_n$ Optimization}
We have computed a grid of DFTB Hamiltonian and overlap matrix SK tables that span values of $3.2\le R_\Psi \le 5.0$ au and  $6.0\le R_n \le 17.0$ au, for a total of 30 different combinations of $R_\Psi$ and $R_n$. These parameterizations were then passed through our workflow in order to create a unique ChIMES repulsive energy for each set of $\{R_\Psi, R_n\}$. We find that in general that smaller $R_\Psi$ and $R_n$ are associated with smaller force and stress root-mean-squared errors (RMSE). For example, the $\{8,4,0\}$ model set with $w_{\rm \sigma} = 1$ yielded Fe force errors on the order of 8.1 kcal/mol.\AA{}, P force errors 5.4 kcal/mol.\AA{}, and stress tensor errors of 4.5~GPa for $\{R_\Psi=3.2\, \mathrm{au}, R_n=6.0\, \mathrm{au}\}$, in contrast to values of 16.6 kcal/mol.\AA{}, 13.4 kcal/mol.\AA{}, and 4.14~GPa for $\{R_\Psi=5.0\, \mathrm{au}, R_n=17.0\, \mathrm{au}\}$.  The force RMSE exhibits relatively small changes with increasing stress tensor weight, though setting $w_{\rm \sigma} = 50$ decreases the stress errors to $< 1$~GPa. Increasing the polynomial order of the ChIMES model has the overall effect of lowering the RMSE. For example, the $\{16,12,0\}$ model set yields Fe force errors of approximately 7.8 kcal/mol.\AA{}, P forces of 5.1 kcal/mol.\AA{}, and stress tensor errors of 2.7~GPa for $\{R_\Psi=3.2\, \mathrm{au}, R_n=6.0\, \mathrm{au}\}$ and $w_{\rm \sigma }= 1$. (Please see Supporting Information for additional data.). 

The resulting DFTB/ChIMES models were then validated against a number of properties previously computed results from DFT calculations\cite{Dettori_Fe3p} that were not a part of our training set. This includes the Fe$_3$P lattice constants, bulk modulus, surface energies for the (001), (100), and (110) crystal facets, and water bulk absorption and surface adsorption energies. All DFTB/ChIMES validation calculations were performed using the same k-point meshes as the DFT results, namely, bulk systems were run using a $6\times6\times6$ $\mathbf{k}$-point Monkhorst-Pack grid, while a $1\times7\times7$ mesh was used for the surface calculations, with a smaller mesh value in the surface direction. We again used orbital resolved SCC with a charge convergence criteria of $2.72\times10^{-5}$ eV ($10^{-6}$ au) and a force convergence tolerance of 0.01 eV/\AA{} for each atom in each direction. 

We now report on the results from our validation calculations. In general, we find that the inclusion of 4B terms in the $\left\{12,8,4\right\}$ parameterization yields only very change small in accuracy relative to the $\left\{12,8,0\right\}$ set. Hence, for the sake of brevity, we limit our discussion to the $\{8,4,0\}$ and $\{16,12,0\}$ parameterizations, which are the lowest and highest limits of ChIMES complexity in our study (the complete set of results can be found in the Supporting Information).   

\paragraph{Lattice constants.}

\begin{figure}
\begin{subfigure}{\textwidth}
  \centering
  \includegraphics[width=1.\linewidth]{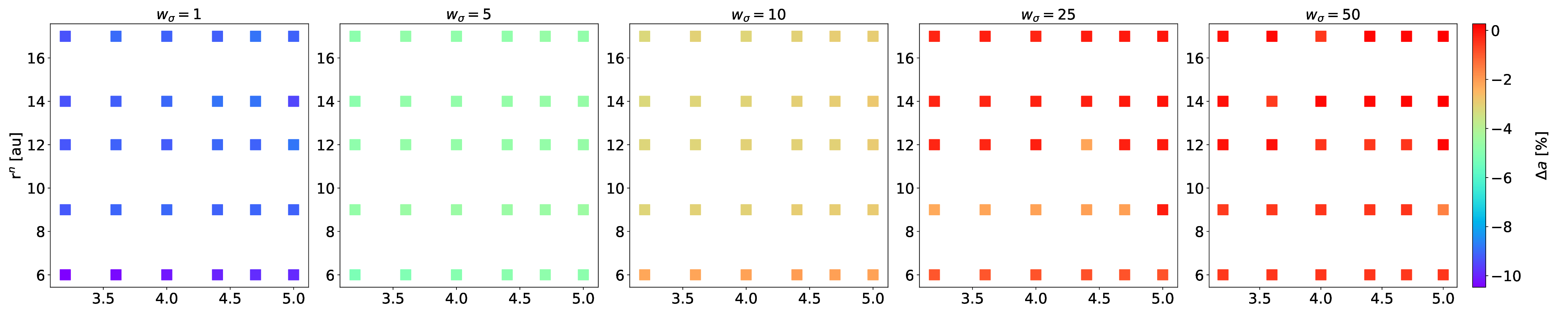}
  \caption{ }
  \label{fig:8_4_latconst_a}
\end{subfigure}
\begin{subfigure}{\textwidth}
  \centering
  \includegraphics[width=1.\linewidth]{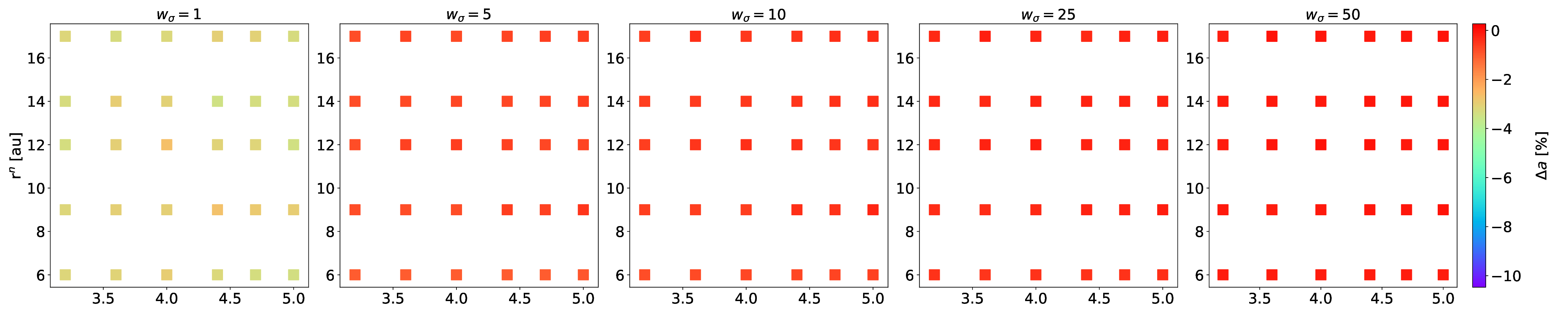}
  \caption{ }
  \label{fig:16_12_latconst_a}
\end{subfigure}
\caption{Percentage deviation from DFT values of the schreibersite $a$ lattice constant for (a) \{8,4,0\}, and (b) \{16,12,0\} model sets, as a function of the stress tensor weights (reported above the plots). Results for the $c$ lattice constant are included in the Supporting Information.}
\label{fig:latconst}
\end{figure}

Our results for the relative difference between the $\{8,4,0\}$ model and DFT for the schreibersite equilibrium lattice constants (Fig.~\ref{fig:latconst} for the $a$ constant, with results for $c$ included in the Supporting Information) indicate that using equal weights in our optimizations for the forces and stress tensor components ($w_\mathrm{F} = w_{\rm \sigma} = 1$) consistently yields an overly compact lattice, where the values of $a$ and $c$ are approximately $8\%$ and $10\%$ smaller than those from DFT. These higher density structures also optimize to antiferromagnetic spin geometries, rather than the correct ferromagnetic result. In addition, we observe a relatively small dependence on values of $R_\Psi$ and $R_n$. However, the introduction of weights of $w_{\rm \sigma} > 1$ while keeping $w_\mathrm{F}$ at unity improves the results appreciably, where observe the errors improve to $-5\%$ for $w_{\rm \sigma} = 5$ to $\sim 1\%$ for $w_{\rm \sigma}=25, 50$.

Increasing the polynomial order to $\{16,12,0\}$ uniformly improves the lattice constant results relative to DFT. Here, the average error in $a$ and $c$ is $\sim3.5\%$ for $w_{\rm \sigma}=1$, which decreases to $<0.5\%$ for $w_{\rm \sigma} = 25, 50$. We observe only small differences between results for $\{12,8,0\}$ and $\{12,8,4\}$, where the lattice constants both agree within 1\% for values of $w_{\rm \sigma}=10$ and $25$ (see Supporting Information for detailed data). 

\paragraph{Bulk modulus.} 
We now investigate the influence the confining radii on the schreibersite bulk modulus in order to test the validity of our models for high-pressure material property prediction. In this case, we have computed the zero temperature compression curves by isotropically compressing the system pressure to 100 GPa with each DFTB/ChIMES model. We then determine the bulk modulus $K_0$ through optimization to a Rose-Vinet equation of state.\cite{Vinet1987} For reference, we previously obtained a DFT bulk modulus of $K_{\rm DFT}=173.2$ GPa\cite{Dettori_Fe3p}, compared to two available experimental results of $172.0$\cite{Gu} and $157.9$\cite{Scott} GPa. 

\begin{figure}
\begin{subfigure}{1.\textwidth}
  \centering
  \includegraphics[width=1.\linewidth]{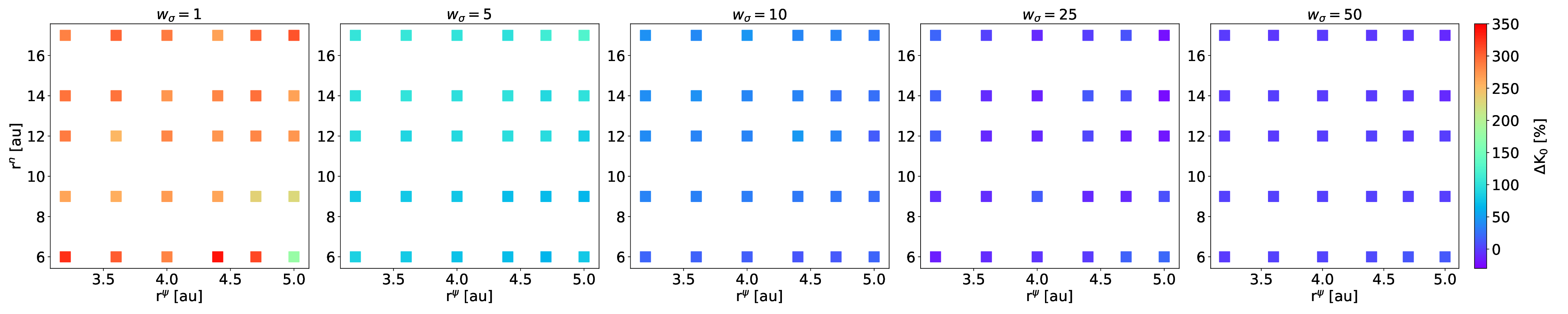}
  \caption{ }
  \label{fig:bulkmod_8_4}
\end{subfigure}
\begin{subfigure}{1.\textwidth}
  \centering
  \includegraphics[width=1.\linewidth]{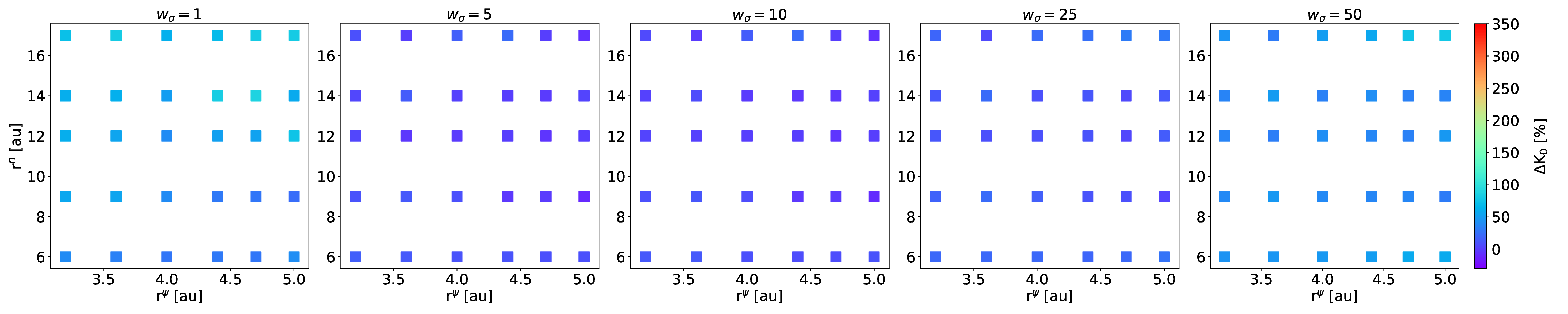}
  \caption{ }
  \label{fig:bulkmod_16_12}
\end{subfigure}
\caption{Percentage deviation from DFT bulk modulus of schreibersite for (a) the $\{8,4,0\}$ set and (b) $\{16,12,0\}$ set a function of stress tensor weights (reported above the plots).}
\label{fig:bulkm}
\end{figure}

Fig.~\ref{fig:bulkm} reports the results in terms of relative difference $\Delta K_0$ with respect to DFT. We find that all results for $w_{\rm \sigma}=1$ yield bulk moduli that are up to 400\% too high for the $\{8,4,0\}$ set, and over 100\% too high for the $\{16,12,0\}$ set. This is due to the excessively compact antiferromagnetic structure present in all of these DFTB/ChIMES models. We observe that applying values of $w_{\rm \sigma}$ greater than one yields improved results. The $\{8,4,0\}$ model set yields errors of approximately 100\% for $w_{\rm \sigma}=5$, and errors approaching zero for $w_{\rm \sigma}=25$ and greater. Our results for $\{16,12,0\}$ initially converge more rapidly to low error results, where we observe a number of DFTB/ChIMES optimizations that yield errors of close to zero for $w_{\rm \sigma} = 5$ and $10$. However, the relative values of $\Delta K_0$ increase with higher values of $w_{\rm \sigma}$. This is likely because the large stress tensor weights combined with the higher complexity of the model yield higher errors in the forces overall. In general, we find that the $\{16,12,0\}$ model set generally yields the highest level of accuracy for our validation studies. Hence, we choose to focus on this model set for the remainder of our confining radii study.

\paragraph{Surface Energies.}
Next, we have investigated the surface formation energy for the  $(001)$, $(100)$, and $(110)$ low index surface facets. The surface formation energy is defined as 

\begin{equation}
    E_{\rm surface} = \dfrac{E_{\rm slab} - n\left( \tfrac{E_{\rm bulk}}{n_{\rm bulk}}\right)}{2A}.
\end{equation}

\noindent Here, $E_{\rm slab}$ is the total electronic energy of the slab, $\left(E_{\rm bulk}/n_{\rm bulk}\right)$ is the energy per atom of the bulk cell, $n$ is the number of atoms contained in the slab, $A$ is the surface area of a given facet, and the factor of two accounts for both surfaces present in a periodically replicated system. We use our previously computed surface structures from DFT as our initial configuration for the DFTB/ChIMES calculations. In this case, each surface slab is six atomic layers thick (corresponding to a total of 96 atoms) and with a vacuum region of 20 \AA{} in each direction perpendicular to the slab surface in order to minimize nonphysical interactions between periodic images. Furthermore, due to the heterogeneity of the system, each facet was constructed with different surface terminations (i.e., the number of P or Fe atoms at the surface). By enforcing both symmetry (top and bottom layers are identical) and stoichiometry (i.e., the system maintains a 3:1 iron to phosphorus ratio), we were able to build two different surface terminations for each facet, where either a majority of iron (labeled `type 1') or phosphorus atoms (labeled `type 2') is present. The errors for the (001) and (100) surfaces were similar for all sets of confining radii. Hence, for our discussion, we choose to focus on both types of the (001) and (110) surfaces, only. 

\begin{figure}
\begin{subfigure}{1.\textwidth}
  \centering
\includegraphics[width=1.\linewidth]{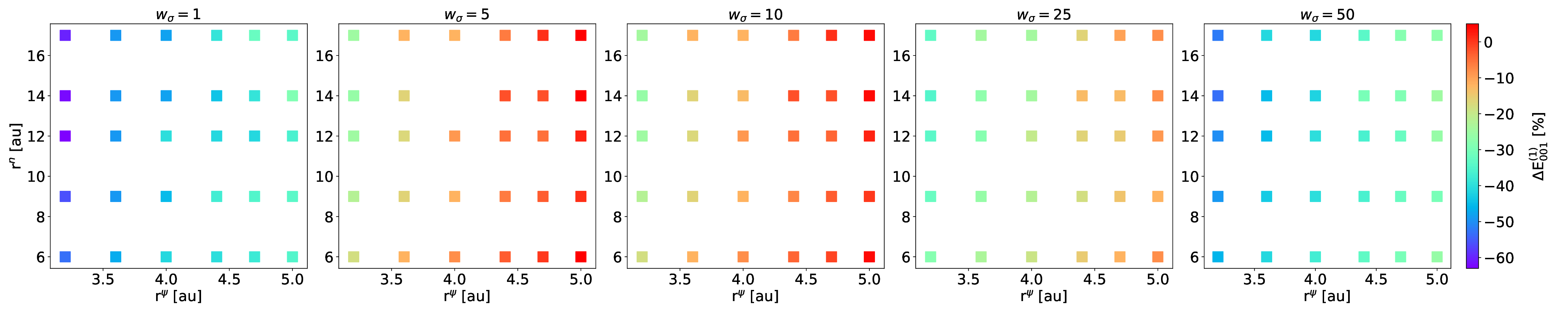}
  \caption{ }
\label{fig:E001_1_16_12}
\end{subfigure}
\begin{subfigure}{1.\textwidth}
  \centering
\includegraphics[width=1.\linewidth]{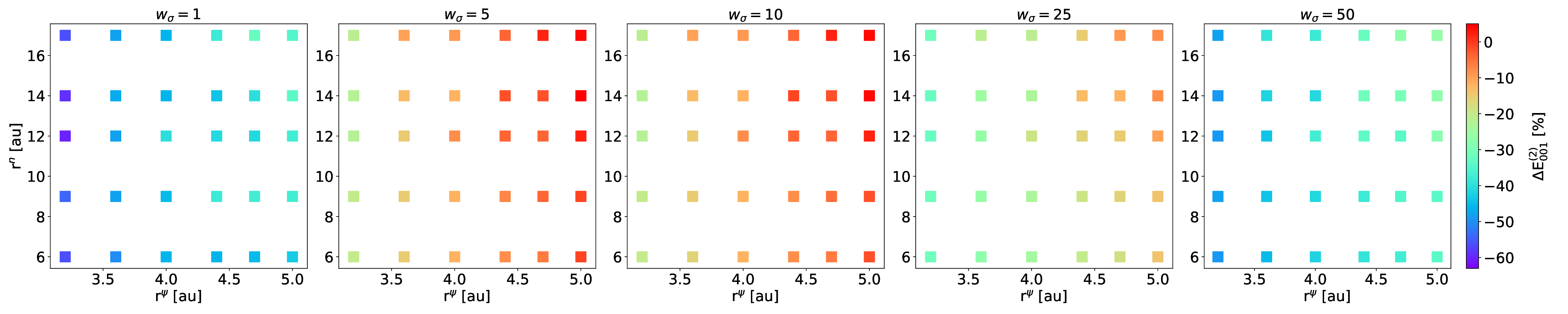}
  \caption{ }
\label{fig:E001_2_16_12}
\end{subfigure}
\caption{Percentage deviation from DFT values of schreibersite surface energy formation for $(001)$ facets of termination type 1 (a) and type 2 (b) obtained for $\{16,12,0\}$.}
\label{fig:s001_energy}
\end{figure}

The complete set of formation energies is reported in the Supporting Information. Our results for $\{16,12,0\}$ show that the energies for both type~1 and type~2 (001) surfaces are highly sensitive to the value of $w_{\rm \sigma}$ (Fig.~\ref{fig:s001_energy}). When no weights are applied the stress tensor components, the surface energy can be up to $60\%$ larger than DFT (first panel on the left of Fig.~\ref{fig:E001_1_16_12}), with the highest value of $E^{(1)}_{\rm 001}=3.66$ J/m$^2$ from $R_\Psi=3.2$ au and $R_n=12$ au compression radii for the type~1 iron dominated surface, compared to the DFT value of $2.24$ J/m$^2$. We find that increasing the $w_{\rm \sigma}$ to 5 or 10 improves the results significantly, where surface energies with $R_\Psi > 4.4$ au deviate from DFT for both type~1 and~2 surfaces by less than 5\%, with values of $E^{(1)}_{\rm 001}$ of $2.15$ and $2.33$ J/m$^2$, each. Increasing $w_{\rm \sigma}$ to even larger values yields significantly larger errors, where the $w_{\rm \sigma} = 25$ set yields surface energies that are 20-30\% higher for $R_\Psi \le 4.4$ au, with values of $E^{(1)}_{\rm 001}$ between $2.67$ and $3.02$ J/m$^2$. This is again likely due to large errors in the forces for this set, which results in large errors in optimized geometries. 

\begin{figure}
\begin{subfigure}{1.\textwidth}
  \centering
\includegraphics[width=1.\linewidth]{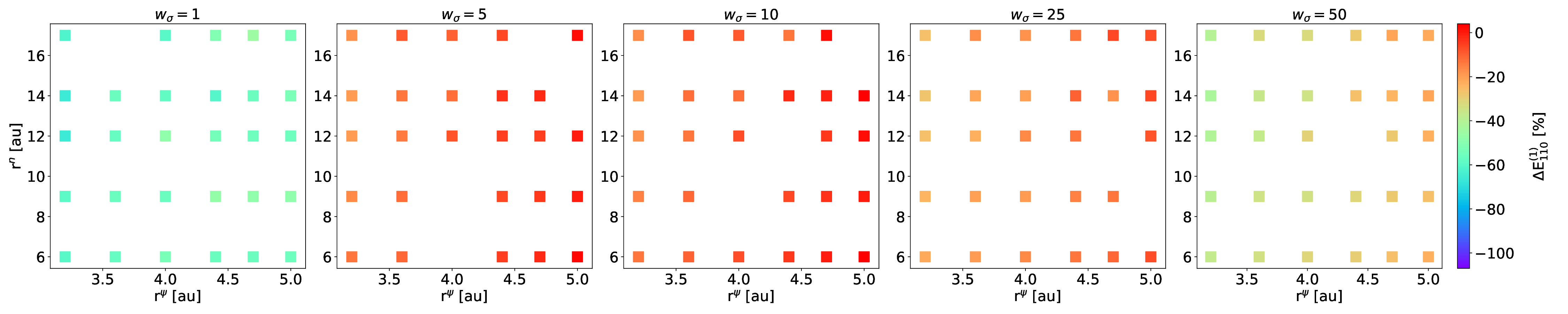}
  \caption{ }
\label{fig:E110_1_12_8}
\end{subfigure}
\begin{subfigure}{1.\textwidth}
  \centering
\includegraphics[width=1.\linewidth]{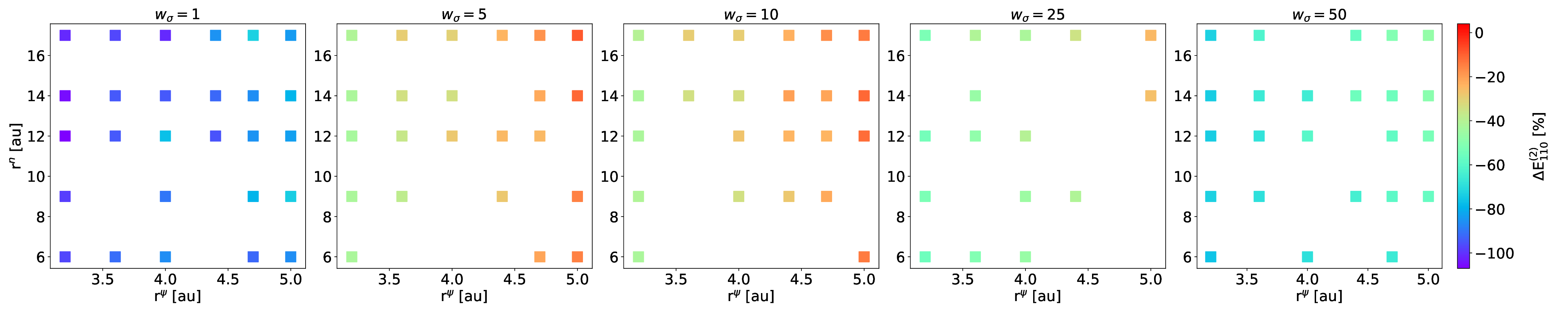}
  \caption{ }
\label{fig:E110_2_16_12}
\end{subfigure}
\caption{Percentage deviation from DFT values of schreibersite surface energy formation for (110) facets of termination type 1 (a) and type 2 (b) obtained for $\{16,12,0\}$ as a function of stress weights.}
\label{fig:s110_energy}
\end{figure}

In contrast, the effect of different confining radii on the (110) surface energy was a bit more difficult to characterize (Fig.~\ref{fig:s110_energy}), as our set of DFTB/ChIMES models for the $\{16,12,0\}$ set were more likely to be overfit. This resulted in a lack of convergence of a number of our surface optimizations, as indicated by the lack of data points for some combinations of $\{R_\Psi, R_n\}$. However, we still observe a similar trend for both type~1 and type~2 surfaces, where values of $R_\Psi > 4.0$ and $w_{\rm \sigma} = 5$ or $10$ yielded the smallest deviations from DFT. For example, the DFTB/ChIMES model obtained with $\{R_\Psi = 4.7\ \mathrm{au}, R_n = 9.0\ \mathrm{au}\}$  and $w_{\rm \sigma} = 10$ yielded an error of $\sim3\%$ for the type 1 surface and $22\%$ of the type 2 surface, with values $2.61$ and $1.88$ J/m$^2$, compared to the DFT results of $2.53$ and $1.54$ J/m$^2$, respectively. It is likely the difficulties with convergence for this set would be improved with additional training data, which is the subject of future work.

\begin{figure}
\begin{subfigure}{.48\textwidth}
  \centering
\includegraphics[width=1.\linewidth]{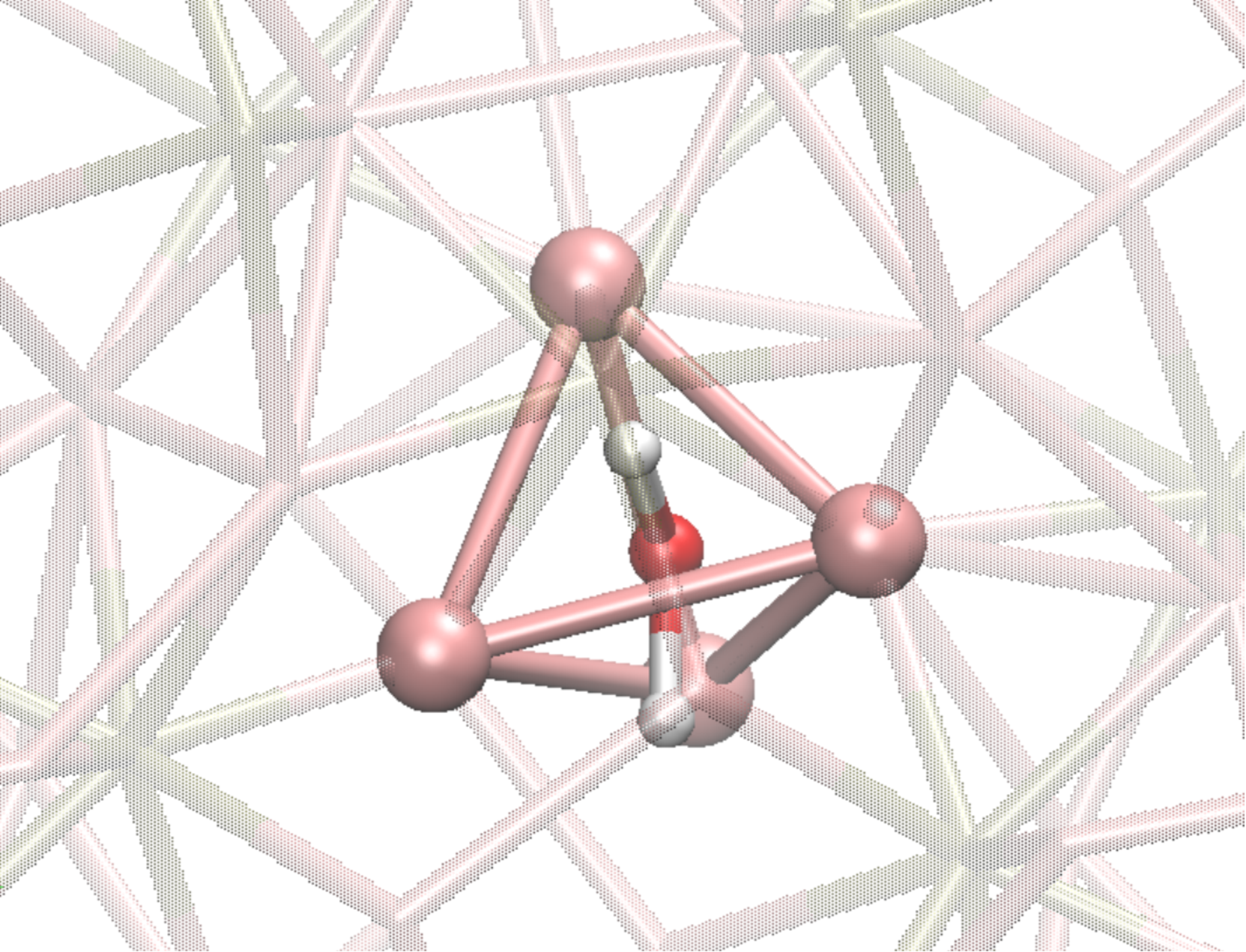}
  \caption{ }
\label{fig:interstitial_1}
\end{subfigure}
\begin{subfigure}{.48\textwidth}
  \centering
\includegraphics[width=1.\linewidth]{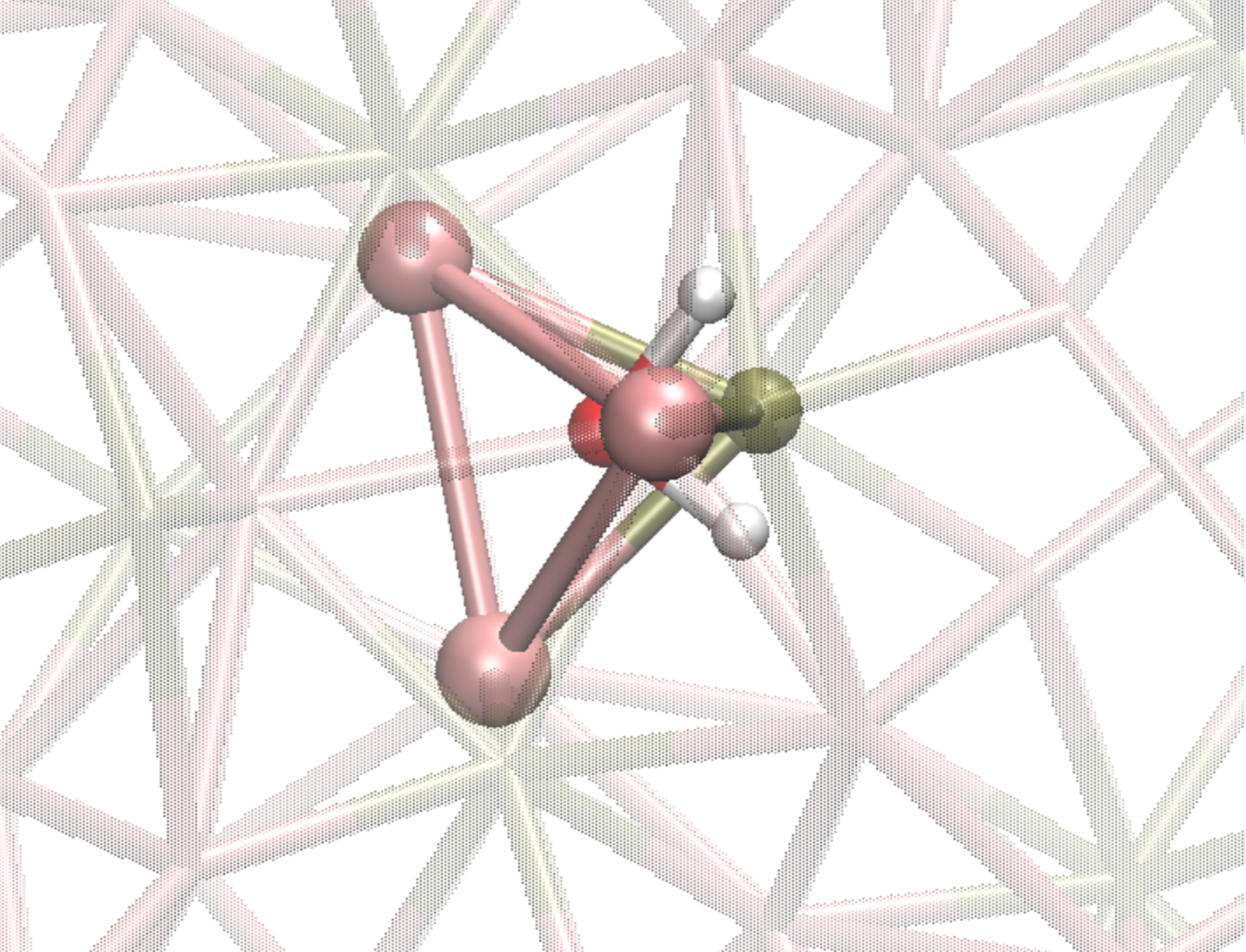}
  \caption{ }
\label{fig:interstitial_2}
\end{subfigure}
\caption{(Bulk interstitial sites for water absorption in schreibersite for (a) site 1 with only Fe atoms, (b) site 2 with one P atom as a vertex of the tetrahedral cage.}
\label{fig:interstitial}
\end{figure}

\paragraph{Bulk interstitial water absorption.}
We now examine the interaction of schreibersite with water molecules in the bulk. We have used the PyMatGen code\cite{pymatgen_code} to identify two distinct interstitial sites within the lattice (Fig.~\ref{fig:interstitial}). Specifically, the first type of interstitial site (site~1) consists of a tetrahedral cage composed of iron atoms. The second type of interstitial site (site~2) is formed by a tetrahedron that incorporates a phosphorus atom at one of its vertices. The formation energies for these sites are calculated via the equation $E_{\rm def}=E_{\rm Fe_3P+H_2O}-(E_{\rm bulk}+E_{\rm H_2O})$, where $E_{\rm Fe_3P+H_2O}$ represents the total energy of the relaxed defected structure, $E_{\rm bulk}$ is the total energy of the bulk schreibersite, and $E_{\rm H_2O}$ denotes the energy of an isolated water molecule in vacuum. In this case, both DFT and all DFTB/ChIMES calculations were initiated from the same initial configurations, which resulted in dissociation of the water molecule upon optimization. For reference, the values computed via DFT are $E_{\rm site 1}=-3.56$ eV and $E_{\rm site 2}=-3.49$ eV, respectively. 

\begin{figure}
\begin{subfigure}{1.\textwidth}
  \centering
\includegraphics[width=1.\linewidth]{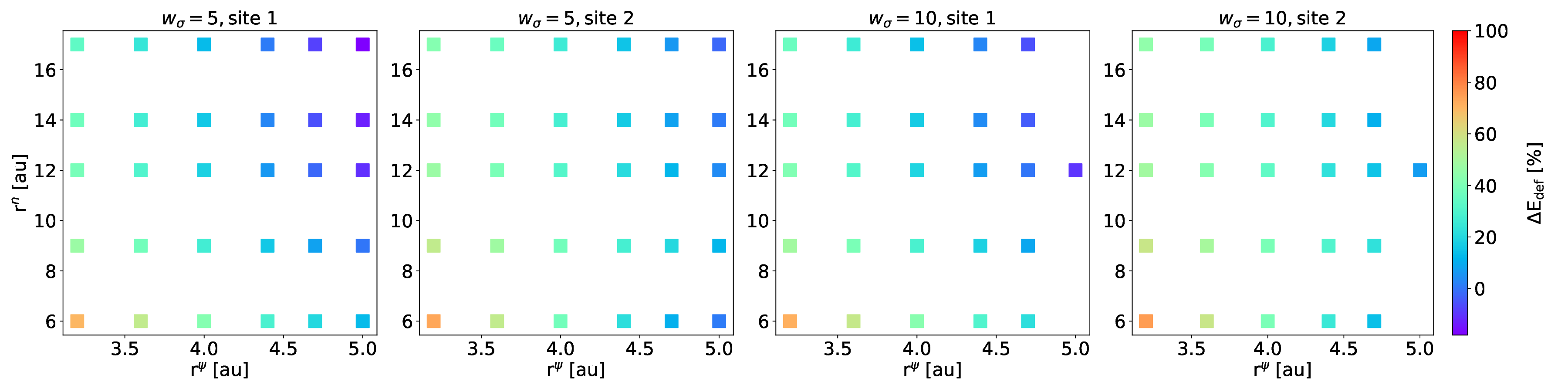}
\end{subfigure}
\caption{Percentage deviation from DFT values of schreibersite bulk water absorption for $\{16,12,0\}$ obtained for both interstitial sites and for $w=5$ and $w=10$.}
\label{fig:bulk_int_energy}
\end{figure}

Our sweep tests of confining radii for the $\{16,12,0\}$/$w_{\rm \sigma} = 5,10$ model sets indicate that in general these dissociative absorption energies are more accurately computed when considering longer confining radii (Fig.~\ref{fig:bulk_int_energy}). For example, errors relative to DFT are minimized for the $w_{\rm \sigma} = 5$ set when $R_\Psi = 5.0$ au and $R_n = 17.0$ au, with values of $-3.49$ and $-3.88$~eV, respectively for site 1 and site 2. However, we observe that smaller confining radii such as $R_\Psi = 4.7$ au and $R_n = 9.0$ au yield relatively small errors as well, with values of $-3.64$ and $-3.22$~eV for site 1 and site 2, respectively, when $w_{\rm \sigma} = 5$. It is worth noting that, although the $w_{\rm \sigma} = 10$ in general also exhibits high accuracy, it tends to be overfit when $R_\Psi = 5.0$ (indicated by the missing points in Fig.~\ref{fig:bulk_int_energy}). The increased error with higher stress tensor weighting is commensurate with the higher error in predicted forces, discussed above. The errors are also generally highest for the $\{R_\Psi = 3.2\ \mathrm{au}, R_n = 6.0\ \mathrm{au}\}$ models due to over compression of the wavefunction and density.

\paragraph{Optimal Hyperparameters.} Our findings thus far provide a broad picture of how the ChIMES hyperparameters and DFTB configuration influence the accuracy of our model. Given our results, we have chosen $\{R_\Psi = 4.7\, \mathrm{au}, R_n = 9.0\, \mathrm{au}\}$, $\{n_{\rm 2B}=16, n_{\rm 3B}=12, n_{\rm 4B}=0\}$ as our optimal set of DFTB/ChIMES hyperparameters, with $w_\sigma=5$ for the weights assigned to the stress tensor components (final results shown in Table~\ref{tab:chimes_summary}). This model yielded the highest overall degree of accuracy while avoiding overfitting for our validation calcuations. This model is able to predict the correct energetic ordering of absorption sites ($E_{\rm site 2} > E_{\rm site 1}$) with a somewhat large $\Delta E$  of 0.41~eV relative to our DFT calculations, though that value likely changes with the choice of functional and basis set. 

\begin{table}[]
\caption{}
\label{tab:chimes_summary}
\begin{tabular}{cccc}
\multicolumn{1}{l}{}                                                 & \multicolumn{1}{l}{} & \begin{tabular}[c]{@{}c@{}}DFTB/\\ ChIMES\end{tabular} & DFT   \\ \hline
\begin{tabular}[c]{@{}c@{}}Lattice\\      constant  [\AA]\end{tabular} & $a$                    & 9.008                                                  & 9.045 \\ \cline{2-4} 
                                                                     & $c$                    & 4.366                                                  & 4.380 \\ \hline
$K_0$ [GPa]                                                             &                      & 170.3                                                    & 173.2 \\ \hline
\multirow{2}{*}{(001) [J/m$^2$]}                                        & type 1               & 2.32                                                   & 2.24  \\ \cline{2-4} 
                                                                     & type 2               & 2.10                                                   & 2.00  \\ \hline
\multirow{2}{*}{(100) [J/m$^2$]}                                        & type 1               & 2.20                                                   & 2.05  \\ \cline{2-4} 
                                                                     & type 2               & 2.05                                                   & 1.95  \\ \hline
\multirow{2}{*}{(110) [J/m$^2$]}                                        & type 1               & 2.61                                                   & 2.53  \\ \cline{2-4} 
                                                                     & type 2               & 1.88                                                   & 1.54  \\ \hline
\multirow{2}{*}{$E_{\rm def}$ [eV]}                                           & Site 1               & -3.64                                                  & -3.78 \\ \cline{2-4} 
                                                                     & Site 2               & -3.22                                                  & -3.49 \\ \hline
\end{tabular}
\end{table}

\subsection{Water Surface Absorption Energies} 
\begin{figure}
\center
\includegraphics[width=.4\linewidth]{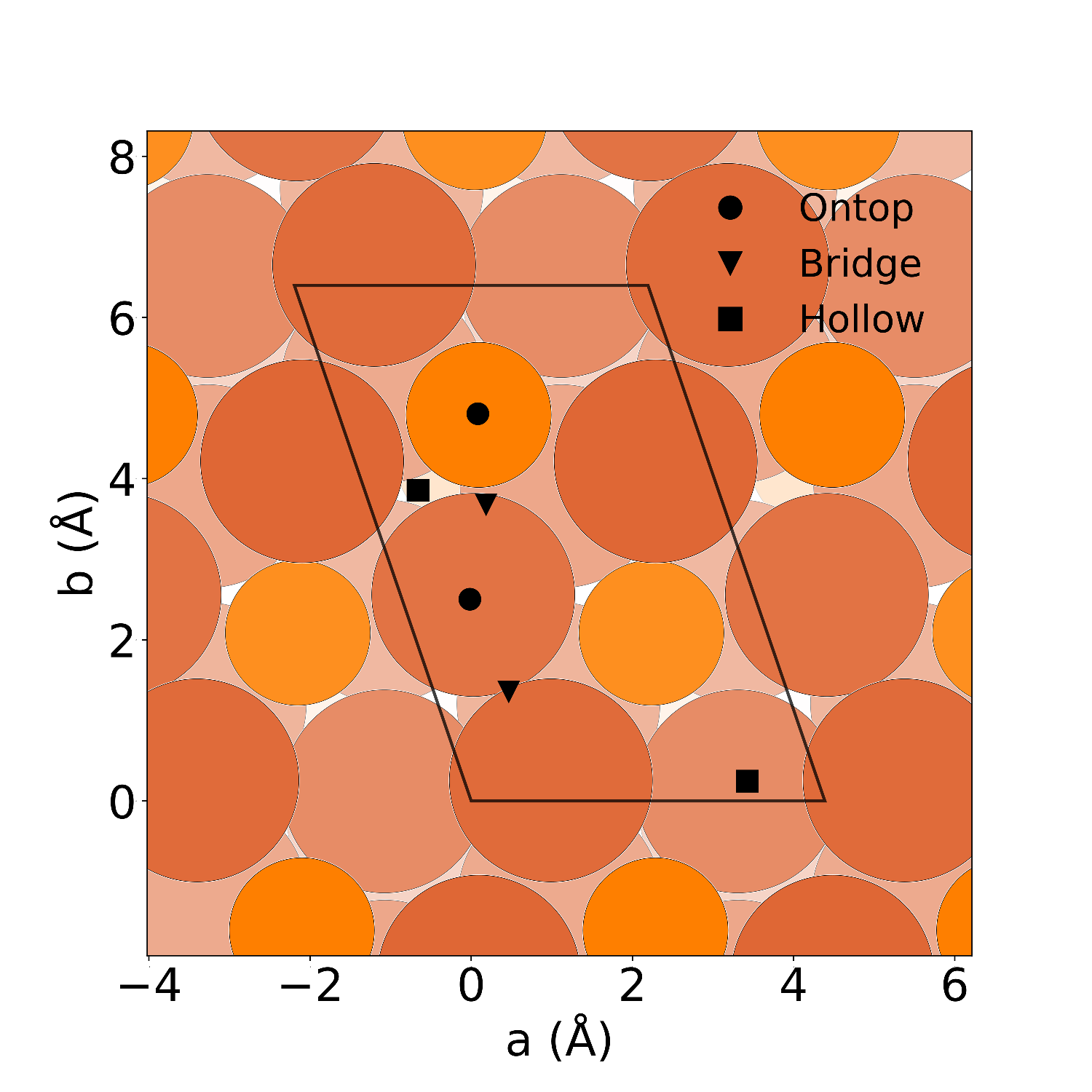}
\caption{Different adsorption sites on $(110)$ surface. In this image, the big brown circles represent iron atoms, while the smaller orange circles stand for the phosphorous atoms. Reproduced from Ref.~\citenum{Dettori_Fe3p}. Copyright 2022 American Chemical Society.}
\label{fig:absorption_sites}
\end{figure}

We now test our optimal DFTB/ChIMES model for water absorption on the most stable facet, i.e., the $(110)$ type 2 (phosphorus terminated) surface. Similarly to interstitial defects, we defined the adsorption energies as $E_{\rm ads}=E_{\rm slab+H_2O}-(E_{\rm slab}+E_{\rm H_2O})$ where $E_{\rm slab+H_2O}$ is the energy of the Fe$_3$P slab with the adsorbed molecule in its final state, $E_{\rm slab}$ is the energy of the pristine slab, and again $E_{\rm H_2O}$ is the energy of an isolated water molecule in vacuum. In our previous work\cite{Dettori_Fe3p}, we identified three different adsorption sites, i.e., `ontop', `bridge' and `hollow' (Fig.~\ref{fig:absorption_sites}). Each type of site can involve different combinations of Fe or P atoms, resulting in two different site types for each classification. We thus validate our model against results for ontop sites on an iron or on a phosphorous atom, bridge sites between two iron atoms or one iron and one phosphorous, and two different hollow sites, with or without phosphorus, depending on the packing of the surface atoms. In addition, the computed adsorption energies for each site depend on the initial orientation of the water molecule, i.e., whether one of the hydrogen atoms from the water molecule is pointing up (`$u$') into the vacuum, down into the slab (`$d$'), or the H$_2$O molecular plane is parallel to the schreibersite surface (`$p$').\cite{Pantaleone2022,Dettori_Fe3p}. Hence, each adsorption site is examined for its $u,p, \mathrm{and}\, d$ dependence as well, for a total of 18 adsorption energies (Table~\ref{tab:Eads}). 

\begin{table}[]
\caption{Adsorption energies for the different sites shown in Fig.~\ref{fig:absorption_sites}, and for the initial molecular configurations investigated here. We note that `p' stands for planar, `u' stands for upward and `d' for downward. Values for $u, p, \mathrm{and}\, d$ that are nearly isoenergetic indicate optimization to a similar final water molecule geometry. }
\label{tab:Eads}
\begin{tabular}{ccccc}
                        &                                 &     & \begin{tabular}[c]{@{}c@{}}DFTB/\\      ChIMES\end{tabular} & DFT    \\ \hline
\multirow{6}{*}{Bridge} & \multirow{3}{*}{Fe - Fe}        & (p) & -0.649                                                      & -0.429 \\
                        &                                 & (u) & -0.694                                                      & -0.444 \\
                        &                                 & (d) & -0.693                                                      & -0.445 \\ \cline{2-5} 
                        & \multirow{3}{*}{Fe   - P}       & (p) & -0.152                                                      & -0.211 \\
                        &                                 & (u) & -0.156                                                      & -0.168 \\
                        &                                 & (d) & -0.156                                                      & -0.184 \\ \hline
\multirow{6}{*}{Hollow} & \multirow{3}{*}{Fe - Fe - P}    & (p) & -0.710                                                      & -0.331 \\
                        &                                 & (u) & -0.710                                                      & -0.338 \\
                        &                                 & (d) & -0.728                                                      & -0.334 \\ \cline{2-5} 
                        & \multirow{3}{*}{Fe   - Fe - Fe} & (p) & -0.188                                                      & -0.195 \\
                        &                                 & (u) & -0.729                                                      & -0.484 \\
                        &                                 & (d) & -0.725                                                      & -0.484 \\ \hline
\multirow{6}{*}{Ontop}  & \multirow{3}{*}{Fe}             & (p) & -0.799                                                      & -0.227 \\
                        &                                 & (u) & -0.720                                                      & -0.269 \\
                        &                                 & (d) & -0.702                                                      & -0.268 \\ \cline{2-5} 
                        & \multirow{3}{*}{P}              & (p) & -0.145                                                      & -0.173 \\
                        &                                 & (u) & -0.131                                                      & -0.162 \\
                        &                                 & (d) & -0.125                                                      & -0.164 \\ \hline
\end{tabular}
\end{table}

We observe that the strength of the adsorption energies involving iron atoms are generally overestimated by DFTB/ChIMES relative to DFT, yielding values that are $\sim$0.22-0.25~eV too negative for the Fe-Fe bridge site, $\sim$0.38-0.40~eV too negative for the Fe-Fe-P hollow site, and 0.43-0.57~eV too negative for the Fe ontop site. One exception is the Fe-Fe-Fe $p$ hollow site, where our model agrees with DFT within 0.07~eV. 

In contrast, the adsorption energies from DFTB/ChIMES predominantly involving phosphorus atoms tend to be more accurate, where the Fe-P bridge site results differ from DFT by less than 0.05~eV and the P ontop site by less than 0.04~eV. These discrepancies in the Fe results are likely due in part to the fact that the Fe-O and Fe-H pairwise repulsive energies were determined by matching to the trans3d parameter set, \red{which were optimized for molecular data, whereas inclusion of periodic solids as well as chemical surface interactions can require additional optimizations.}\cite{Goldman2021}  However, the mio-1-1 parameters used for O, H, and P interactions were optimized to a properties of a relatively large series different molecules, yielding improved surface-molecule interactions. 

\subsection{Magentization as a Function of Pressure}

\begin{figure}
  \centering
\includegraphics[width=.6\linewidth]{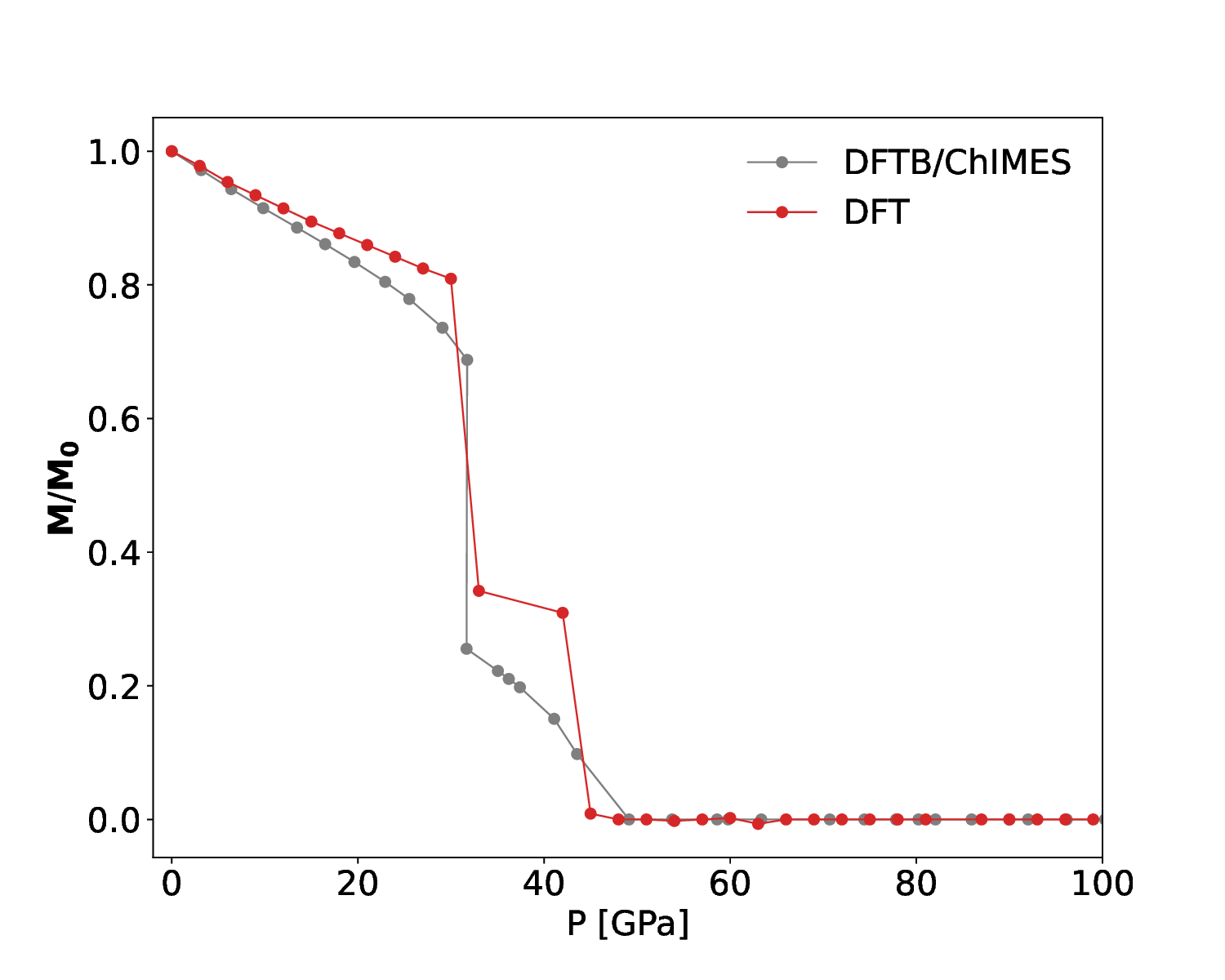}
\caption{Normalized magnetization of Fe$_3$P as a function of external pressure.}
\label{fig:magnetization}
\end{figure}

Given one of our objectives of developing a model suitable for high-pressure impact studies, we consider the variation of magnetization in response to external pressure. Here, we have computed the relative magnetization in scheibersite as it is compressed at zero Kelvin from zero to 100~GPa (Fig.~\ref{fig:magnetization}). Results from DFTB/ChIMES and DFT are very similar throughout the entire pressure range, with a gradual decrease up to 30~GPa, where we observe a rapid drop that corresponds to a potential first-order phase transition.\cite{Scott,Gu,Scott2008} This diminishing trend culminates in a non-magnetic state at $P>50$ GPa in both DFT and DFTB/ChIMES, where the iron atoms have largely lost their magnetic moments, as previously noted by high-pressure experiments.\cite{Scott2008}

\subsection{\red{Phonon Density of States}}

\begin{figure}
  \centering
\includegraphics[width=.6\linewidth]{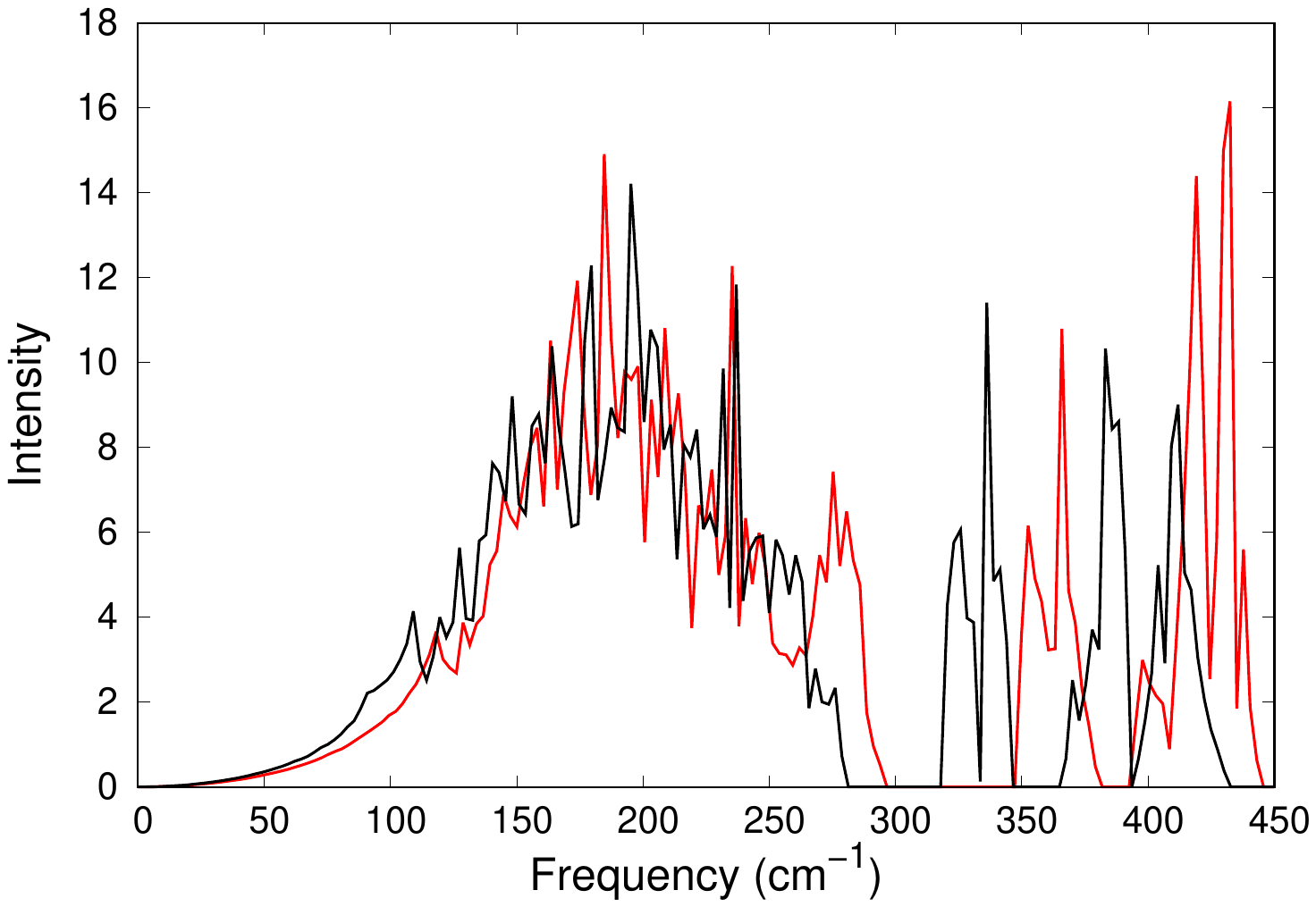}
\caption{\red{Phonon density of states, with results from DFTB/ChIMES in red and those from DFT in black.}}
\label{fig:phonon_dos}
\end{figure}

\red{We now examine the accuracy of the vibrational frequency from DFTB/ChIMES by computing the total phonon density of states (Fig.~\ref{fig:phonon_dos}). These were computed from a zero-temperature optimized primitive cell with the phonopy code,\cite{phonopy-phono3py-JPCM,phonopy-phono3py-JPSJ} using a $2 \times 2 \times 2$ supercell and a $31 \times 31\times 31$ q-point (wavevector) mesh. Overall, we observe reasonable agreement between DFTB/ChIMES and DFT. The manifold of vibrational states between zero and $\sim$300~cm$^{-1}$ shows close agreement up to 250~cm$^{-1}$, though DFTB/ChIMES shows an enhanced density of states between 250-300~cm$^{-1}$ that is not present in the DFT results. Both DFT and DFTB/ChIMES exhibit a doublet above 300~cm$^{-1}$, though the DFTB/ChIMES result occurs $\sim$50~cm$^{-1}$ higher. The remaining manifold of peaks from DFT occurs roughly between 360--430~cm$^{-1}$, and is somewhat different in shape compared to the DFTB/ChIMES result, which spans approximately 390--445~cm$^{-1}$. These deviations are relatively in magnitude, given that a difference of 50~cm$^{-1}$, for example, corresponds to a change in energy of $\sim 6 \times 10^{-4}$~eV.}

\subsection{Molecular Dynamics}
In order to test the dynamics of our model under ambient and extreme conditions, we have performed molecular dynamics (MD) calculations using DFTB/ChIMES at pressures and temperatures from ambient up to 50~GPa and 1500~K. Both DFTB/ChIMES and DFT $NVT$ simulations were initiated using an optimized 32-atom schreibersite bulk simulation cell, run for 5~ps and using a timestep of $1$ fs. Here, we have used a $2\times 2\times 2$ $\mathbf{k}$-point Monkhorst-Pack grid for both sets of simulations. 

\begin{figure}
\begin{subfigure}{.6\textwidth}
  \centering
\includegraphics[width=1.\linewidth]{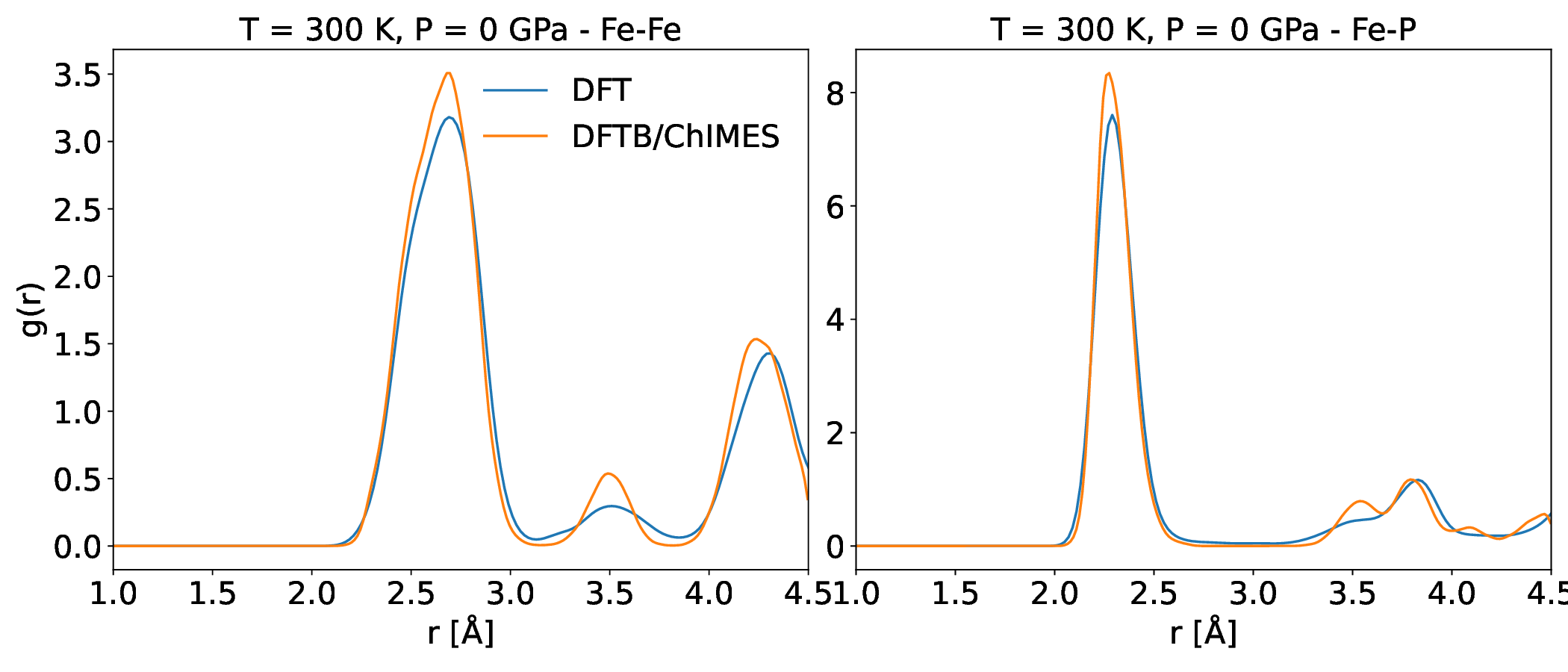}
  \caption{ }
\label{fig:T300_P0GPa}
\end{subfigure}
\begin{subfigure}{.6\textwidth}
  \centering
\includegraphics[width=1.\linewidth]{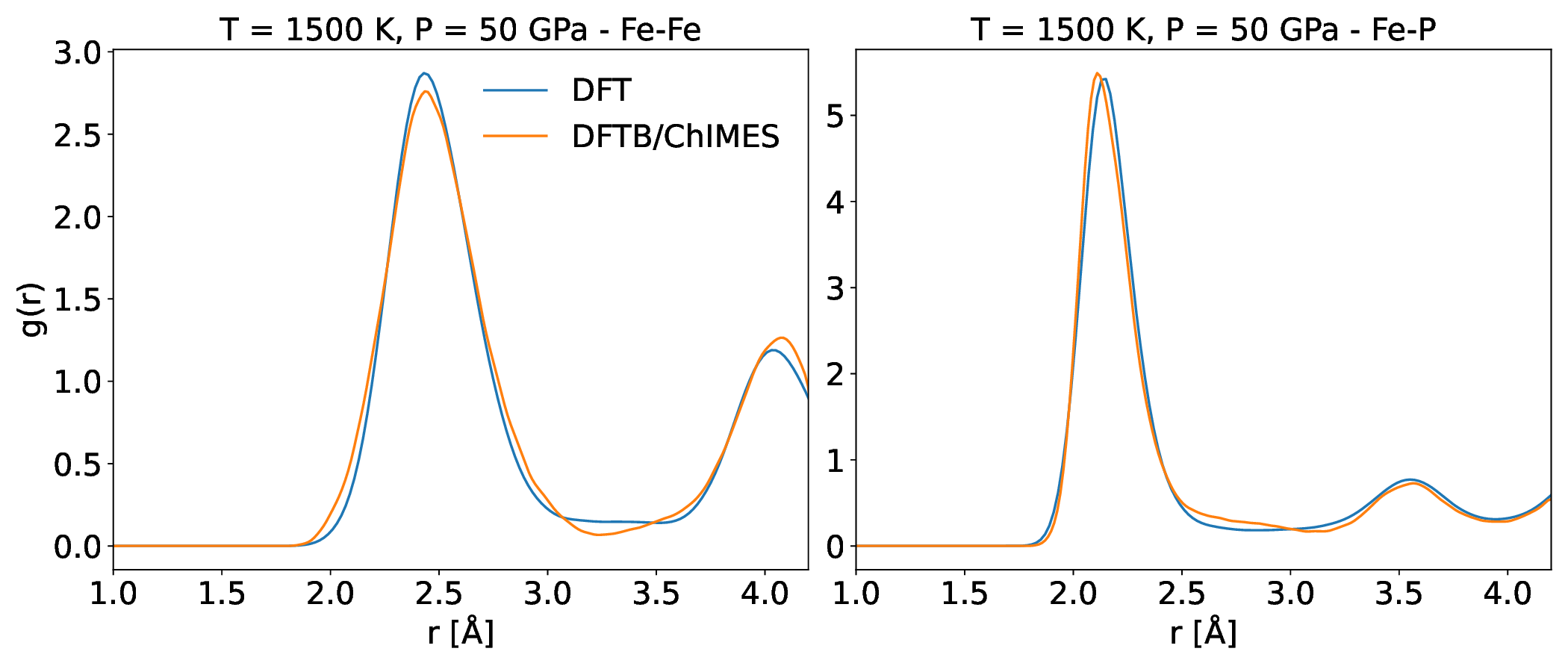}
  \caption{ }
\label{fig:T1500_P50GPa}
\end{subfigure}
\caption{Radial distribution functions obtained for Fe-Fe and Fe-P interatomic distances in schreibersite, as a function of temperature and pressure. The top row (a) shows results for $T=300$K and $P=0$~GPa, while the bottom row (b) corresponds to $T=1500$K and $P=50$~GPa.}
\label{fig:MD_runs}
\end{figure}

Our results for the Fe-Fe and Fe-P RDFs indicate overall very good agreement between the DFTB/ChIMES and DFT datasets (Fig.~\ref{fig:MD_runs}), where we have excluded the P-P RDFs for the sake of brevity. Simulations under ambient conditions (Fig.~\ref{fig:T300_P0GPa}) indicate that DFTB/ChIMES can effectively capture the first coordination shells, where we observe a close match with DFT for the first peak in both the Fe-Fe and Fe-P RDFs. DFTB/ChIMES reveals a slightly less structured atomic distribution for the second and third coordination shells, though the differences are relatively small. At elevated \red{temperatures and pressures (Fig.~\ref{fig:T1500_P50GPa})}, we observe strong agreement with DFT for all solvation shells captured within our simulations. Notably, configurations at $P=50$ GPa were absent from our dataset, indicative of the ability of our model to accurately model state points beyond our training regime. 

\subsection{Transferability of the model}
Finally, we examine the transferability of our DFTB/ChIMES model through lattice optimizations of materials other than Fe$_3$P schreibersite. All optimizations were performed using the material unit cells and the DFTB$+$ convergence criteria discussed above. We first performed calculations on two other iron phosphide structures, FeP and Fe$_2$P. Iron monophosphide (FeP) belongs to a family of transition-metal monopnictides, crystallizing in the MnP structure type (orthorhombic, space group Pnma). Recently, FeP has garnered attention due to its distinctive magnetic and superconducting properties, including the formation of a helical magnetic structure\cite{Chernyavskii2020}. Di-iron phosphide (Fe$_2$), has been investigated as an anode material in sodium-ion batteries.\cite{Huang2016} Fe$_2$P crystallizes in a hexagonal crystal structure that can be described as layers of Fe atoms in a hexagonal close-packed arrangement with P atoms occupying half of the octahedral interstices. 

\begin{table}[]
\caption{Comparison between DFTB/ChIMES, DFT, and experiments for the FeP and Fe$_2$P lattice constants.}
\label{tab:FeP_structures}
\begin{tabular}{ccccccc}
      & \multicolumn{3}{c}{FeP}                                             & \multicolumn{3}{c}{Fe$_2$P} \\ \cline{2-7} 
      & \begin{tabular}[c]{@{}c@{}}DFTB/\\      ChIMES\end{tabular} & DFT   & expt\cite{Chernyavskii2020}  & \begin{tabular}[c]{@{}c@{}}DFTB/\\      ChIMES\end{tabular} & DFT                    & expt\cite{Dera2008}                   \\ \hline
a [\AA] & 3.340                                                       & 3.034 & 3.089 & \multirow{2}{*}{5.828}                                      & \multirow{2}{*}{5.802} & \multirow{2}{*}{5.876} \\ \cline{1-4}
b [\AA] & 5.356                                                       & 5.133 & 5.174 &                                                             &                        &                        \\ \hline
c [\AA] & 5.745                                                       & 5.694 & 5.772 & 3.340                                                       & 3.408                  & 3.449                  \\ \hline
\end{tabular}
\end{table}

Our optimization results for these two iron phosphides are summarized in Table~\ref{tab:FeP_structures}. The $b$ and $c$ lattice constants for FeP compare very well with DFT and experiment, with errors of $\sim$1\% and 4\%, respectively. The $a$ lattice constant differs by roughly $10\%$, though DFTB/ChIMES is able to retain the orthorhombic primitive cell. The Fe$_2$P optimization yields a somewhat closer comparison with DFT and experiment overall, where both $a$ and $c$ lattice constants agree within $\sim$1-2\%.

We have also performed transferability studies on three allotropic forms of iron metal, including $\alpha$-Fe, $\gamma$-Fe, and $\epsilon$-Fe. Ferrite, or $\alpha$-iron, is the ground state and corresponds to a ferromagnetic body-centered cubic (bcc) structure that remains stable up to 1185 K. Beyond this temperature but at low pressure, $\alpha$-Fe transitions to the face-centered cubic (fcc) $\gamma$-Fe structure, also named austenite. Under pressures exceeding approximately 10-13 GPa and temperatures up to around 700 K, $\alpha$-Fe adopts the hexagonal close-packed (hcp) $\epsilon$-iron structure, also called hexaferrum\cite{Mathon2004}. The results of our calculations indicate strong agreement for the lattice constants of all three phases, with agreement of $\lesssim 1\%$ for both $\alpha$-Fe and $\gamma$-Fe. The agreement for $\epsilon$-Fe is also reasonable, where DFTB/ChIMES differs by $\sim$1\% or the $a$ lattice constant, and by $\sim$4\% for the $c$ lattice constant (Table~\ref{tab:Fe_structures}). However, we observe some disagreement with DFT for the energetic ordering of the allotropes. In this case, DFT yields an energetic ordering of $\alpha$-Fe followed by $\epsilon$-Fe and $\gamma$-Fe, whereas DFTB/ChIMES yields $\epsilon$-Fe as the most stable phase. However, $\epsilon$-Fe is nearly isoenergetic with $\alpha$-Fe for both sets of calculations, and the energetic difference between $\alpha$-Fe and $\gamma$-Fe is similar, with a value of 0.13~eV from DFTB/ChIMES and 0.09~eV from DFT.
Regardless, our results indicate the relatively strong transferability of DFTB/ChIMES to Fe/P-containing materials, given that none of these materials were present in our training set. 

\begin{table}[]
\caption{Comparison of the DFTB/ChIMES with DFT and experimental values of the different iron allotropes' lattice constants and energy per atom. In particular, the energetic ordering is given with respect to the most stable structures, which has a 0.0 eV value in the table.}
\label{tab:Fe_structures}
\begin{tabular}{cccccccccc}
\cline{2-10}
                       & \multicolumn{3}{c}{$\alpha-$Fe}                                                                                  & \multicolumn{3}{c}{$\gamma-$Fe}                                                                                  & \multicolumn{3}{c}{$\epsilon-$Fe}                                                                            \\ \cline{2-10} 
                       & \begin{tabular}[c]{@{}c@{}}DFTB/\\      ChIMES\end{tabular} & DFT                    & expt\cite{Greenwood1996}                   & \begin{tabular}[c]{@{}c@{}}DFTB/\\      ChIMES\end{tabular} & DFT                   & expt\cite{SEKI2005}                    & \begin{tabular}[c]{@{}c@{}}DFTB/\\      ChIMES\end{tabular} & DFT                    & expt\cite{Takahashi1968}               \\ \hline
\multirow{2}{*}{a [\AA]} & \multirow{3}{*}{2.827}                                      & \multirow{3}{*}{2.831} & \multirow{3}{*}{2.856} & \multirow{3}{*}{3.647}                                      & \multirow{3}{*}{3.630} & \multirow{3}{*}{3.6519} & \multirow{2}{*}{2.464}                                      & \multirow{2}{*}{2.484} & \multirow{2}{*}{2.461} \\
                       &                                                             &                        &                        &                                                             &                       &                         &                                                             &                        &                    \\ \cline{1-1} \cline{8-10} 
c [\AA]                  &                                                             &                        &                        &                                                             &                       &                         & 3.715                                                       & 3.898                  & 3.952                  \\
\cline{1-1} \cline{8-10} 
$\rm \Delta E$ [eV]                  &        0.022                                                     &            0.000            &                        &               0.157                                              &         0.093              &                         & 0.000                                                      & 0.072                  &                  \\ \hline
\end{tabular}
\end{table}

\section{Conclusions}

In this work, we have determined an optimal DFTB/ChIMES model in order to conduct quantum simulations of chemical reactivity due to the meteoritic mineral schreibersite. We have used our workflow to optimize the DFTB semi-empirical Hamiltonian and overlap matrices, while simultaneously determining a many-body ChIMES repulsive energy. The advantage to our workflow stems from the rapidity with which the linear ChIMES parameters can be determined for each set of quantum interactions, as well as the relatively small data requirements for our repulsive energy optimizations. Our DFTB/ChIMES workflow can be performed with a semi-automated approach, allowing for a rapid down selection of DFTB confining radii, which can be cumbersome to optimize, otherwise.

Our DFTB/ChIMES model yields accurate results for the validation of our study, indicative of a robust model for simulations of Fe$_3$P degradation due to planetary impact. This included a wide diversity of data, such as the scheribersite lattice constant, bulk modulus, and low-index surface energies. We also compute relatively accurate results for bulk water absorption and water surface adsorption, though these results could be improved further by including these types of systems in our optimization workflow. Regardless, DFTB/ChIMES can determine a highly accurate representation of the magnetization as a function of increasing pressure, where results compare closely with DFT in terms of a schreibersite phase transition at approximately 30~GPa and the absence of magnetization above 50~GPa. Furthermore, our model is shown to be robust for molecular dynamics simulations under extreme pressures and temperatures. Finally. we show that our model shows transferability to different materials, including iron monophosphide, di-iron phosphide, and three allotropes of iron metal. Future work can involve further refinement of our Fe/P DFTB model by including training data from additional materials, phases, and thermodynamic state points. Ultimately, our model will help with simulations of prebiotic synthesis, where the role of aqueous schreibersite systems in phosphate production is difficult to discern, and there is subsequently a need for quantum simulations for predictions of chemical and physical properties to help elucidate future experiments.

\begin{acknowledgement}
This work was performed under the auspices of the U.S. Department of Energy by Lawrence Livermore National Laboratory under Contract DE-AC52-07NA27344. The authors gratefully thank the Exobiology Program Element NNH20ZDA001N-EXO for support (proposal \#20-EXO20-0149). 
\end{acknowledgement}

\begin{suppinfo}

Additional validation data for all sets of confining radii and ChIMES repulsive energies included in our study.

\end{suppinfo}

\section{\red{Data Availability}}

\red{The SKF interaction files and the ChIMES parameter files for our optimal DFTB/ChIMES model are available in the following public repository: https://github.com/rdettori/dftb\_FeP\_Erep (last accessed November 6, 2024).}

\newpage
\bibliography{acs-achemso}

\providecommand{\latin}[1]{#1}
\providecommand*\mcitethebibliography{\thebibliography}
\csname @ifundefined\endcsname{endmcitethebibliography}
  {\let\endmcitethebibliography\endthebibliography}{}
\begin{mcitethebibliography}{84}
\providecommand*\natexlab[1]{#1}
\providecommand*\mciteSetBstSublistMode[1]{}
\providecommand*\mciteSetBstMaxWidthForm[2]{}
\providecommand*\mciteBstWouldAddEndPuncttrue
  {\def\EndOfBibitem{\unskip.}}
\providecommand*\mciteBstWouldAddEndPunctfalse
  {\let\EndOfBibitem\relax}
\providecommand*\mciteSetBstMidEndSepPunct[3]{}
\providecommand*\mciteSetBstSublistLabelBeginEnd[3]{}
\providecommand*\EndOfBibitem{}
\mciteSetBstSublistMode{f}
\mciteSetBstMaxWidthForm{subitem}{(\alph{mcitesubitemcount})}
\mciteSetBstSublistLabelBeginEnd
  {\mcitemaxwidthsubitemform\space}
  {\relax}
  {\relax}

\bibitem[Pasek \latin{et~al.}(2017)Pasek, Gull, and Herschy]{Pasek2017}
Pasek,~M.~A.; Gull,~M.; Herschy,~B. Phosphorylation on the early earth.
  \emph{Chemical Geology} \textbf{2017}, \emph{475}, 149 -- 170\relax
\mciteBstWouldAddEndPuncttrue
\mciteSetBstMidEndSepPunct{\mcitedefaultmidpunct}
{\mcitedefaultendpunct}{\mcitedefaultseppunct}\relax
\EndOfBibitem
\bibitem[Gulick(1955)]{Gulick}
Gulick,~A. Phosphorus as a factor in the origin of life. \emph{American
  Scientist} \textbf{1955}, \emph{43}, 479--489\relax
\mciteBstWouldAddEndPuncttrue
\mciteSetBstMidEndSepPunct{\mcitedefaultmidpunct}
{\mcitedefaultendpunct}{\mcitedefaultseppunct}\relax
\EndOfBibitem
\bibitem[Pasek and Kee(2011)Pasek, and Kee]{Pasek2011}
Pasek,~M.~A.; Kee,~T.~P. \emph{Origins of Life: The Primal Self-Organization};
  Springer Berlin Heidelberg, 2011; pp 57--84\relax
\mciteBstWouldAddEndPuncttrue
\mciteSetBstMidEndSepPunct{\mcitedefaultmidpunct}
{\mcitedefaultendpunct}{\mcitedefaultseppunct}\relax
\EndOfBibitem
\bibitem[Schwartz(2006)]{Schwartz2006}
Schwartz,~A.~W. Phosphorus in prebiotic chemistry. \emph{Philos. Trans. R. Soc.
  Lond. B Biol. Sci.} \textbf{2006}, \emph{361}, 1743--1749\relax
\mciteBstWouldAddEndPuncttrue
\mciteSetBstMidEndSepPunct{\mcitedefaultmidpunct}
{\mcitedefaultendpunct}{\mcitedefaultseppunct}\relax
\EndOfBibitem
\bibitem[Pasek(2017)]{Pasek2017b}
Pasek,~M.~A. Schreibersite on the early Earth: Scenarios for prebiotic
  phosphorylation. \emph{Geoscience Frontiers} \textbf{2017}, \emph{8},
  329--335\relax
\mciteBstWouldAddEndPuncttrue
\mciteSetBstMidEndSepPunct{\mcitedefaultmidpunct}
{\mcitedefaultendpunct}{\mcitedefaultseppunct}\relax
\EndOfBibitem
\bibitem[Pasek and Lauretta(2005)Pasek, and Lauretta]{Pasek2005}
Pasek,~M.~A.; Lauretta,~D.~A. Aqueous Corrosion of Phosphide Minerals From Iron
  Meteorites: A Highly Reactive Source of Prebiotic Phosphorus on the Surface
  of the Early Earth. \emph{Astrobiology} \textbf{2005}, \emph{5},
  515--535\relax
\mciteBstWouldAddEndPuncttrue
\mciteSetBstMidEndSepPunct{\mcitedefaultmidpunct}
{\mcitedefaultendpunct}{\mcitedefaultseppunct}\relax
\EndOfBibitem
\bibitem[Pasek \latin{et~al.}(2008)Pasek, Kee, Bryant, Pavlov, and
  Lunine]{Pasek2008}
Pasek,~M.; Kee,~T.; Bryant,~D.; Pavlov,~A.; Lunine,~J. Production of
  Potentially Prebiotic Condensed Phosphates by Phosphorus Redox Chemistry.
  \emph{Angewandte Chemie International Edition} \textbf{2008}, \emph{47},
  7918--7920\relax
\mciteBstWouldAddEndPuncttrue
\mciteSetBstMidEndSepPunct{\mcitedefaultmidpunct}
{\mcitedefaultendpunct}{\mcitedefaultseppunct}\relax
\EndOfBibitem
\bibitem[Gull(2014)]{Gull2014}
Gull,~M. Prebiotic Phosphorylation Reactions on the Early Earth.
  \emph{Challenges} \textbf{2014}, \emph{5}, 193--212\relax
\mciteBstWouldAddEndPuncttrue
\mciteSetBstMidEndSepPunct{\mcitedefaultmidpunct}
{\mcitedefaultendpunct}{\mcitedefaultseppunct}\relax
\EndOfBibitem
\bibitem[Gull \latin{et~al.}(2015)Gull, Mojica, Fern\'andez, Gaul, Orlando,
  Liotta, and Pasek]{Gull2015}
Gull,~M.; Mojica,~M.~A.; Fern\'andez,~F.~M.; Gaul,~D.~A.; Orlando,~T.~M.;
  Liotta,~C.~L.; Pasek,~M.~A. Nucleoside phosphorylation by the mineral
  schreibersite. \emph{Scientific Reports} \textbf{2015}, \emph{4}, 17198\relax
\mciteBstWouldAddEndPuncttrue
\mciteSetBstMidEndSepPunct{\mcitedefaultmidpunct}
{\mcitedefaultendpunct}{\mcitedefaultseppunct}\relax
\EndOfBibitem
\bibitem[Cruz \latin{et~al.}(2016)Cruz, Qasim, Abbott-Lyon, Pirim, McKee,
  Orlando, Gull, Lindsaya, and Pasek]{Cruz2016}
Cruz,~N. L.~L.; Qasim,~D.; Abbott-Lyon,~H.; Pirim,~C.; McKee,~A.~D.;
  Orlando,~T.; Gull,~M.; Lindsaya,~D.; Pasek,~M.~A. The evolution of the
  surface of the mineral schreibersite in prebiotic chemistry. \emph{Phys.
  Chem. Chem. Phys.} \textbf{2016}, \emph{18}, 20160--20167\relax
\mciteBstWouldAddEndPuncttrue
\mciteSetBstMidEndSepPunct{\mcitedefaultmidpunct}
{\mcitedefaultendpunct}{\mcitedefaultseppunct}\relax
\EndOfBibitem
\bibitem[Miller(1953)]{Miller53}
Miller,~S.~L. A production of amino acids under possible primitive Earth
  conditions. \emph{Science} \textbf{1953}, \emph{117}, 528\relax
\mciteBstWouldAddEndPuncttrue
\mciteSetBstMidEndSepPunct{\mcitedefaultmidpunct}
{\mcitedefaultendpunct}{\mcitedefaultseppunct}\relax
\EndOfBibitem
\bibitem[Manaa \latin{et~al.}(2009)Manaa, Reed, Fried, and Goldman]{Manaa09}
Manaa,~M.~R.; Reed,~E.~J.; Fried,~L.~E.; Goldman,~N. Nitrogen-rich heterocycles
  as reactivity retardants in shocked insensitive explosives. \emph{J. Am.
  Chem. Soc.} \textbf{2009}, \emph{131}, 5493--5487\relax
\mciteBstWouldAddEndPuncttrue
\mciteSetBstMidEndSepPunct{\mcitedefaultmidpunct}
{\mcitedefaultendpunct}{\mcitedefaultseppunct}\relax
\EndOfBibitem
\bibitem[Goldman \latin{et~al.}(2010)Goldman, Reed, Fried, Kuo, and
  Maiti]{Goldman10}
Goldman,~N.; Reed,~E.~J.; Fried,~L.~E.; Kuo,~I.-F.~W.; Maiti,~A. Synthesis of
  glycine-containing complexes in impacts of comets on early Earth. \emph{Nat.
  Chem.} \textbf{2010}, \emph{2}, 949--954\relax
\mciteBstWouldAddEndPuncttrue
\mciteSetBstMidEndSepPunct{\mcitedefaultmidpunct}
{\mcitedefaultendpunct}{\mcitedefaultseppunct}\relax
\EndOfBibitem
\bibitem[Goldman and Tamblyn(2013)Goldman, and Tamblyn]{Goldman13}
Goldman,~N.; Tamblyn,~I. Prebiotic chemistry within a simple impacting icy
  mixture. \emph{J. Phys. Chem. A} \textbf{2013}, \emph{117}, 5124 --
  5131\relax
\mciteBstWouldAddEndPuncttrue
\mciteSetBstMidEndSepPunct{\mcitedefaultmidpunct}
{\mcitedefaultendpunct}{\mcitedefaultseppunct}\relax
\EndOfBibitem
\bibitem[Dettori and Goldman(2022)Dettori, and Goldman]{Dettori_Fe3p}
Dettori,~R.; Goldman,~N. First-Principles Surface Characterization and Water
  Adsorption of Fe3P Schreibersite. \emph{ACS Earth and Space Chemistry}
  \textbf{2022}, \emph{6}, 512--520\relax
\mciteBstWouldAddEndPuncttrue
\mciteSetBstMidEndSepPunct{\mcitedefaultmidpunct}
{\mcitedefaultendpunct}{\mcitedefaultseppunct}\relax
\EndOfBibitem
\bibitem[Pantaleone \latin{et~al.}(2021)Pantaleone, Corno, Rimola, Balucani,
  and Ugliengo]{Pantaleone2021}
Pantaleone,~S.; Corno,~M.; Rimola,~A.; Balucani,~N.; Ugliengo,~P. Ab Initio
  Computational Study on Fe$_2$NiP Schreibersite: Bulk and Surface
  Characterization. \emph{ACS Earth and Space Chemistry} \textbf{2021},
  \emph{5}, 1741--1751\relax
\mciteBstWouldAddEndPuncttrue
\mciteSetBstMidEndSepPunct{\mcitedefaultmidpunct}
{\mcitedefaultendpunct}{\mcitedefaultseppunct}\relax
\EndOfBibitem
\bibitem[Pantaleone \latin{et~al.}(2022)Pantaleone, Corno, Rimola, Balucani,
  and Ugliengo]{Pantaleone2022}
Pantaleone,~S.; Corno,~M.; Rimola,~A.; Balucani,~N.; Ugliengo,~P. Water
  Interaction with Fe2NiP Schreibersite (110) Surface: a Quantum Mechanical
  Atomistic Perspective. \emph{The Journal of Physical Chemistry C}
  \textbf{2022}, \emph{126}, 2243--2252\relax
\mciteBstWouldAddEndPuncttrue
\mciteSetBstMidEndSepPunct{\mcitedefaultmidpunct}
{\mcitedefaultendpunct}{\mcitedefaultseppunct}\relax
\EndOfBibitem
\bibitem[Cannella and Goldman(2015)Cannella, and Goldman]{Cannella15}
Cannella,~C.; Goldman,~N. Carbyne fiber synthesis within evaporating metallic
  liquid carbon. \emph{J. Phys. Chem. C} \textbf{2015}, \emph{119},
  21605--21611\relax
\mciteBstWouldAddEndPuncttrue
\mciteSetBstMidEndSepPunct{\mcitedefaultmidpunct}
{\mcitedefaultendpunct}{\mcitedefaultseppunct}\relax
\EndOfBibitem
\bibitem[Lindsey \latin{et~al.}(2021)Lindsey, Bastea, Goldman, and
  Fried]{Lindsey_DNTF_2020}
Lindsey,~R.~K.; Bastea,~S.; Goldman,~N.; Fried,~L.~E. Investigating
  3,4-bis(3-nitrofurazan-4-yl)furoxan detonation with a rapidly tuned density
  functional tight binding model. \emph{J. Chem. Phys.} \textbf{2021},
  \emph{154}, 164115\relax
\mciteBstWouldAddEndPuncttrue
\mciteSetBstMidEndSepPunct{\mcitedefaultmidpunct}
{\mcitedefaultendpunct}{\mcitedefaultseppunct}\relax
\EndOfBibitem
\bibitem[Armstrong \latin{et~al.}(2020)Armstrong, Lindsey, Goldman, Nielsen,
  Stavrou, Fried, Zaug, and Bastea]{Armstrong2020}
Armstrong,~M.~R.; Lindsey,~R.~K.; Goldman,~N.; Nielsen,~M.~H.; Stavrou,~E.;
  Fried,~L.~E.; Zaug,~J.~M.; Bastea,~S. Ultrafast shock synthesis of nanocarbon
  from a liquid precursor. \emph{Nature Communications} \textbf{2020},
  \emph{11}, 353\relax
\mciteBstWouldAddEndPuncttrue
\mciteSetBstMidEndSepPunct{\mcitedefaultmidpunct}
{\mcitedefaultendpunct}{\mcitedefaultseppunct}\relax
\EndOfBibitem
\bibitem[Lindsey \latin{et~al.}(2022)Lindsey, Goldman, Fried, and
  Bastea]{Lindsey_CO_2022}
Lindsey,~R.~K.; Goldman,~N.; Fried,~L.~E.; Bastea,~S. Chemistry-mediated
  Ostwald ripening in carbonrich C/O systems at extreme conditions.
  \emph{Nature Communications} \textbf{2022}, \emph{13}, 1424\relax
\mciteBstWouldAddEndPuncttrue
\mciteSetBstMidEndSepPunct{\mcitedefaultmidpunct}
{\mcitedefaultendpunct}{\mcitedefaultseppunct}\relax
\EndOfBibitem
\bibitem[Nomora \latin{et~al.}(2007)Nomora, Kalia, Nakano, Vashista, van Duin,
  and Goddard]{vanduin07}
Nomora,~K.; Kalia,~R.~K.; Nakano,~A.; Vashista,~P.; van Duin,~A. C.~T.;
  Goddard,~W.~A. Dynamic transition in the shock structure of an energetic
  material. \emph{Phys. Rev. Lett.} \textbf{2007}, \emph{99}, 148303\relax
\mciteBstWouldAddEndPuncttrue
\mciteSetBstMidEndSepPunct{\mcitedefaultmidpunct}
{\mcitedefaultendpunct}{\mcitedefaultseppunct}\relax
\EndOfBibitem
\bibitem[Drautz(2019)]{Drautz_ACE1}
Drautz,~R. Atomic cluster expansion for accurate and transferable interatomic
  potentials. \emph{Phys. Rev. B} \textbf{2019}, \emph{99}, 014104\relax
\mciteBstWouldAddEndPuncttrue
\mciteSetBstMidEndSepPunct{\mcitedefaultmidpunct}
{\mcitedefaultendpunct}{\mcitedefaultseppunct}\relax
\EndOfBibitem
\bibitem[Smith \latin{et~al.}(2019)Smith, Nebgen, Zubatyuk, Lubbers, Devereux,
  Barros, Tretiak, Isayev, and Roitberg]{ANI-1ccx}
Smith,~J.~S.; Nebgen,~B.~T.; Zubatyuk,~R.; Lubbers,~N.; Devereux,~C.;
  Barros,~K.; Tretiak,~S.; Isayev,~O.; Roitberg,~A.~E. Approaching coupled
  cluster accuracy with a general-purpose neural network potential through
  transfer learning. \emph{Nat. Commun.} \textbf{2019}, \emph{10}, 1--8\relax
\mciteBstWouldAddEndPuncttrue
\mciteSetBstMidEndSepPunct{\mcitedefaultmidpunct}
{\mcitedefaultendpunct}{\mcitedefaultseppunct}\relax
\EndOfBibitem
\bibitem[Kov\'acs \latin{et~al.}(2021)Kov\'acs, van~der Oord, Kucera, Allen,
  Cole, Ortner, and Cs\'anyi]{Csanyi_ACE_organic}
Kov\'acs,~D.~P.; van~der Oord,~C.; Kucera,~J.; Allen,~A. E.~A.; Cole,~D.~J.;
  Ortner,~C.; Cs\'anyi,~G. Linear Atomic Cluster Expansion Force Fields for
  Organic Molecules: Beyond RMSE. \emph{J. Chem. Theory. Comput.}
  \textbf{2021}, \emph{17}, 7696--7711\relax
\mciteBstWouldAddEndPuncttrue
\mciteSetBstMidEndSepPunct{\mcitedefaultmidpunct}
{\mcitedefaultendpunct}{\mcitedefaultseppunct}\relax
\EndOfBibitem
\bibitem[Pham \latin{et~al.}(2021)Pham, Lindsey, Fried, and
  Goldman]{Pham_HN3_2021}
Pham,~C.~H.; Lindsey,~R.~K.; Fried,~L.~E.; Goldman,~N. Calculation of the
  detonation state of HN$_3$ with quantum accuracy. \emph{J. Chem. Phys.}
  \textbf{2021}, \emph{153}, 224102\relax
\mciteBstWouldAddEndPuncttrue
\mciteSetBstMidEndSepPunct{\mcitedefaultmidpunct}
{\mcitedefaultendpunct}{\mcitedefaultseppunct}\relax
\EndOfBibitem
\bibitem[Elstner \latin{et~al.}(1998)Elstner, Porezag, Jungnickel, Elsner,
  Haugk, Frauenheim, Suhai, and Seifert]{Elstner}
Elstner,~M.; Porezag,~D.; Jungnickel,~G.; Elsner,~J.; Haugk,~M.;
  Frauenheim,~T.; Suhai,~S.; Seifert,~G. Self-consistent-charge
  density-functional tight-binding method for simulations of complex materials
  properties. \emph{Phys. Rev. B} \textbf{1998}, \emph{58}, 7260--7268\relax
\mciteBstWouldAddEndPuncttrue
\mciteSetBstMidEndSepPunct{\mcitedefaultmidpunct}
{\mcitedefaultendpunct}{\mcitedefaultseppunct}\relax
\EndOfBibitem
\bibitem[Gaus \latin{et~al.}(2011)Gaus, Cui, and Elstner]{Gaus_mioPP}
Gaus,~M.; Cui,~Q.; Elstner,~M. DFTB3: Extension of the Self-Consistent-Charge
  Density-Functional Tight-Binding Method (SCC-DFTB). \emph{Journal of Chemical
  Theory and Computation} \textbf{2011}, \emph{7}, 931--948\relax
\mciteBstWouldAddEndPuncttrue
\mciteSetBstMidEndSepPunct{\mcitedefaultmidpunct}
{\mcitedefaultendpunct}{\mcitedefaultseppunct}\relax
\EndOfBibitem
\bibitem[Hourahine \latin{et~al.}(2020)Hourahine, Aradi, Blum, Bonaf\'e,
  Buccheri, Camacho, Cevallos, Deshaye, Dumitric\u{a}, Dominguez, Ehlert,
  Elstner, van~der Heide, Hermann, Irle, Kranz, K\"{o}hler, Kowalczyk,
  Kuba\v{r}, Lee, Lutsker, Maurer, Min, Mitchell, Negre, Niehaus, Niklasson,
  Page, Pecchia, Penazzi, Persson, \v{R}ezi\'{a}\v{c}, S\'{a}nchez, Sternberg,
  St\"{o}hr, Stuckenberg, Tkatchenko, Yu, and Frauenheim]{DFTB+_current}
Hourahine,~B.; Aradi,~B.; Blum,~V.; Bonaf\'e,~F.; Buccheri,~A.; Camacho,~C.;
  Cevallos,~C.; Deshaye,~M.~Y.; Dumitric\u{a},~T.; Dominguez,~A. \latin{et~al.}
   DFTB+, a software package for efficient approximate density functional
  theory based atomistic simulations. \emph{The Journal of Chemical Physics}
  \textbf{2020}, \emph{152}, 124101\relax
\mciteBstWouldAddEndPuncttrue
\mciteSetBstMidEndSepPunct{\mcitedefaultmidpunct}
{\mcitedefaultendpunct}{\mcitedefaultseppunct}\relax
\EndOfBibitem
\bibitem[Goldman \latin{et~al.}(2015)Goldman, Koziol, and Fried]{Goldman15-fm}
Goldman,~N.; Koziol,~L.; Fried,~L.~E. Using force-matched potentials to improve
  the accuracy of density functional tight binding for reactive conditions.
  \emph{J. Chem. Theory Comput.} \textbf{2015}, \emph{11}, 4530--4535\relax
\mciteBstWouldAddEndPuncttrue
\mciteSetBstMidEndSepPunct{\mcitedefaultmidpunct}
{\mcitedefaultendpunct}{\mcitedefaultseppunct}\relax
\EndOfBibitem
\bibitem[Kandy \latin{et~al.}(2021)Kandy, Wadbro, Aradi, Broqvist, and
  Kullgren]{Kullgren_CCS_2021}
Kandy,~A. K.~A.; Wadbro,~E.; Aradi,~B.; Broqvist,~P.; Kullgren,~J. Curvature
  Constrained Splines for DFTB Repulsive Potential Parametrization. \emph{J.
  Chem. Theory Comput.} \textbf{2021}, \emph{1771-1781}, 21\relax
\mciteBstWouldAddEndPuncttrue
\mciteSetBstMidEndSepPunct{\mcitedefaultmidpunct}
{\mcitedefaultendpunct}{\mcitedefaultseppunct}\relax
\EndOfBibitem
\bibitem[Wengert \latin{et~al.}(2021)Wengert, Cs\'anyi, Reuter, and
  Margraf]{Margraf_GPR_2021}
Wengert,~S.; Cs\'anyi,~G.; Reuter,~K.; Margraf,~J.~T. Data-efficient machine
  learning for molecular crystal structure prediction. \emph{Chem. Sci.}
  \textbf{2021}, \emph{12}, 4536\relax
\mciteBstWouldAddEndPuncttrue
\mciteSetBstMidEndSepPunct{\mcitedefaultmidpunct}
{\mcitedefaultendpunct}{\mcitedefaultseppunct}\relax
\EndOfBibitem
\bibitem[Goldman \latin{et~al.}(2023)Goldman, Fried, Lindsey, Pham, and
  Dettori]{Goldman_SEQM_2023}
Goldman,~N.; Fried,~L.~E.; Lindsey,~R.~K.; Pham,~C.~H.; Dettori,~R. Enhancing
  the accuracy of density functional tight binding models through ChIMES
  many-body interaction potentials. \emph{J. Chem. Phys.} \textbf{2023},
  \emph{158}, 144112\relax
\mciteBstWouldAddEndPuncttrue
\mciteSetBstMidEndSepPunct{\mcitedefaultmidpunct}
{\mcitedefaultendpunct}{\mcitedefaultseppunct}\relax
\EndOfBibitem
\bibitem[St{\"o}hr \latin{et~al.}(2020)St{\"o}hr, Medrano~Sandonas, and
  Tkatchenko]{ISO34_DFTB_DTNN}
St{\"o}hr,~M.; Medrano~Sandonas,~L.; Tkatchenko,~A. Accurate Many-Body
  Repulsive Potentials for Density-Functional Tight Binding from Deep Tensor
  Neural Networks. \emph{J. Phys. Chem. Lett.} \textbf{2020}, \emph{11},
  6835--6843\relax
\mciteBstWouldAddEndPuncttrue
\mciteSetBstMidEndSepPunct{\mcitedefaultmidpunct}
{\mcitedefaultendpunct}{\mcitedefaultseppunct}\relax
\EndOfBibitem
\bibitem[Yang \latin{et~al.}(2008)Yang, Yu, York, Elstner, and Cui]{Yang2008}
Yang,~Y.; Yu,~H.; York,~D.; Elstner,~M.; Cui,~Q. Description of Phosphate
  Hydrolysis Reactions with the Self-Consistent-Charge
  Density-Functional-Tight-Binding ({SCC}-{DFTB}) Theory. 1. Parameterization.
  \emph{Journal of Chemical Theory and Computation} \textbf{2008}, \emph{4},
  2067--2084\relax
\mciteBstWouldAddEndPuncttrue
\mciteSetBstMidEndSepPunct{\mcitedefaultmidpunct}
{\mcitedefaultendpunct}{\mcitedefaultseppunct}\relax
\EndOfBibitem
\bibitem[Niehaus \latin{et~al.}(2001)Niehaus, Elstner, Frauenheim, and
  Suhai]{Niehaus}
Niehaus,~T.; Elstner,~M.; Frauenheim,~T.; Suhai,~S. Application of an
  approximate density-functional method to sulfur containing compounds.
  \emph{Journal of Molecular Structure: THEOCHEM} \textbf{2001}, \emph{541},
  185--194\relax
\mciteBstWouldAddEndPuncttrue
\mciteSetBstMidEndSepPunct{\mcitedefaultmidpunct}
{\mcitedefaultendpunct}{\mcitedefaultseppunct}\relax
\EndOfBibitem
\bibitem[Guimar{\~{a}}es \latin{et~al.}(2007)Guimar{\~{a}}es, Enyashin,
  Frenzel, Heine, Duarte, and Seifert]{Guimares2007}
Guimar{\~{a}}es,~L.; Enyashin,~A.~N.; Frenzel,~J.; Heine,~T.; Duarte,~H.~A.;
  Seifert,~G. Imogolite Nanotubes: Stability, Electronic, and Mechanical
  Properties. \emph{{ACS} Nano} \textbf{2007}, \emph{1}, 362--368\relax
\mciteBstWouldAddEndPuncttrue
\mciteSetBstMidEndSepPunct{\mcitedefaultmidpunct}
{\mcitedefaultendpunct}{\mcitedefaultseppunct}\relax
\EndOfBibitem
\bibitem[Luschtinetz \latin{et~al.}(2008)Luschtinetz, Oliveira, Frenzel,
  Joswig, Seifert, and Duarte]{Luschtinetz2008}
Luschtinetz,~R.; Oliveira,~A.~F.; Frenzel,~J.; Joswig,~J.-O.; Seifert,~G.;
  Duarte,~H.~A. Adsorption of phosphonic and ethylphosphonic acid on aluminum
  oxide surfaces. \emph{Surface Science} \textbf{2008}, \emph{602},
  1347--1359\relax
\mciteBstWouldAddEndPuncttrue
\mciteSetBstMidEndSepPunct{\mcitedefaultmidpunct}
{\mcitedefaultendpunct}{\mcitedefaultseppunct}\relax
\EndOfBibitem
\bibitem[Sieck \latin{et~al.}(2003)Sieck, Frauenheim, and Jackson]{Sieck2003}
Sieck,~A.; Frauenheim,~T.; Jackson,~K.~A. Shape transition of medium-sized
  neutral silicon clusters. \emph{physica status solidi (b)} \textbf{2003},
  \emph{240}, 537--548\relax
\mciteBstWouldAddEndPuncttrue
\mciteSetBstMidEndSepPunct{\mcitedefaultmidpunct}
{\mcitedefaultendpunct}{\mcitedefaultseppunct}\relax
\EndOfBibitem
\bibitem[K\"{o}hler \latin{et~al.}(2005)K\"{o}hler, Seifert, and
  Frauenheim]{Khler2005}
K\"{o}hler,~C.; Seifert,~G.; Frauenheim,~T. Density functional based
  calculations for Fen (n$\le$32). \emph{Chemical Physics} \textbf{2005},
  \emph{309}, 23--31\relax
\mciteBstWouldAddEndPuncttrue
\mciteSetBstMidEndSepPunct{\mcitedefaultmidpunct}
{\mcitedefaultendpunct}{\mcitedefaultseppunct}\relax
\EndOfBibitem
\bibitem[Gaus \latin{et~al.}(2012)Gaus, Goez, and Elstner]{Gaus2012}
Gaus,~M.; Goez,~A.; Elstner,~M. Parametrization and Benchmark of {DFTB}3 for
  Organic Molecules. \emph{Journal of Chemical Theory and Computation}
  \textbf{2012}, \emph{9}, 338--354\relax
\mciteBstWouldAddEndPuncttrue
\mciteSetBstMidEndSepPunct{\mcitedefaultmidpunct}
{\mcitedefaultendpunct}{\mcitedefaultseppunct}\relax
\EndOfBibitem
\bibitem[Kubillus \latin{et~al.}(2014)Kubillus, Kuba{\v{r}}, Gaus,
  {\v{R}}ez{\'{a}}{\v{c}}, and Elstner]{Kubillus2014}
Kubillus,~M.; Kuba{\v{r}},~T.; Gaus,~M.; {\v{R}}ez{\'{a}}{\v{c}},~J.;
  Elstner,~M. Parameterization of the {DFTB}3 Method for Br, Ca, Cl, F, I, K,
  and Na in Organic and Biological Systems. \emph{Journal of Chemical Theory
  and Computation} \textbf{2014}, \emph{11}, 332--342\relax
\mciteBstWouldAddEndPuncttrue
\mciteSetBstMidEndSepPunct{\mcitedefaultmidpunct}
{\mcitedefaultendpunct}{\mcitedefaultseppunct}\relax
\EndOfBibitem
\bibitem[Gaus \latin{et~al.}(2014)Gaus, Lu, Elstner, and Cui]{Gaus2014}
Gaus,~M.; Lu,~X.; Elstner,~M.; Cui,~Q. Parameterization of {DFTB}3/3OB for
  Sulfur and Phosphorus for Chemical and Biological Applications. \emph{Journal
  of Chemical Theory and Computation} \textbf{2014}, \emph{10},
  1518--1537\relax
\mciteBstWouldAddEndPuncttrue
\mciteSetBstMidEndSepPunct{\mcitedefaultmidpunct}
{\mcitedefaultendpunct}{\mcitedefaultseppunct}\relax
\EndOfBibitem
\bibitem[Zheng \latin{et~al.}(2007)Zheng, Witek, Bobadova-Parvanova, Irle,
  Musaev, Prabhakar, Morokuma, Lundberg, Elstner, K\"{o}hler, and
  Frauenheim]{Zheng2007}
Zheng,~G.; Witek,~H.~A.; Bobadova-Parvanova,~P.; Irle,~S.; Musaev,~D.~G.;
  Prabhakar,~R.; Morokuma,~K.; Lundberg,~M.; Elstner,~M.; K\"{o}hler,~C.
  \latin{et~al.}  Parameter Calibration of Transition-Metal Elements for the
  Spin-Polarized Self-Consistent-Charge Density-Functional Tight-Binding
  ({DFTB}) Method:{\hspace{0.167em}} Sc, Ti, Fe, Co, and Ni. \emph{Journal of
  Chemical Theory and Computation} \textbf{2007}, \emph{3}, 1349--1367\relax
\mciteBstWouldAddEndPuncttrue
\mciteSetBstMidEndSepPunct{\mcitedefaultmidpunct}
{\mcitedefaultendpunct}{\mcitedefaultseppunct}\relax
\EndOfBibitem
\bibitem[Lindsey \latin{et~al.}(2020)Lindsey, Goldman, Fried, and
  Bastea]{Lindsey2017}
Lindsey,~R.~K.; Goldman,~N.; Fried,~L.~E.; Bastea,~S. {Many-body reactive force
  field development for carbon condensation in C/O systems under extreme
  conditions}. \emph{The Journal of Chemical Physics} \textbf{2020},
  \emph{153}, 054103\relax
\mciteBstWouldAddEndPuncttrue
\mciteSetBstMidEndSepPunct{\mcitedefaultmidpunct}
{\mcitedefaultendpunct}{\mcitedefaultseppunct}\relax
\EndOfBibitem
\bibitem[Lindsey \latin{et~al.}(2020)Lindsey, Fried, Goldman, and
  Bastea]{Lindsey2020}
Lindsey,~R.~K.; Fried,~L.~E.; Goldman,~N.; Bastea,~S. Active learning for
  robust, high-complexity reactive atomistic simulations. \emph{The Journal of
  Chemical Physics} \textbf{2020}, \emph{153}, 134117\relax
\mciteBstWouldAddEndPuncttrue
\mciteSetBstMidEndSepPunct{\mcitedefaultmidpunct}
{\mcitedefaultendpunct}{\mcitedefaultseppunct}\relax
\EndOfBibitem
\bibitem[Goldman \latin{et~al.}(2023)Goldman, Fried, Lindsey, Pham, and
  Dettori]{Goldman2023}
Goldman,~N.; Fried,~L.~E.; Lindsey,~R.~K.; Pham,~C.~H.; Dettori,~R. Enhancing
  the accuracy of density functional tight binding models through {ChIMES}
  many-body interaction potentials. \emph{The Journal of Chemical Physics}
  \textbf{2023}, \emph{158}, 144112\relax
\mciteBstWouldAddEndPuncttrue
\mciteSetBstMidEndSepPunct{\mcitedefaultmidpunct}
{\mcitedefaultendpunct}{\mcitedefaultseppunct}\relax
\EndOfBibitem
\bibitem[Goldman \latin{et~al.}(2021)Goldman, Kweon, Sadigh, Heo, Lindsey,
  Pham, Fried, Aradi, Holliday, Jeffries, and Wood]{Goldman_TiH2}
Goldman,~N.; Kweon,~K.~E.; Sadigh,~B.; Heo,~T.-W.; Lindsey,~R.~K.; Pham,~C.~H.;
  Fried,~L.~E.; Aradi,~B.; Holliday,~K.; Jeffries,~J.~R. \latin{et~al.}
  Semi-Automated Creation of Density Functional Tight Binding Models through
  Leveraging Chebyshev Polynomial-Based Force Fields. \emph{J. Chem. Theory
  Comput.} \textbf{2021}, \emph{17}, 4435--4448\relax
\mciteBstWouldAddEndPuncttrue
\mciteSetBstMidEndSepPunct{\mcitedefaultmidpunct}
{\mcitedefaultendpunct}{\mcitedefaultseppunct}\relax
\EndOfBibitem
\bibitem[Goldman \latin{et~al.}(2018)Goldman, Aradi, Lindsey, and
  Fried]{Goldman_DFTB_H_Pu}
Goldman,~N.; Aradi,~B.; Lindsey,~R.~K.; Fried,~L.~E. Development of a
  Multicenter Density Functional Tight Binding Model for Plutonium Surface
  Hydriding. \emph{J. Chem. Theory. Comput.} \textbf{2018}, \emph{14},
  2652--2660\relax
\mciteBstWouldAddEndPuncttrue
\mciteSetBstMidEndSepPunct{\mcitedefaultmidpunct}
{\mcitedefaultendpunct}{\mcitedefaultseppunct}\relax
\EndOfBibitem
\bibitem[Pham \latin{et~al.}(2022)Pham, Lindsel, Fried, and
  Goldman]{Pham_DFTB_JPCL}
Pham,~C.~H.; Lindsel,~R.~K.; Fried,~L.~E.; Goldman,~N. High-Accuracy
  Semiempirical Quantum Models Based on a Minimal Training Set. \emph{J. Phys.
  Chem. Lett.} \textbf{2022}, \emph{13}, 2934--2942\relax
\mciteBstWouldAddEndPuncttrue
\mciteSetBstMidEndSepPunct{\mcitedefaultmidpunct}
{\mcitedefaultendpunct}{\mcitedefaultseppunct}\relax
\EndOfBibitem
\bibitem[Kresse and Hafner(1993)Kresse, and Hafner]{Kresse1}
Kresse,~G.; Hafner,~J. Ab initio molecular dynamics for liquid metals.
  \emph{Phys. Rev. B} \textbf{1993}, \emph{47}, 558--561\relax
\mciteBstWouldAddEndPuncttrue
\mciteSetBstMidEndSepPunct{\mcitedefaultmidpunct}
{\mcitedefaultendpunct}{\mcitedefaultseppunct}\relax
\EndOfBibitem
\bibitem[Kresse and Hafner(1994)Kresse, and Hafner]{Kresse2}
Kresse,~G.; Hafner,~J. Ab initio molecular-dynamics simulation of the
  liquid-metal--amorphous-semiconductor transition in germanium. \emph{Phys.
  Rev. B} \textbf{1994}, \emph{49}, 14251--14269\relax
\mciteBstWouldAddEndPuncttrue
\mciteSetBstMidEndSepPunct{\mcitedefaultmidpunct}
{\mcitedefaultendpunct}{\mcitedefaultseppunct}\relax
\EndOfBibitem
\bibitem[Kresse and Furthm\"uller(1996)Kresse, and Furthm\"uller]{Kresse3}
Kresse,~G.; Furthm\"uller,~J. Efficient iterative schemes for ab initio
  total-energy calculations using a plane-wave basis set. \emph{Phys. Rev. B}
  \textbf{1996}, \emph{54}, 11169--11186\relax
\mciteBstWouldAddEndPuncttrue
\mciteSetBstMidEndSepPunct{\mcitedefaultmidpunct}
{\mcitedefaultendpunct}{\mcitedefaultseppunct}\relax
\EndOfBibitem
\bibitem[Kresse and Joubert(1999)Kresse, and Joubert]{Kresse4}
Kresse,~G.; Joubert,~D. From ultrasoft pseudopotentials to the projector
  augmented-wave method. \emph{Phys. Rev. B} \textbf{1999}, \emph{59},
  1758--1775\relax
\mciteBstWouldAddEndPuncttrue
\mciteSetBstMidEndSepPunct{\mcitedefaultmidpunct}
{\mcitedefaultendpunct}{\mcitedefaultseppunct}\relax
\EndOfBibitem
\bibitem[Bl\"ochl(1994)]{Blochl}
Bl\"ochl,~P.~E. Projector augmented-wave method. \emph{Phys. Rev. B}
  \textbf{1994}, \emph{50}, 17953--17979\relax
\mciteBstWouldAddEndPuncttrue
\mciteSetBstMidEndSepPunct{\mcitedefaultmidpunct}
{\mcitedefaultendpunct}{\mcitedefaultseppunct}\relax
\EndOfBibitem
\bibitem[Perdew \latin{et~al.}(1996)Perdew, Burke, and Ernzerhof]{Perdew}
Perdew,~J.~P.; Burke,~K.; Ernzerhof,~M. Generalized Gradient Approximation Made
  Simple. \emph{Phys. Rev. Lett.} \textbf{1996}, \emph{77}, 3865--3868\relax
\mciteBstWouldAddEndPuncttrue
\mciteSetBstMidEndSepPunct{\mcitedefaultmidpunct}
{\mcitedefaultendpunct}{\mcitedefaultseppunct}\relax
\EndOfBibitem
\bibitem[Methfessel and Paxton(1989)Methfessel, and Paxton]{Methfessel}
Methfessel,~M.; Paxton,~A.~T. High-precision sampling for Brillouin-zone
  integration in metals. \emph{Phys. Rev. B} \textbf{1989}, \emph{40},
  3616--3621\relax
\mciteBstWouldAddEndPuncttrue
\mciteSetBstMidEndSepPunct{\mcitedefaultmidpunct}
{\mcitedefaultendpunct}{\mcitedefaultseppunct}\relax
\EndOfBibitem
\bibitem[Monkhorst and Pack(1976)Monkhorst, and Pack]{Monkhorst76}
Monkhorst,~H.~J.; Pack,~J.~D. Special points for Brillouin-zone integrations.
  \emph{Phys. Rev. B} \textbf{1976}, \emph{13}, 5188--5192\relax
\mciteBstWouldAddEndPuncttrue
\mciteSetBstMidEndSepPunct{\mcitedefaultmidpunct}
{\mcitedefaultendpunct}{\mcitedefaultseppunct}\relax
\EndOfBibitem
\bibitem[Bedolla \latin{et~al.}(2014)Bedolla, Feldbauer, Wolloch, Eder, Dörr,
  Mohn, Redinger, and Vernes]{Bedolla}
Bedolla,~P.~O.; Feldbauer,~G.; Wolloch,~M.; Eder,~S.~J.; Dörr,~N.; Mohn,~P.;
  Redinger,~J.; Vernes,~A. Effects of van der Waals Interactions in the
  Adsorption of Isooctane and Ethanol on Fe(100) Surfaces. \emph{The Journal of
  Physical Chemistry C} \textbf{2014}, \emph{118}, 17608--17615\relax
\mciteBstWouldAddEndPuncttrue
\mciteSetBstMidEndSepPunct{\mcitedefaultmidpunct}
{\mcitedefaultendpunct}{\mcitedefaultseppunct}\relax
\EndOfBibitem
\bibitem[Elstner \latin{et~al.}(1998)Elstner, Porezag, Jungnickel, Elsner,
  Haugk, Frauenheim, Suhai, and Seifert]{Elstner1998}
Elstner,~M.; Porezag,~D.; Jungnickel,~G.; Elsner,~J.; Haugk,~M.;
  Frauenheim,~T.; Suhai,~S.; Seifert,~G. Self-consistent-charge
  density-functional tight-binding method for simulations of complex materials
  properties. \emph{Physical Review B} \textbf{1998}, \emph{58},
  7260--7268\relax
\mciteBstWouldAddEndPuncttrue
\mciteSetBstMidEndSepPunct{\mcitedefaultmidpunct}
{\mcitedefaultendpunct}{\mcitedefaultseppunct}\relax
\EndOfBibitem
\bibitem[Aradi \latin{et~al.}(2007)Aradi, Hourahine, and Frauenheim]{DFTB+}
Aradi,~B.; Hourahine,~B.; Frauenheim,~T. DFTB+, a sparse matrix-based
  implementation of the DFTB method. \emph{J. Phys. Chem. A} \textbf{2007},
  \emph{111}, 5678--5684\relax
\mciteBstWouldAddEndPuncttrue
\mciteSetBstMidEndSepPunct{\mcitedefaultmidpunct}
{\mcitedefaultendpunct}{\mcitedefaultseppunct}\relax
\EndOfBibitem
\bibitem[Koskinen and M\"{a}kinen(2009)Koskinen, and M\"{a}kinen]{Koskinen09}
Koskinen,~P.; M\"{a}kinen,~V. Density-functional tight-binding for beginners.
  \emph{Comp. Mater. Sci.} \textbf{2009}, \emph{47}, 237--253\relax
\mciteBstWouldAddEndPuncttrue
\mciteSetBstMidEndSepPunct{\mcitedefaultmidpunct}
{\mcitedefaultendpunct}{\mcitedefaultseppunct}\relax
\EndOfBibitem
\bibitem[Gaus \latin{et~al.}(2011)Gaus, Cui, and Elstner]{Gaus11}
Gaus,~M.; Cui,~Q.; Elstner,~M. DFTB3: Extension of the Self-Consistent-Charge
  Density-Functional Tight-Binding Method (SCC-DFTB). \emph{J. Chem. Theory
  Comput.} \textbf{2011}, \emph{7}, 931--948\relax
\mciteBstWouldAddEndPuncttrue
\mciteSetBstMidEndSepPunct{\mcitedefaultmidpunct}
{\mcitedefaultendpunct}{\mcitedefaultseppunct}\relax
\EndOfBibitem
\bibitem[Vuong \latin{et~al.}(2020)Vuong, Madridejos, Aradi, Sumpter, Metha,
  and Irle]{Vuong2020}
Vuong,~V.~Q.; Madridejos,~J. M.~L.; Aradi,~B.; Sumpter,~B.~G.; Metha,~G.~F.;
  Irle,~S. Density-functional tight-binding for phosphine-stabilized nanoscale
  gold clusters. \emph{Chemical Science} \textbf{2020}, \emph{11},
  13113--13128\relax
\mciteBstWouldAddEndPuncttrue
\mciteSetBstMidEndSepPunct{\mcitedefaultmidpunct}
{\mcitedefaultendpunct}{\mcitedefaultseppunct}\relax
\EndOfBibitem
\bibitem[Dantanarayana \latin{et~al.}(2020)Dantanarayana, Nematiaram, Vong,
  Anthony, Troisi, Cong, Goldman, Faller, and Moul{\'{e}}]{Dantanarayana2020}
Dantanarayana,~V.; Nematiaram,~T.; Vong,~D.; Anthony,~J.~E.; Troisi,~A.;
  Cong,~K.~N.; Goldman,~N.; Faller,~R.; Moul{\'{e}},~A.~J. Predictive Model of
  Charge Mobilities in Organic Semiconductor Small Molecules with Force-Matched
  Potentials. \emph{Journal of Chemical Theory and Computation} \textbf{2020},
  \emph{16}, 3494--3503\relax
\mciteBstWouldAddEndPuncttrue
\mciteSetBstMidEndSepPunct{\mcitedefaultmidpunct}
{\mcitedefaultendpunct}{\mcitedefaultseppunct}\relax
\EndOfBibitem
\bibitem[Chou \latin{et~al.}(2016)Chou, Nishimura, Fan, Mazur, Irle, and
  Witek]{DFTB_transferability_1}
Chou,~C.-P.; Nishimura,~Y.; Fan,~C.-C.; Mazur,~G.; Irle,~S.; Witek,~H.~A.
  Automatized parameterization of DFTB using particle swarm optimization.
  \emph{J. Chem. Theory Comput.} \textbf{2016}, \emph{12}, 53--64\relax
\mciteBstWouldAddEndPuncttrue
\mciteSetBstMidEndSepPunct{\mcitedefaultmidpunct}
{\mcitedefaultendpunct}{\mcitedefaultseppunct}\relax
\EndOfBibitem
\bibitem[Goldman \latin{et~al.}(2021)Goldman, Kweon, Sadigh, Heo, Lindsey,
  Pham, Fried, Aradi, Holliday, Jeffries, and Wood]{Goldman2021}
Goldman,~N.; Kweon,~K.~E.; Sadigh,~B.; Heo,~T.~W.; Lindsey,~R.~K.; Pham,~C.~H.;
  Fried,~L.~E.; Aradi,~B.; Holliday,~K.; Jeffries,~J.~R. \latin{et~al.}
  Semi-Automated Creation of Density Functional Tight Binding Models through
  Leveraging Chebyshev Polynomial-Based Force Fields. \emph{Journal of Chemical
  Theory and Computation} \textbf{2021}, \emph{17}, 4435--4448\relax
\mciteBstWouldAddEndPuncttrue
\mciteSetBstMidEndSepPunct{\mcitedefaultmidpunct}
{\mcitedefaultendpunct}{\mcitedefaultseppunct}\relax
\EndOfBibitem
\bibitem[Cui \latin{et~al.}(2024)Cui, Reuter, and Margraf]{Margraf-DFTB-2024}
Cui,~M.; Reuter,~K.; Margraf,~J. Obtaining Robust Density Functional
  Tight-Binding Parameters for Solids across the Periodic Table. \emph{J. Chem.
  Theory Comput.} \textbf{2024}, \emph{20}, 5276--5290\relax
\mciteBstWouldAddEndPuncttrue
\mciteSetBstMidEndSepPunct{\mcitedefaultmidpunct}
{\mcitedefaultendpunct}{\mcitedefaultseppunct}\relax
\EndOfBibitem
\bibitem[Tersoff(1988)]{Tersoff88}
Tersoff,~J. Empirical interatomic potential for carbon, with application to
  amorphous-carbon. \emph{Phys. Rev. Lett.} \textbf{1988}, \emph{61},
  2879\relax
\mciteBstWouldAddEndPuncttrue
\mciteSetBstMidEndSepPunct{\mcitedefaultmidpunct}
{\mcitedefaultendpunct}{\mcitedefaultseppunct}\relax
\EndOfBibitem
\bibitem[Vinet \latin{et~al.}(1987)Vinet, Smith, Ferrante, and Rose]{Vinet1987}
Vinet,~P.; Smith,~J.~R.; Ferrante,~J.; Rose,~J.~H. Temperature effects on the
  universal equation of state of solids. \emph{Physical Review B}
  \textbf{1987}, \emph{35}, 1945--1953\relax
\mciteBstWouldAddEndPuncttrue
\mciteSetBstMidEndSepPunct{\mcitedefaultmidpunct}
{\mcitedefaultendpunct}{\mcitedefaultseppunct}\relax
\EndOfBibitem
\bibitem[Gu(2014)]{Gu}
High-pressure behavior of Fe$_3$P and the role of phosphorus in planetary
  cores. \emph{Earth and Planetary Science Letters} \textbf{2014}, \emph{390},
  296--303\relax
\mciteBstWouldAddEndPuncttrue
\mciteSetBstMidEndSepPunct{\mcitedefaultmidpunct}
{\mcitedefaultendpunct}{\mcitedefaultseppunct}\relax
\EndOfBibitem
\bibitem[Scott \latin{et~al.}(2007)Scott, Huggins, Frank, Maglio, Martin, Meng,
  Santillán, and Williams]{Scott}
Scott,~H.~P.; Huggins,~S.; Frank,~M.~R.; Maglio,~S.~J.; Martin,~C.~D.;
  Meng,~Y.; Santillán,~J.; Williams,~Q. Equation of state and high-pressure
  stability of Fe$_3$P-schreibersite: Implications for phosphorus storage in
  planetary cores. \emph{Geophysical Research Letters} \textbf{2007},
  \emph{34}, L06302\relax
\mciteBstWouldAddEndPuncttrue
\mciteSetBstMidEndSepPunct{\mcitedefaultmidpunct}
{\mcitedefaultendpunct}{\mcitedefaultseppunct}\relax
\EndOfBibitem
\bibitem[pym()]{pymatgen_code}
https://pymatgen.org (last accessed November 7, 2024)\relax
\mciteBstWouldAddEndPuncttrue
\mciteSetBstMidEndSepPunct{\mcitedefaultmidpunct}
{\mcitedefaultendpunct}{\mcitedefaultseppunct}\relax
\EndOfBibitem
\bibitem[Scott \latin{et~al.}(2008)Scott, Kiefer, Martin, Boateng, Frank, and
  Meng]{Scott2008}
Scott,~H.; Kiefer,~B.; Martin,~C.~D.; Boateng,~N.; Frank,~M.; Meng,~Y. P-V
  equation of state for Fe2P and pressure-induced phase transition in Fe3P.
  \emph{High Pressure Research} \textbf{2008}, \emph{28}, 375--384\relax
\mciteBstWouldAddEndPuncttrue
\mciteSetBstMidEndSepPunct{\mcitedefaultmidpunct}
{\mcitedefaultendpunct}{\mcitedefaultseppunct}\relax
\EndOfBibitem
\bibitem[Togo \latin{et~al.}(2023)Togo, Chaput, Tadano, and
  Tanaka]{phonopy-phono3py-JPCM}
Togo,~A.; Chaput,~L.; Tadano,~T.; Tanaka,~I. Implementation strategies in
  phonopy and phono3py. \emph{J. Phys. Condens. Matter} \textbf{2023},
  \emph{35}, 353001\relax
\mciteBstWouldAddEndPuncttrue
\mciteSetBstMidEndSepPunct{\mcitedefaultmidpunct}
{\mcitedefaultendpunct}{\mcitedefaultseppunct}\relax
\EndOfBibitem
\bibitem[Togo(2023)]{phonopy-phono3py-JPSJ}
Togo,~A. First-principles Phonon Calculations with Phonopy and Phono3py.
  \emph{J. Phys. Soc. Jpn.} \textbf{2023}, \emph{92}, 012001\relax
\mciteBstWouldAddEndPuncttrue
\mciteSetBstMidEndSepPunct{\mcitedefaultmidpunct}
{\mcitedefaultendpunct}{\mcitedefaultseppunct}\relax
\EndOfBibitem
\bibitem[Chernyavskii \latin{et~al.}(2020)Chernyavskii, Nikitin, Onykiienko,
  Inosov, Stahl, Geck, Hong, Hess, Gass, Wolter, Wolf, Lubk, Efremov,
  Yokaichiya, Aswartham, B\"{u}chner, and Morozov]{Chernyavskii2020}
Chernyavskii,~I.~O.; Nikitin,~S.~E.; Onykiienko,~Y.~A.; Inosov,~D.~S.;
  Stahl,~Q.; Geck,~J.; Hong,~X.~C.; Hess,~C.; Gass,~S.; Wolter,~A. U.~B.
  \latin{et~al.}  Incommensurate magnet iron monophosphide {FeP}: Crystal
  growth and characterization. \emph{Physical Review Materials} \textbf{2020},
  \emph{4}\relax
\mciteBstWouldAddEndPuncttrue
\mciteSetBstMidEndSepPunct{\mcitedefaultmidpunct}
{\mcitedefaultendpunct}{\mcitedefaultseppunct}\relax
\EndOfBibitem
\bibitem[Huang \latin{et~al.}(2016)Huang, Yu, Yang, Han, Zhao, Li, Liu, and
  Qiu]{Huang2016}
Huang,~H.; Yu,~C.; Yang,~J.; Han,~X.; Zhao,~C.; Li,~S.; Liu,~Z.; Qiu,~J.
  Ultrasmall diiron phosphide nanodots anchored on graphene sheets with
  enhanced electrocatalytic activity for hydrogen production via
  high-efficiency water splitting. \emph{Journal of Materials Chemistry A}
  \textbf{2016}, \emph{4}, 16028--16035\relax
\mciteBstWouldAddEndPuncttrue
\mciteSetBstMidEndSepPunct{\mcitedefaultmidpunct}
{\mcitedefaultendpunct}{\mcitedefaultseppunct}\relax
\EndOfBibitem
\bibitem[Dera \latin{et~al.}(2008)Dera, Lavina, Borkowski, Prakapenka, Sutton,
  Rivers, Downs, Boctor, and Prewitt]{Dera2008}
Dera,~P.; Lavina,~B.; Borkowski,~L.~A.; Prakapenka,~V.~B.; Sutton,~S.~R.;
  Rivers,~M.~L.; Downs,~R.~T.; Boctor,~N.~Z.; Prewitt,~C.~T. High-pressure
  polymorphism of Fe$_2$P and its implications for meteorites and Earth's core.
  \emph{Geophysical Research Letters} \textbf{2008}, \emph{35}, L10301\relax
\mciteBstWouldAddEndPuncttrue
\mciteSetBstMidEndSepPunct{\mcitedefaultmidpunct}
{\mcitedefaultendpunct}{\mcitedefaultseppunct}\relax
\EndOfBibitem
\bibitem[Mathon \latin{et~al.}(2004)Mathon, Baudelet, Iti{\'{e}}, Polian,
  d{\textquotesingle}Astuto, Chervin, and Pascarelli]{Mathon2004}
Mathon,~O.; Baudelet,~F.; Iti{\'{e}},~J.~P.; Polian,~A.;
  d{\textquotesingle}Astuto,~M.; Chervin,~J.~C.; Pascarelli,~S. Dynamics of the
  Magnetic and Structural $\alpha-\epsilon$ Phase Transition in Iron.
  \emph{Physical Review Letters} \textbf{2004}, \emph{93}, 255503\relax
\mciteBstWouldAddEndPuncttrue
\mciteSetBstMidEndSepPunct{\mcitedefaultmidpunct}
{\mcitedefaultendpunct}{\mcitedefaultseppunct}\relax
\EndOfBibitem
\bibitem[Gre(1996)]{Greenwood1996}
\emph{Introductory Chemistry for the Environmental Sciences}; Cambridge
  University Press, 1996; pp 148--209\relax
\mciteBstWouldAddEndPuncttrue
\mciteSetBstMidEndSepPunct{\mcitedefaultmidpunct}
{\mcitedefaultendpunct}{\mcitedefaultseppunct}\relax
\EndOfBibitem
\bibitem[SEKI and NAGATA(2005)SEKI, and NAGATA]{SEKI2005}
SEKI,~I.; NAGATA,~K. Lattice Constant of Iron and Austenite Including Its
  Supersaturation Phase of Carbon. \emph{{ISIJ} International} \textbf{2005},
  \emph{45}, 1789--1794\relax
\mciteBstWouldAddEndPuncttrue
\mciteSetBstMidEndSepPunct{\mcitedefaultmidpunct}
{\mcitedefaultendpunct}{\mcitedefaultseppunct}\relax
\EndOfBibitem
\bibitem[Takahashi \latin{et~al.}(1968)Takahashi, Bassett, and
  Mao]{Takahashi1968}
Takahashi,~T.; Bassett,~W.~A.; Mao,~H.-K. Isothermal compression of the alloys
  of iron up to 300 kilobars at room temperature: Iron-nickel alloys.
  \emph{Journal of Geophysical Research} \textbf{1968}, \emph{73},
  4717--4725\relax
\mciteBstWouldAddEndPuncttrue
\mciteSetBstMidEndSepPunct{\mcitedefaultmidpunct}
{\mcitedefaultendpunct}{\mcitedefaultseppunct}\relax
\EndOfBibitem
\end{mcitethebibliography}


\providecommand{\latin}[1]{#1}
\providecommand*\mcitethebibliography{\thebibliography}
\csname @ifundefined\endcsname{endmcitethebibliography}
  {\let\endmcitethebibliography\endthebibliography}{}
\begin{mcitethebibliography}{0}
\providecommand*\natexlab[1]{#1}
\providecommand*\mciteSetBstSublistMode[1]{}
\providecommand*\mciteSetBstMaxWidthForm[2]{}
\providecommand*\mciteBstWouldAddEndPuncttrue
  {\def\EndOfBibitem{\unskip.}}
\providecommand*\mciteBstWouldAddEndPunctfalse
  {\let\EndOfBibitem\relax}
\providecommand*\mciteSetBstMidEndSepPunct[3]{}
\providecommand*\mciteSetBstSublistLabelBeginEnd[3]{}
\providecommand*\EndOfBibitem{}
\mciteSetBstSublistMode{f}
\mciteSetBstMaxWidthForm{subitem}{(\alph{mcitesubitemcount})}
\mciteSetBstSublistLabelBeginEnd
  {\mcitemaxwidthsubitemform\space}
  {\relax}
  {\relax}

\end{mcitethebibliography}

\end{document}

\newpage
\section{Root-mean-squared errors (RMSE) for forces and stress tensor}

\begin{figure}
\centering
\includegraphics[width=.95\linewidth]{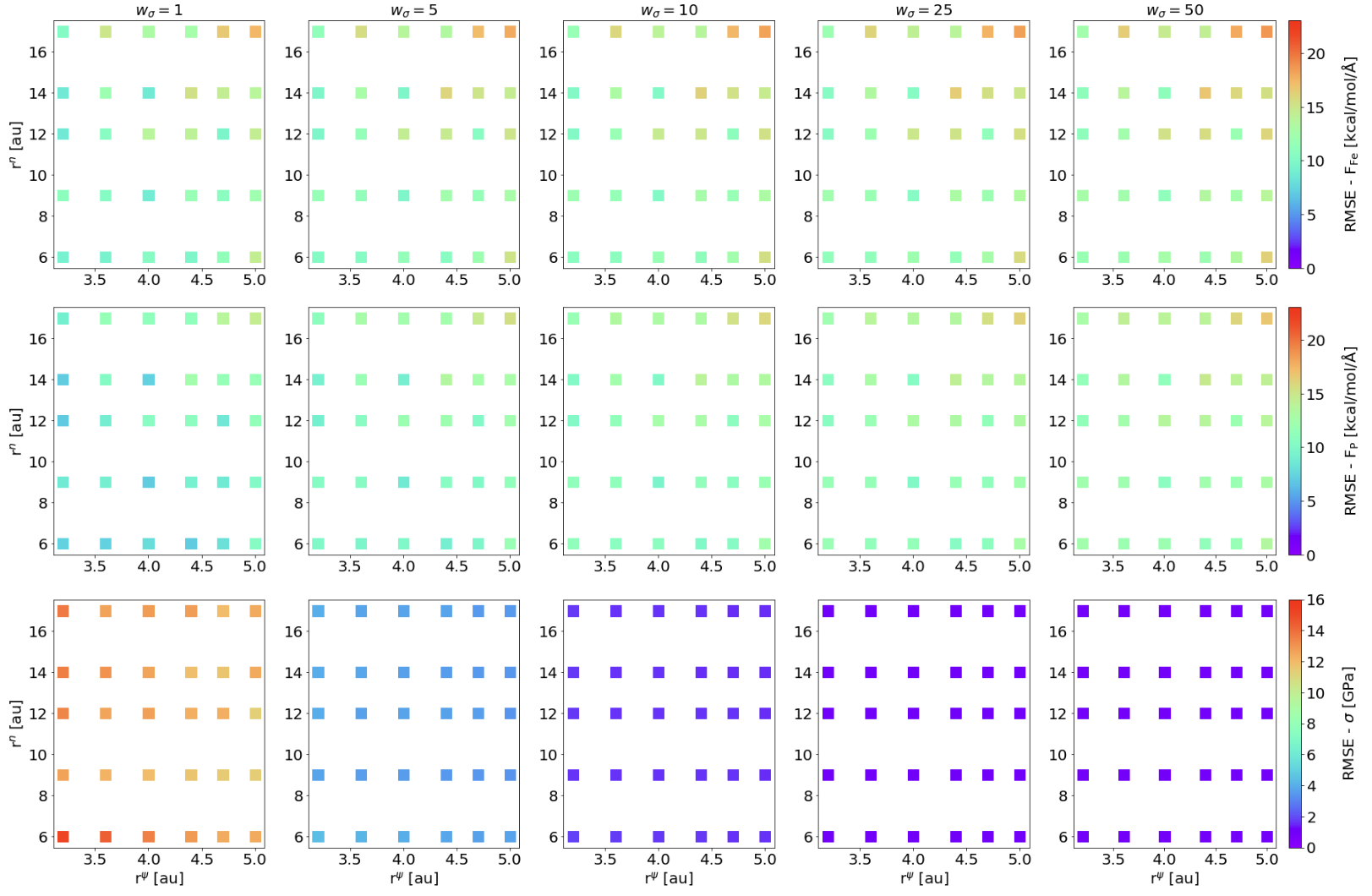}
\caption{RMSE for the \{8,4,0\} set. Top panel: forces on iron atoms, central panel: forces on phosphorous atoms, bottom panel: stress tensor.}
\label{fig:RMSE_8-4-0}
\end{figure}

\begin{table}[]
\tiny
\caption{RMSE for the \{8,4,0\} set. RMSE on forces are expressed in kcal/mol/\AA{}, RMSE on stress tensor components are expressed in GPa}
\label{tab:RMSE_8-4-0}
\begin{tabular}{ccccccccccccccccc}
     &    & \multicolumn{3}{c}{$w_\sigma$=0} & \multicolumn{3}{c}{$w_\sigma$=5} & \multicolumn{3}{c}{$w_\sigma$=10} & \multicolumn{3}{c}{$w_\sigma$=25} & \multicolumn{3}{c}{$w_\sigma$=50} \\ \cline{3-17} 
$R_\Psi$ & \multicolumn{1}{c}{$R_n$} & Fe    & P     & $\sigma$   & Fe     & P     & $\sigma$  & Fe     & P      & $\sigma$  & Fe     & P      & $\sigma$  & Fe     & P      & $\sigma$  \\ \hline
3.2  & 6  & 9.223  & 7.288  & 15.219 & 10.318  & 9.991  & 4.455 & 10.801  & 11.184  & 1.824 & 11.162  & 11.762  & 1.171 & 11.803  & 12.377  & 0.997 \\
3.2  & 9  & 10.901 & 8.743  & 12.671 & 11.683  & 10.387 & 3.787 & 12.015  & 11.199  & 1.669 & 12.262  & 11.649  & 1.15  & 12.778  & 12.256  & 0.981 \\
3.2  & 12 & 8.711  & 7.014  & 13.389 & 9.803   & 9.157  & 3.965 & 10.239  & 10.15   & 1.698 & 10.526  & 10.678  & 1.144 & 11.078  & 11.33   & 0.981 \\
3.2  & 14 & 8.689  & 7.043  & 13.565 & 9.793   & 9.229  & 3.995 & 10.234  & 10.232  & 1.7   & 10.519  & 10.759  & 1.141 & 11.06   & 11.396  & 0.979 \\
3.2  & 17 & 10.323 & 9.108  & 13.701 & 11.281  & 10.908 & 4.028 & 11.679  & 11.78   & 1.704 & 11.937  & 12.24   & 1.139 & 12.421  & 12.797  & 0.978 \\

3.6  & 6  & 9.426  & 7.237  & 14.388 & 10.441  & 9.688  & 4.241 & 10.888  & 10.783  & 1.778 & 11.244  & 11.336  & 1.169 & 11.902  & 11.983  & 0.991 \\
3.6  & 9  & 10.446 & 9.024  & 12.203 & 11.279  & 10.457 & 3.648 & 11.618  & 11.186  & 1.639 & 11.877  & 11.606  & 1.147 & 12.427  & 12.214  & 0.977 \\
3.6  & 12 & 10.583 & 9.005  & 12.578 & 11.389  & 10.559 & 3.747 & 11.723  & 11.335  & 1.659 & 11.96   & 11.778  & 1.151 & 12.469  & 12.375  & 0.987 \\
3.6  & 14 & 12.008 & 10.314 & 13.114 & 12.812  & 11.776 & 3.858 & 13.138  & 12.519  & 1.67  & 13.371  & 12.929  & 1.137 & 13.828  & 13.468  & 0.971 \\
3.6  & 17 & 15.031 & 11.463 & 12.682 & 15.629  & 12.746 & 3.779 & 15.882  & 13.412  & 1.649 & 16.063  & 13.788  & 1.129 & 16.415  & 14.294  & 0.966 \\

4    & 6  & 10.537 & 7.909  & 13.519 & 11.38   & 9.953  & 4.018 & 11.755  & 10.899  & 1.733 & 12.076  & 11.407  & 1.167 & 12.697  & 12.066  & 0.985 \\
4    & 9  & 8.893  & 7.034  & 11.912 & 9.827   & 8.739  & 3.582 & 10.203  & 9.565   & 1.631 & 10.497  & 10.048  & 1.149 & 11.121  & 10.765  & 0.975 \\
4    & 12 & 13.788 & 10.612 & 12.69  & 14.443  & 11.979 & 3.767 & 14.718  & 12.677  & 1.651 & 14.914  & 13.074  & 1.132 & 15.31   & 13.609  & 0.967 \\
4    & 14 & 8.74   & 7.22   & 12.6   & 9.756   & 9.069  & 3.717 & 10.151  & 9.943   & 1.643 & 10.432  & 10.431  & 1.137 & 11      & 11.098  & 0.97  \\
4    & 17 & 12.983 & 11.106 & 12.855 & 13.727  & 12.419 & 3.805 & 14.03   & 13.102  & 1.664 & 14.25   & 13.492  & 1.137 & 14.686  & 14.024  & 0.967 \\

4.4  & 6  & 9.761  & 7.034  & 12.897 & 10.648  & 9.057  & 3.823 & 11.027  & 9.973   & 1.69  & 11.368  & 10.478  & 1.162 & 12.053  & 11.185  & 0.98  \\
4.4  & 9  & 11.757 & 9.26   & 11.721 & 12.453  & 10.572 & 3.528 & 12.741  & 11.241  & 1.631 & 12.978  & 11.65   & 1.16  & 13.51   & 12.273  & 0.986 \\
4.4  & 12 & 14.067 & 10.891 & 12.364 & 14.689  & 12.158 & 3.69  & 14.949  & 12.811  & 1.649 & 15.157  & 13.199  & 1.143 & 15.588  & 13.763  & 0.967 \\
4.4  & 14 & 15.349 & 12.427 & 11.866 & 15.966  & 13.438 & 3.566 & 16.211  & 13.98   & 1.622 & 16.406  & 14.316  & 1.137 & 16.805  & 14.836  & 0.962 \\
4.4  & 17 & 12.625 & 11.117 & 12.757 & 13.409  & 12.404 & 3.754 & 13.719  & 13.064  & 1.655 & 13.941  & 13.447  & 1.139 & 14.387  & 13.984  & 0.969 \\

4.7  & 6  & 11.33  & 8.07   & 12.417 & 12.107  & 9.739  & 3.709 & 12.433  & 10.539  & 1.67  & 12.753  & 10.99   & 1.162 & 13.396  & 11.666  & 0.978 \\
4.7  & 9  & 10.768 & 8.759  & 11.445 & 11.478  & 10.075 & 3.475 & 11.773  & 10.748  & 1.628 & 12.026  & 11.179  & 1.166 & 12.603  & 11.858  & 0.988 \\
4.7  & 12 & 9.648  & 8.593  & 12.333 & 10.582  & 10.121 & 3.662 & 10.952  & 10.88   & 1.646 & 11.24   & 11.322  & 1.146 & 11.836  & 11.962  & 0.97  \\
4.7  & 14 & 14.439 & 11.578 & 11.711 & 15.063  & 12.631 & 3.505 & 15.308  & 13.187  & 1.608 & 15.495  & 13.538  & 1.136 & 15.889  & 14.084  & 0.962 \\
4.7  & 17 & 16.601 & 13.84  & 11.902 & 17.176  & 14.747 & 3.539 & 17.4    & 15.236  & 1.614 & 17.574  & 15.54   & 1.136 & 17.935  & 16.007  & 0.966 \\

5    & 6  & 14.448 & 10.439 & 12.482 & 15.14   & 11.773 & 3.751 & 15.427  & 12.462  & 1.685 & 15.699  & 12.857  & 1.162 & 16.248  & 13.445  & 0.971 \\
5    & 9  & 12.221 & 10.151 & 11.383 & 12.858  & 11.282 & 3.43  & 13.114  & 11.87   & 1.622 & 13.315  & 12.264  & 1.168 & 13.812  & 12.888  & 0.991 \\
5    & 12 & 14.497 & 11.333 & 11.374 & 15.112  & 12.335 & 3.434 & 15.353  & 12.872  & 1.61  & 15.554  & 13.229  & 1.146 & 15.981  & 13.82   & 0.963 \\
5    & 14 & 13.834 & 11.687 & 12.455 & 14.53   & 12.828 & 3.478 & 14.774  & 13.354  & 1.598 & 14.968  & 13.688  & 1.145 & 15.408  & 14.238  & 0.968 \\
5    & 17 & 17.429 & 14.519 & 12.444 & 18.009  & 15.454 & 3.523 & 18.216  & 15.917  & 1.613 & 18.367  & 16.216  & 1.143 & 18.702  & 16.7    & 0.964 \\

\end{tabular}
\end{table}

\begin{figure}
\centering
\includegraphics[width=.95\linewidth]{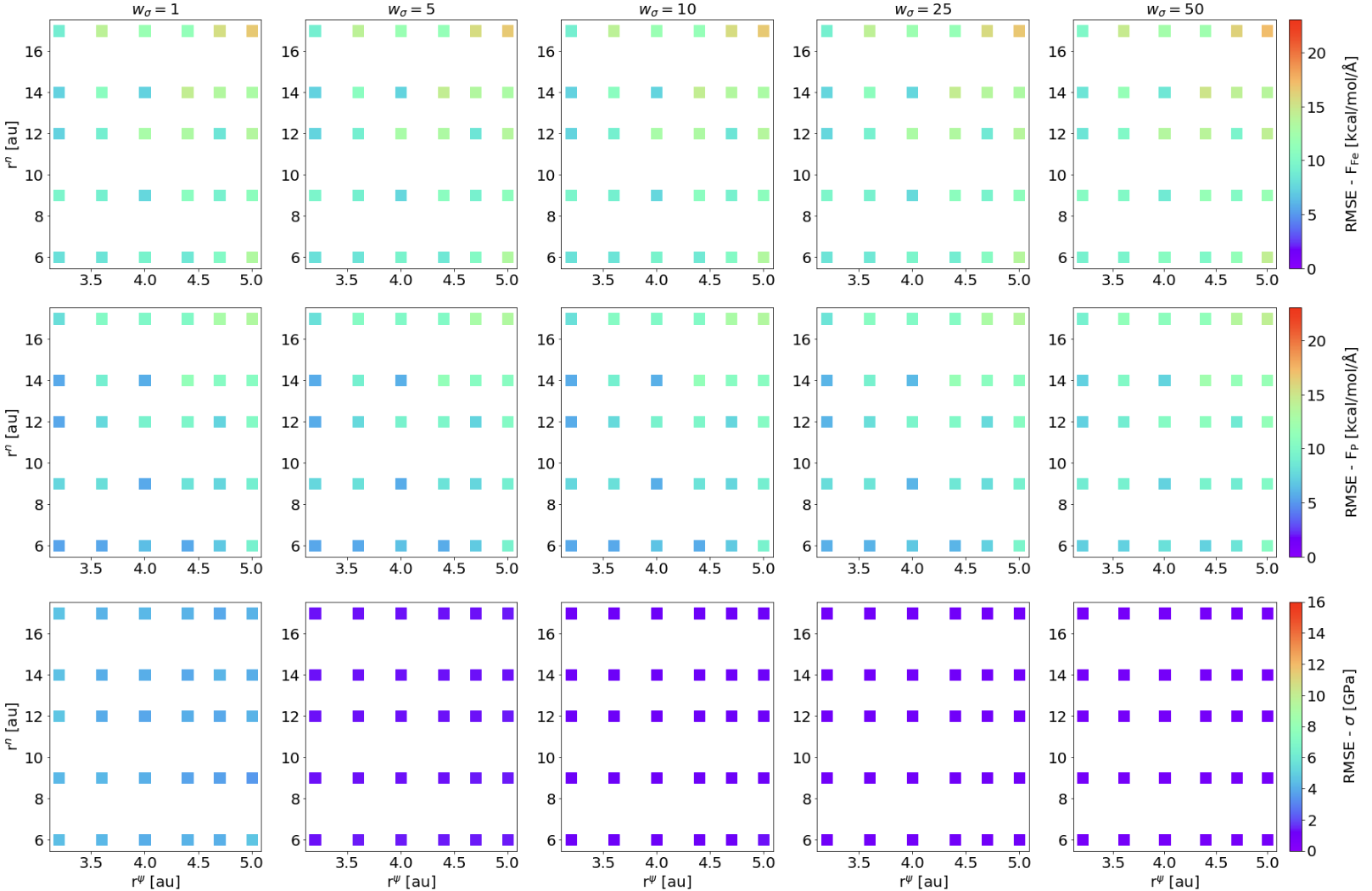}
\caption{RMSE for the \{12,8,0\} set. Top panel: forces on iron atoms, central panel: forces on phosphorous atoms, bottom panel: stress tensor.}
\label{fig:RMSE_12-8-0}
\end{figure}

\begin{table}[]
\tiny
\caption{RMSE for the \{12,8,0\} set. RMSE on forces are expressed in kcal/mol/\AA{}, RMSE on stress tensor components are expressed in GPa}
\label{tab:RMSE_12-8-0}
\begin{tabular}{ccccccccccccccccc}
     &    & \multicolumn{3}{c}{$w_\sigma$=0} & \multicolumn{3}{c}{$w_\sigma$=5} & \multicolumn{3}{c}{$w_\sigma$=10} & \multicolumn{3}{c}{$w_\sigma$=25} & \multicolumn{3}{c}{$w_\sigma$=50} \\ \cline{3-17} 
$R_\Psi$ & \multicolumn{1}{c}{$R_n$} & Fe    & P     & $\sigma$   & Fe     & P     & $\sigma$  & Fe     & P      & $\sigma$  & Fe     & P      & $\sigma$  & Fe     & P      & $\sigma$  \\ \hline
3.2  & 6    & 8.083   & 5.436  & 4.476 & 8.101   & 5.561  & 1.372 & 8.119   & 5.588   & 1.279 & 8.366   & 5.915   & 1.114 & 9.35    & 7.172   & 0.905 \\
3.2  & 9    & 9.648   & 7.537  & 4.272 & 9.653   & 7.617  & 1.35  & 9.655   & 7.641   & 1.266 & 9.796   & 7.892   & 1.108 & 10.523  & 8.861   & 0.909 \\
3.2  & 12   & 7.358   & 5.566  & 4.647 & 7.393   & 5.684  & 1.349 & 7.398   & 5.719   & 1.254 & 7.573   & 6.054   & 1.098 & 8.458   & 7.253   & 0.903 \\
3.2  & 14   & 7.325   & 5.586  & 4.545 & 7.353   & 5.689  & 1.339 & 7.356   & 5.723   & 1.248 & 7.525   & 6.059   & 1.095 & 8.402   & 7.252   & 0.901 \\
3.2  & 17   & 8.983   & 7.994  & 4.617 & 9.011   & 8.063  & 1.339 & 9.014   & 8.086   & 1.246 & 9.167   & 8.323   & 1.092 & 9.926   & 9.219   & 0.9   \\

3.6  & 6    & 8.29    & 5.522  & 4.259 & 8.305   & 5.63   & 1.376 & 8.325   & 5.655   & 1.288 & 8.581   & 5.983   & 1.12  & 9.574   & 7.241   & 0.908 \\
3.6  & 9    & 9.139   & 7.969  & 4.267 & 9.165   & 8.028  & 1.357 & 9.172   & 8.048   & 1.272 & 9.35    & 8.295   & 1.11  & 10.176  & 9.215   & 0.911 \\
3.6  & 12   & 9.076   & 7.682  & 3.965 & 9.098   & 7.735  & 1.348 & 9.101   & 7.758   & 1.27  & 9.248   & 8.015   & 1.114 & 10.011  & 8.972   & 0.919 \\
3.6  & 14   & 10.724  & 9.06   & 4     & 10.748  & 9.103  & 1.332 & 10.754  & 9.121   & 1.252 & 10.901  & 9.332   & 1.095 & 11.58   & 10.148  & 0.9   \\
3.6  & 17   & 14.1    & 10.356 & 4.183 & 14.113  & 10.405 & 1.325 & 14.12   & 10.421  & 1.238 & 14.224  & 10.603  & 1.084 & 14.697  & 11.307  & 0.894 \\

4    & 6    & 9.466   & 6.486  & 4.058 & 9.476   & 6.574  & 1.382 & 9.492   & 6.594   & 1.297 & 9.715   & 6.892   & 1.123 & 10.607  & 8.029   & 0.909 \\
4    & 9    & 7.572   & 5.673  & 4.115 & 7.597   & 5.758  & 1.361 & 7.606   & 5.787   & 1.278 & 7.821   & 6.126   & 1.114 & 8.781   & 7.354   & 0.909 \\
4    & 12   & 12.849  & 9.394  & 3.858 & 12.869  & 9.438  & 1.326 & 12.873  & 9.456   & 1.247 & 12.986  & 9.657   & 1.089 & 13.53   & 10.433  & 0.897 \\
4    & 14   & 7.371   & 5.824  & 4.059 & 7.401   & 5.901  & 1.336 & 7.407   & 5.929   & 1.255 & 7.595   & 6.253   & 1.097 & 8.505   & 7.428   & 0.899 \\
4    & 17   & 11.815  & 9.956  & 4.265 & 11.839  & 10.002 & 1.346 & 11.846  & 10.021  & 1.258 & 11.982  & 10.224  & 1.095 & 12.609  & 10.984  & 0.897 \\
4.4  & 6    & 8.628   & 5.612  & 4.137 & 8.644   & 5.713  & 1.387 & 8.667   & 5.735   & 1.299 & 8.932   & 6.069   & 1.124 & 9.924   & 7.325   & 0.909 \\
4.4  & 9    & 10.731  & 8.228  & 3.781 & 10.744  & 8.269  & 1.371 & 10.749  & 8.288   & 1.296 & 10.901  & 8.541   & 1.128 & 11.62   & 9.478   & 0.923 \\
4.4  & 12   & 13.2    & 9.752  & 4.047 & 13.217  & 9.804  & 1.357 & 13.225  & 9.822   & 1.271 & 13.354  & 10.026  & 1.105 & 13.93   & 10.82   & 0.9   \\
4.4  & 14   & 14.436  & 11.39  & 3.884 & 14.458  & 11.426 & 1.359 & 14.464  & 11.441  & 1.275 & 14.589  & 11.622  & 1.103 & 15.126  & 12.316  & 0.897 \\
4.4  & 17   & 11.426  & 10.16  & 4.182 & 11.452  & 10.2   & 1.349 & 11.458  & 10.217  & 1.266 & 11.596  & 10.421  & 1.102 & 12.247  & 11.186  & 0.9   \\

4.7  & 6    & 10.242  & 6.87   & 4.081 & 10.271  & 6.95   & 1.394 & 10.29   & 6.969   & 1.303 & 10.518  & 7.247   & 1.125 & 11.394  & 8.319   & 0.909 \\
4.7  & 9    & 9.294   & 7.49   & 3.744 & 9.31    & 7.537  & 1.382 & 9.317   & 7.56    & 1.307 & 9.497   & 7.845   & 1.135 & 10.329  & 8.882   & 0.926 \\
4.7  & 12   & 8.425   & 7.419  & 3.902 & 8.452   & 7.469  & 1.357 & 8.462   & 7.49    & 1.278 & 8.664   & 7.76    & 1.11  & 9.56    & 8.763   & 0.903 \\
4.7  & 14   & 13.482  & 10.519 & 4.195 & 13.507  & 10.564 & 1.359 & 13.512  & 10.581  & 1.27  & 13.63   & 10.783  & 1.099 & 14.164  & 11.526  & 0.895 \\
4.7  & 17   & 15.57   & 12.857 & 3.902 & 15.595  & 12.88  & 1.348 & 15.601  & 12.892  & 1.267 & 15.711  & 13.053  & 1.099 & 16.205  & 13.65   & 0.899 \\

5    & 6    & 13.384  & 9.341  & 4.651 & 13.409  & 9.421  & 1.424 & 13.426  & 9.439   & 1.317 & 13.62   & 9.662   & 1.13  & 14.341  & 10.501  & 0.909 \\
5    & 9    & 10.934  & 9.037  & 3.556 & 10.949  & 9.072  & 1.384 & 10.95   & 9.092   & 1.311 & 11.075  & 9.341   & 1.138 & 11.754  & 10.232  & 0.931 \\
5    & 12   & 13.586  & 10.299 & 4.176 & 13.609  & 10.344 & 1.384 & 13.615  & 10.362  & 1.295 & 13.745  & 10.581  & 1.113 & 14.302  & 11.375  & 0.899 \\
5    & 14   & 12.98   & 10.624 & 4.009 & 13.02   & 10.655 & 1.37  & 13.027  & 10.671  & 1.284 & 13.159  & 10.873  & 1.111 & 13.747  & 11.635  & 0.904 \\
5    & 17   & 16.624  & 13.393 & 4.14  & 16.661  & 13.421 & 1.376 & 16.663  & 13.436  & 1.29  & 16.746  & 13.606  & 1.11  & 17.189  & 14.242  & 0.9   

\end{tabular}
\end{table}

\begin{figure}
\centering
\includegraphics[width=.95\linewidth]{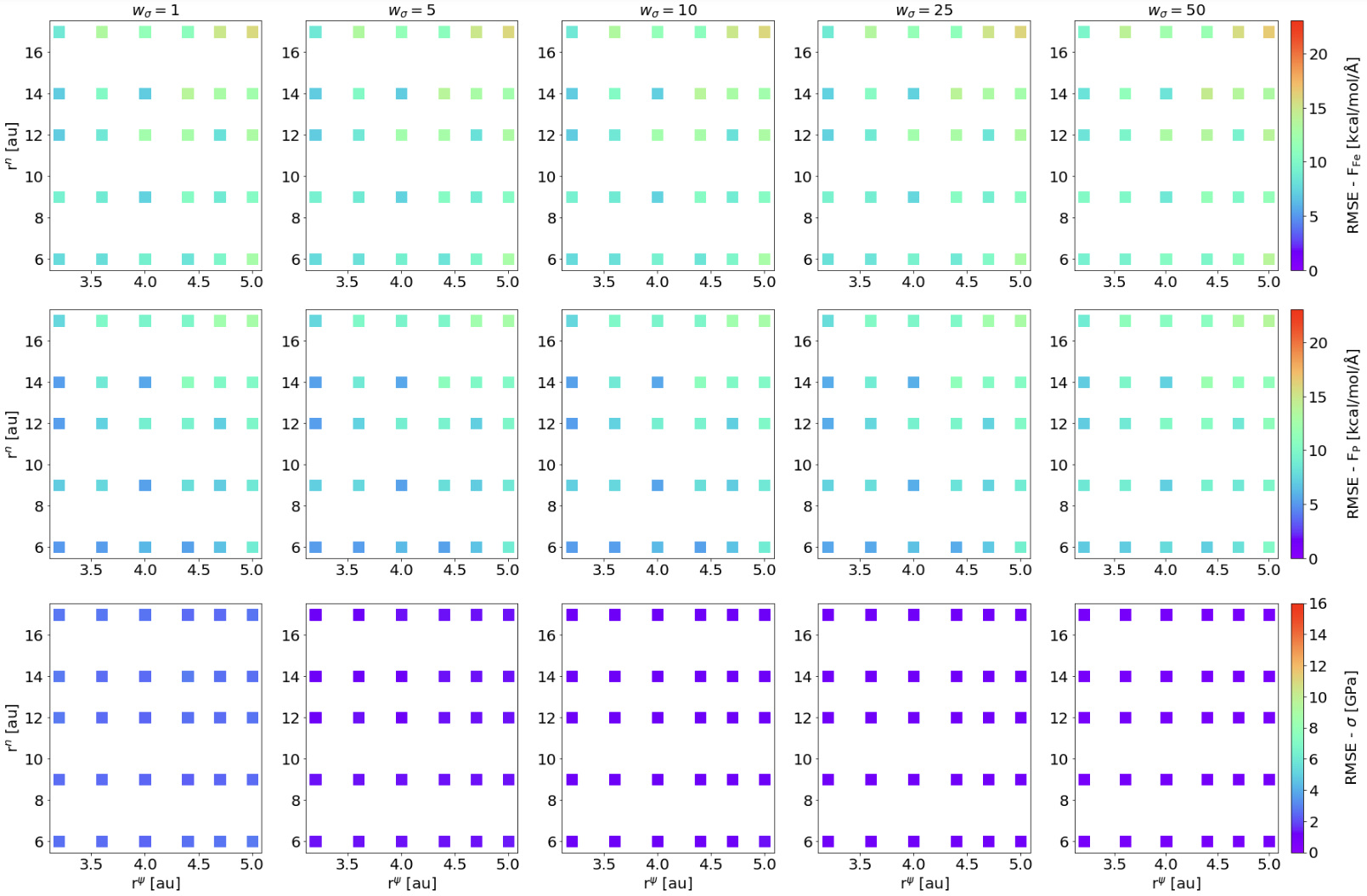}
\caption{RMSE for the \{16,12,0\} set. Top panel: forces on iron atoms, central panel: forces on phosphorous atoms, bottom panel: stress tensor.}
\label{fig:RMSE_16-12-0}
\end{figure}

\begin{table}[]
\tiny
\caption{RMSE for the \{16,12,0\} set. RMSE on forces are expressed in kcal/mol/\AA{}, and RMSE on stress tensor components are expressed in GPa}
\label{tab:RMSE_16-12-0}
\begin{tabular}{ccccccccccccccccc}
     &    & \multicolumn{3}{c}{$w_\sigma$=0} & \multicolumn{3}{c}{$w_\sigma$=5} & \multicolumn{3}{c}{$w_\sigma$=10} & \multicolumn{3}{c}{$w_\sigma$=25} & \multicolumn{3}{c}{$w_\sigma$=50} \\ \cline{3-17} 
$R_\Psi$ & \multicolumn{1}{c}{$R_n$} & Fe    & P     & $\sigma$   & Fe     & P     & $\sigma$  & Fe     & P      & $\sigma$  & Fe     & P      & $\sigma$  & Fe     & P      & $\sigma$  \\ \hline
3.2  & 6    & 7.806   & 5.113  & 2.679 & 7.81    & 5.135  & 1.333 & 7.83    & 5.153   & 1.281 & 8.121   & 5.533   & 1.103 & 9.19    & 6.958   & 0.873 \\
3.2  & 9    & 9.241   & 7.104  & 2.625 & 9.245   & 7.123  & 1.315 & 9.249   & 7.139   & 1.264 & 9.433   & 7.423   & 1.094 & 10.257  & 8.505   & 0.875 \\
3.2  & 12   & 7.092   & 5.261  & 2.86  & 7.1     & 5.287  & 1.309 & 7.103   & 5.312   & 1.255 & 7.317   & 5.692   & 1.087 & 8.311   & 7.051   & 0.87  \\
3.2  & 14   & 7.065   & 5.288  & 2.817 & 7.077   & 5.317  & 1.303 & 7.079   & 5.341   & 1.25  & 7.286   & 5.717   & 1.084 & 8.268   & 7.068   & 0.868 \\
3.2  & 17   & 8.592   & 7.613  & 2.77  & 8.601   & 7.631  & 1.303 & 8.607   & 7.647   & 1.25  & 8.803   & 7.899   & 1.084 & 9.675   & 8.898   & 0.869 \\

3.6  & 6    & 8.015   & 5.214  & 2.576 & 8.02    & 5.232  & 1.341 & 8.042   & 5.249   & 1.289 & 8.34    & 5.627   & 1.108 & 9.411   & 7.044   & 0.875 \\
3.6  & 9    & 8.746   & 7.614  & 2.54  & 8.755   & 7.625  & 1.327 & 8.764   & 7.639   & 1.275 & 8.983   & 7.899   & 1.101 & 9.91    & 8.925   & 0.879 \\
3.6  & 12   & 8.594   & 7.256  & 2.546 & 8.602   & 7.268  & 1.322 & 8.605   & 7.285   & 1.272 & 8.789   & 7.568   & 1.102 & 9.652   & 8.652   & 0.884 \\
3.6  & 14   & 10.257  & 8.665  & 2.572 & 10.265  & 8.674  & 1.312 & 10.274  & 8.686   & 1.262 & 10.457  & 8.916   & 1.09  & 11.253  & 9.834   & 0.87  \\
3.6  & 17   & 13.383  & 9.953  & 2.804 & 13.386  & 9.964  & 1.302 & 13.392  & 9.977   & 1.249 & 13.521  & 10.186  & 1.078 & 14.085  & 11.001  & 0.862 \\

4    & 6    & 9.172   & 6.215  & 2.479 & 9.175   & 6.228  & 1.35  & 9.193   & 6.242   & 1.298 & 9.452   & 6.574   & 1.111 & 10.417  & 7.834   & 0.875 \\
4    & 9    & 7.314   & 5.387  & 2.497 & 7.322   & 5.403  & 1.332 & 7.332   & 5.425   & 1.28  & 7.589   & 5.804   & 1.102 & 8.642   & 7.178   & 0.876 \\
4    & 12   & 12.361  & 9.052  & 2.484 & 12.364  & 9.06   & 1.308 & 12.37   & 9.073   & 1.258 & 12.514  & 9.302   & 1.084 & 13.143  & 10.192  & 0.865 \\
4    & 14   & 7.109   & 5.533  & 2.518 & 7.116   & 5.548  & 1.308 & 7.122   & 5.569   & 1.259 & 7.354   & 5.933   & 1.087 & 8.372   & 7.256   & 0.866 \\
4    & 17   & 11.387  & 9.553  & 2.701 & 11.393  & 9.563  & 1.325 & 11.402  & 9.575   & 1.271 & 11.584  & 9.793   & 1.092 & 12.316  & 10.646  & 0.867 \\
4.4  & 6    & 8.349   & 5.341  & 2.642 & 8.356   & 5.361  & 1.355 & 8.381   & 5.376   & 1.3   & 8.69    & 5.755   & 1.111 & 9.75    & 7.154   & 0.874 \\
4.4  & 9    & 10.421  & 7.969  & 2.333 & 10.427  & 7.975  & 1.354 & 10.435  & 7.991   & 1.303 & 10.62   & 8.271   & 1.118 & 11.398  & 9.31    & 0.89  \\
4.4  & 12   & 12.717  & 9.408  & 2.671 & 12.722  & 9.421  & 1.339 & 12.733  & 9.434   & 1.285 & 12.908  & 9.667   & 1.1   & 13.59   & 10.565  & 0.868 \\
4.4  & 14   & 13.804  & 10.903 & 2.904 & 13.814  & 10.914 & 1.345 & 13.823  & 10.926  & 1.286 & 13.98   & 11.128  & 1.096 & 14.611  & 11.912  & 0.864 \\
4.4  & 17   & 10.951  & 9.735  & 2.532 & 10.957  & 9.744  & 1.321 & 10.966  & 9.756   & 1.268 & 11.141  & 9.965   & 1.09  & 11.876  & 10.794  & 0.868 \\

4.7  & 6    & 9.796   & 6.567  & 2.706 & 9.809   & 6.585  & 1.363 & 9.831   & 6.595   & 1.306 & 10.111  & 6.905   & 1.111 & 11.071  & 8.105   & 0.874 \\
4.7  & 9    & 8.823   & 7.074  & 2.444 & 8.832   & 7.086  & 1.361 & 8.839   & 7.103   & 1.308 & 9.055   & 7.419   & 1.121 & 9.981   & 8.58    & 0.891 \\
4.7  & 12   & 8.168   & 7.15   & 2.608 & 8.179   & 7.163  & 1.336 & 8.19    & 7.179   & 1.282 & 8.438   & 7.47    & 1.098 & 9.426   & 8.581   & 0.869 \\
4.7  & 14   & 12.918  & 10.047 & 2.684 & 12.928  & 10.058 & 1.331 & 12.935  & 10.071  & 1.272 & 13.078  & 10.289  & 1.086 & 13.685  & 11.122  & 0.859 \\
4.7  & 17   & 14.9    & 12.291 & 2.508 & 14.91   & 12.297 & 1.324 & 14.917  & 12.305  & 1.27  & 15.056  & 12.472  & 1.088 & 15.616  & 13.145  & 0.866 \\

5    & 6    & 12.788  & 8.892  & 2.814 & 12.797  & 8.905  & 1.386 & 12.816  & 8.916   & 1.326 & 13.043  & 9.168   & 1.122 & 13.812  & 10.137  & 0.879 \\
5    & 9    & 10.431  & 8.6    & 2.507 & 10.44   & 8.609  & 1.367 & 10.441  & 8.627   & 1.312 & 10.601  & 8.903   & 1.122 & 11.357  & 9.897   & 0.894 \\
5    & 12   & 13.121  & 9.882  & 2.715 & 13.131  & 9.898  & 1.358 & 13.139  & 9.912   & 1.297 & 13.3    & 10.146  & 1.1   & 13.935  & 11.026  & 0.864 \\
5    & 14   & 12.371  & 10.122 & 2.666 & 12.387  & 10.128 & 1.341 & 12.393  & 10.141  & 1.285 & 12.55   & 10.364  & 1.096 & 13.21   & 11.203  & 0.868 \\
5    & 17   & 15.854  & 12.722 & 2.731 & 15.869  & 12.729 & 1.349 & 15.872  & 12.739  & 1.29  & 15.986  & 12.924  & 1.096 & 16.506  & 13.621  & 0.863

\end{tabular}
\end{table}

\begin{figure}
\centering
\includegraphics[width=.95\linewidth]{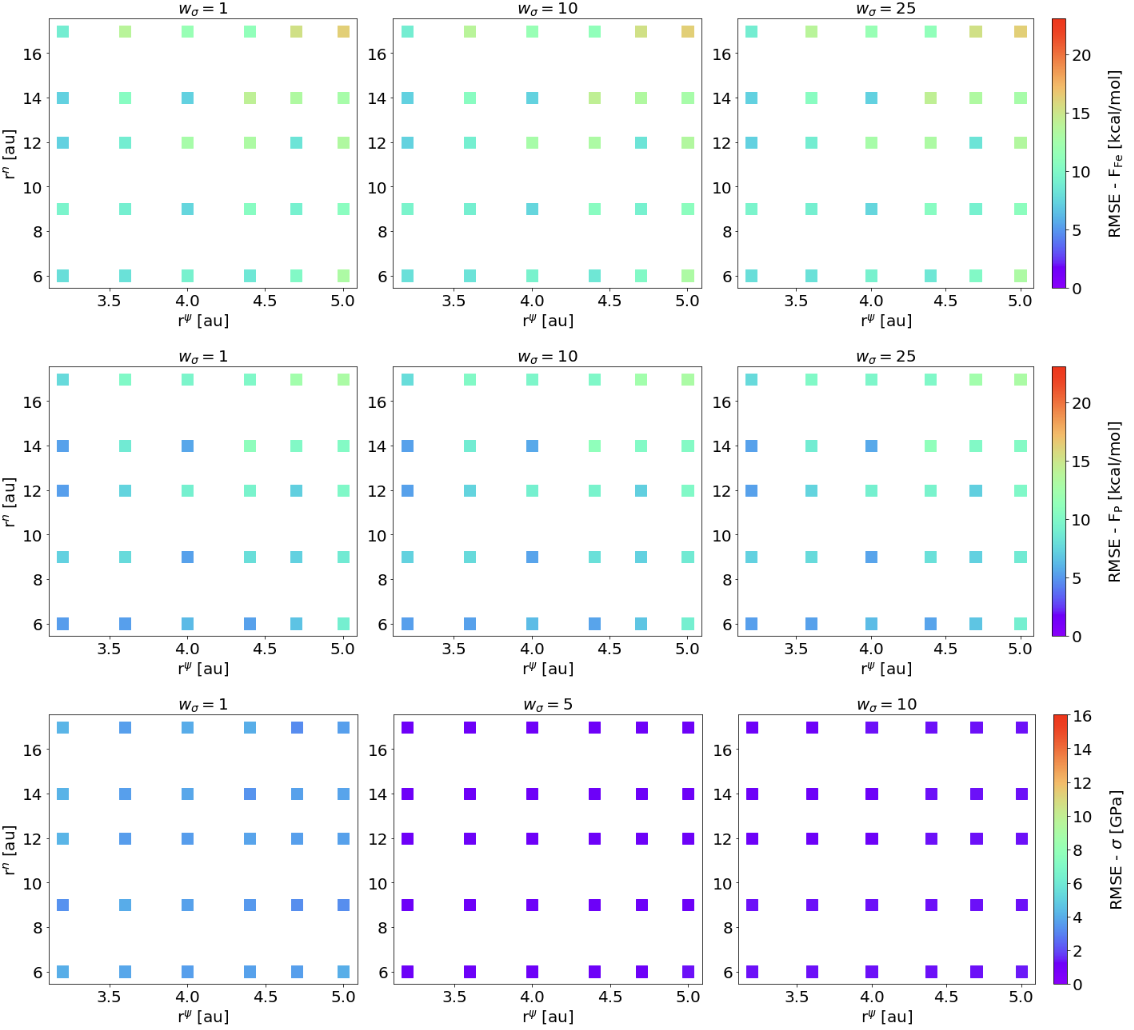}
\caption{RMSE for the \{12,8,4\} set. Top panel: forces on iron atoms, central panel: forces on phosphorous atoms, bottom panel: stress tensor.}
\label{fig:RMSE_12-8-4}
\end{figure}

\begin{table}[]
\tiny
\caption{RMSE for the \{12,8,4\} set. RMSE on forces are expressed in kcal/mol/\AA{}, and RMSE on stress tensor components are expressed in GPa}
\label{tab:RMSE_12-8-4}
\begin{tabular}{ccccccccccc}
     &      & \multicolumn{3}{c}{$w_\sigma$=0} & \multicolumn{3}{c}{$w_\sigma$=5} & \multicolumn{3}{c}{$w_\sigma$=10} \\
$R_\Psi$ & \multicolumn{1}{c}{$R_n$} & Fe    & P     & $\sigma$   & Fe     & P     & $\sigma$  & Fe     & P      & $\sigma$  \\
3.2  & 6    & 7.985   & 5.197  & 4.026 & 8.003   & 5.277  & 1.346 & 8.021   & 5.303   & 1.273 \\
3.2  & 9    & 9.564   & 7.247  & 3.357 & 9.56    & 7.277  & 1.319 & 9.564   & 7.297   & 1.256 \\
3.2  & 12   & 7.277   & 5.306  & 4.236 & 7.294   & 5.369  & 1.312 & 7.3     & 5.398   & 1.237 \\
3.2  & 14   & 7.249   & 5.316  & 4.154 & 7.26    & 5.373  & 1.301 & 7.266   & 5.399   & 1.23  \\
3.2  & 17   & 8.924   & 7.792  & 4.218 & 8.94    & 7.827  & 1.302 & 8.946   & 7.848   & 1.227 \\

3.6  & 6    & 8.18    & 5.279  & 3.83  & 8.193   & 5.346  & 1.353 & 8.213   & 5.368   & 1.281 \\
3.6  & 9    & 9.064   & 7.774  & 3.984 & 9.078   & 7.807  & 1.329 & 9.087   & 7.826   & 1.256 \\
3.6  & 12   & 9.002   & 7.427  & 3.617 & 9.007   & 7.449  & 1.312 & 9.011   & 7.469   & 1.251 \\
3.6  & 14   & 10.63   & 8.815  & 3.668 & 10.644  & 8.837  & 1.305 & 10.651  & 8.855   & 1.238 \\
3.6  & 17   & 13.898  & 10.053 & 3.651 & 13.905  & 10.082 & 1.297 & 13.917  & 10.094  & 1.229 \\

4    & 6    & 9.349   & 6.273  & 3.628 & 9.358   & 6.323  & 1.358 & 9.373   & 6.344   & 1.288 \\
4    & 9    & 7.485   & 5.403  & 3.735 & 7.496   & 5.449  & 1.333 & 7.509   & 5.474   & 1.265 \\
4    & 12   & 12.737  & 9.148  & 3.581 & 12.749  & 9.174  & 1.301 & 12.758  & 9.188   & 1.236 \\
4    & 14   & 7.284   & 5.543  & 3.836 & 7.295   & 5.592  & 1.306 & 7.304   & 5.617   & 1.24  \\
4    & 17   & 11.72   & 9.714  & 3.939 & 11.736  & 9.741  & 1.318 & 11.743  & 9.755   & 1.245 \\
4.4  & 6    & 8.5     & 5.358  & 3.702 & 8.513   & 5.409  & 1.361 & 8.534   & 5.432   & 1.289 \\
4.4  & 9    & 10.621  & 8.011  & 3.521 & 10.628  & 8.034  & 1.347 & 10.636  & 8.053   & 1.28  \\
4.4  & 12   & 13.098  & 9.493  & 3.77  & 13.107  & 9.524  & 1.332 & 13.116  & 9.538   & 1.262 \\
4.4  & 14   & 14.227  & 11.066 & 3.421 & 14.242  & 11.081 & 1.334 & 14.254  & 11.096  & 1.269 \\
4.4  & 17   & 11.297  & 9.912  & 4.028 & 11.318  & 9.936  & 1.321 & 11.328  & 9.952   & 1.248 \\

4.7  & 6    & 10.103  & 6.646  & 3.627 & 10.122  & 6.685  & 1.366 & 10.142  & 6.706   & 1.293 \\
4.7  & 9    & 9.199   & 7.228  & 3.289 & 9.205   & 7.252  & 1.35  & 9.214   & 7.273   & 1.288 \\
4.7  & 12   & 8.326   & 7.152  & 3.684 & 8.345   & 7.182  & 1.327 & 8.358   & 7.2     & 1.259 \\
4.7  & 14   & 13.318  & 10.151 & 3.73  & 13.328  & 10.173 & 1.336 & 13.338  & 10.189  & 1.266 \\
4.7  & 17   & 15.379  & 12.5   & 3.249 & 15.39   & 12.505 & 1.338 & 15.402  & 12.518  & 1.271 \\

5    & 6    & 13.092  & 9.038  & 4.013 & 13.109  & 9.08   & 1.396 & 13.127  & 9.095   & 1.314 \\
5    & 9    & 10.806  & 8.781  & 3.26  & 10.813  & 8.806  & 1.355 & 10.815  & 8.826   & 1.289 \\
5    & 12   & 13.42   & 9.966  & 3.709 & 13.426  & 9.987  & 1.356 & 13.438  & 10.003  & 1.284 \\
5    & 14   & 12.728  & 10.292 & 3.793 & 12.756  & 10.312 & 1.337 & 12.766  & 10.328  & 1.266 \\
5    & 17   & 16.316  & 12.983 & 3.664 & 16.34   & 12.997 & 1.346 & 16.345  & 13.011  & 1.275 
\end{tabular}
\end{table}
\newpage
\section{Lattice constants}

\begin{figure}
\centering
\includegraphics[width=.95\linewidth]{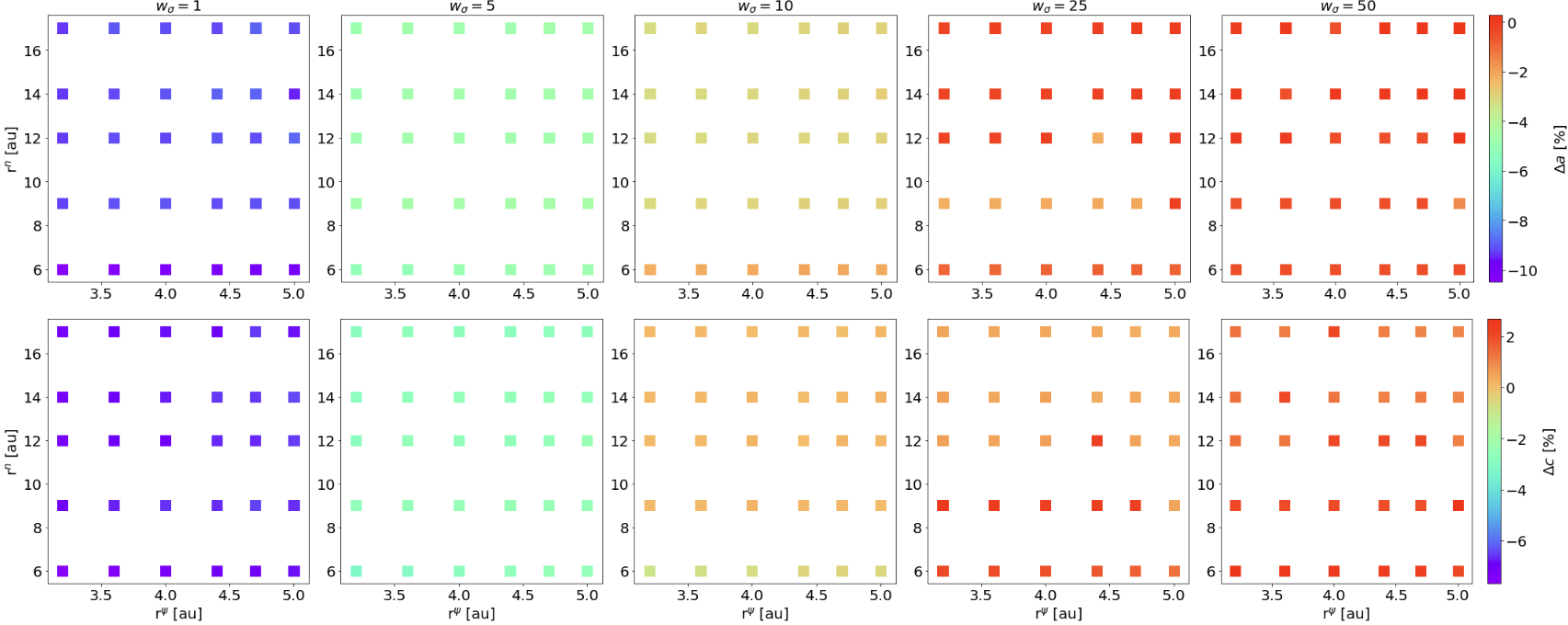}
\caption{Percentage deviation from DFT values of schreibersite lattice constant for the \{8,4,0\} set. Top panel: lattice constant $a$, bottom panel: lattice constant $c$.}
\label{fig:LATCONST_8-4-0}
\end{figure}

\begin{figure}
\centering
\includegraphics[width=.95\linewidth]{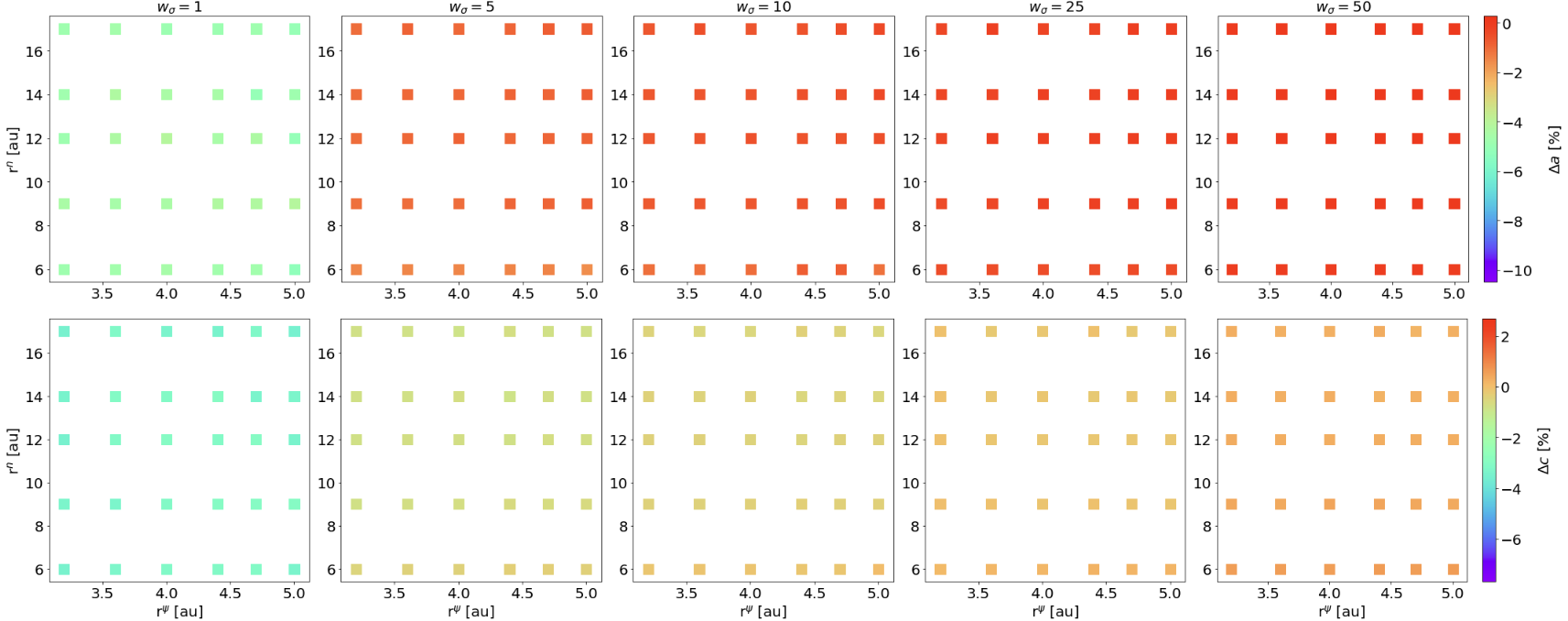}
\caption{Percentage deviation from DFT values of schreibersite lattice constant for the \{12,8,0\} set. Top panel: lattice constant $a$, bottom panel: lattice constant $c$.}
\label{fig:LATCONST_12-8-0}
\end{figure}

\begin{figure}
\centering
\includegraphics[width=.95\linewidth]{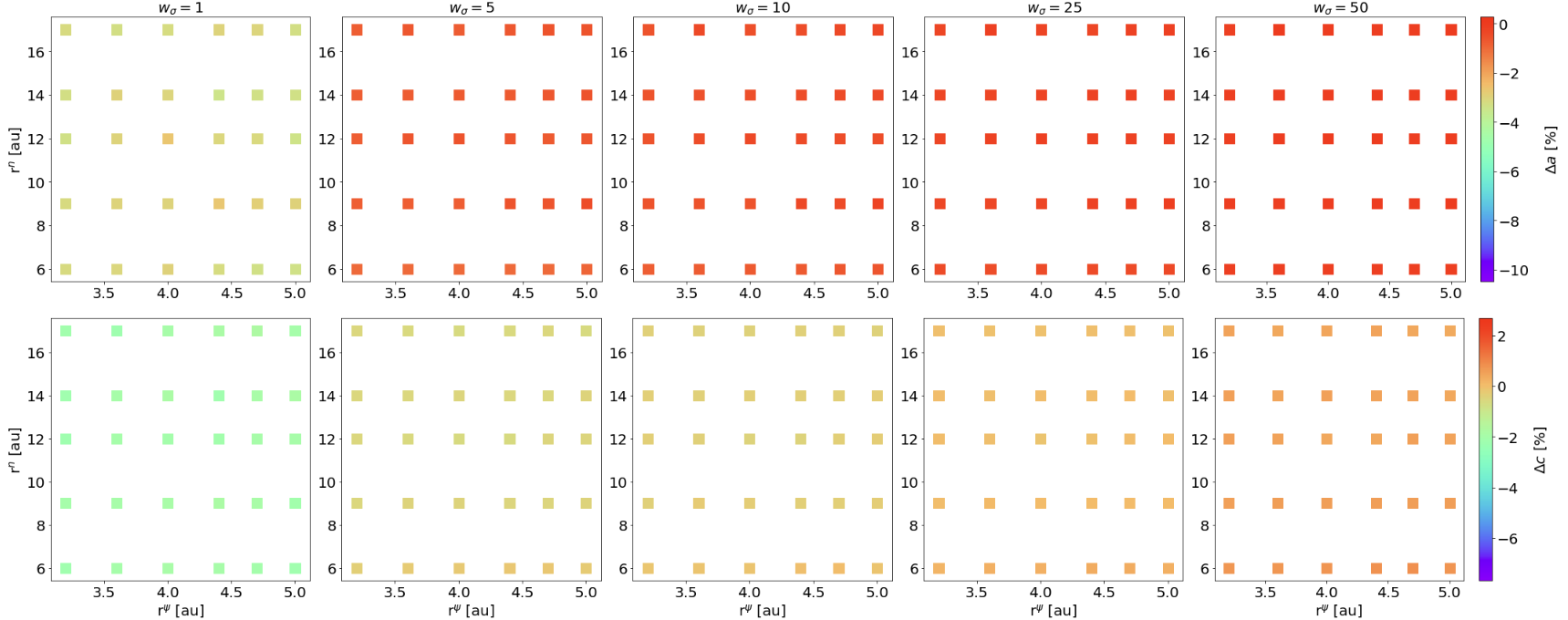}
\caption{Percentage deviation from DFT values of schreibersite lattice constant for the \{16,12,0\} set. Top panel: lattice constant $a$, bottom panel: lattice constant $c$.}
\label{fig:LATCONST_16-12-0}
\end{figure}

\begin{figure}
\centering
\includegraphics[width=.95\linewidth]{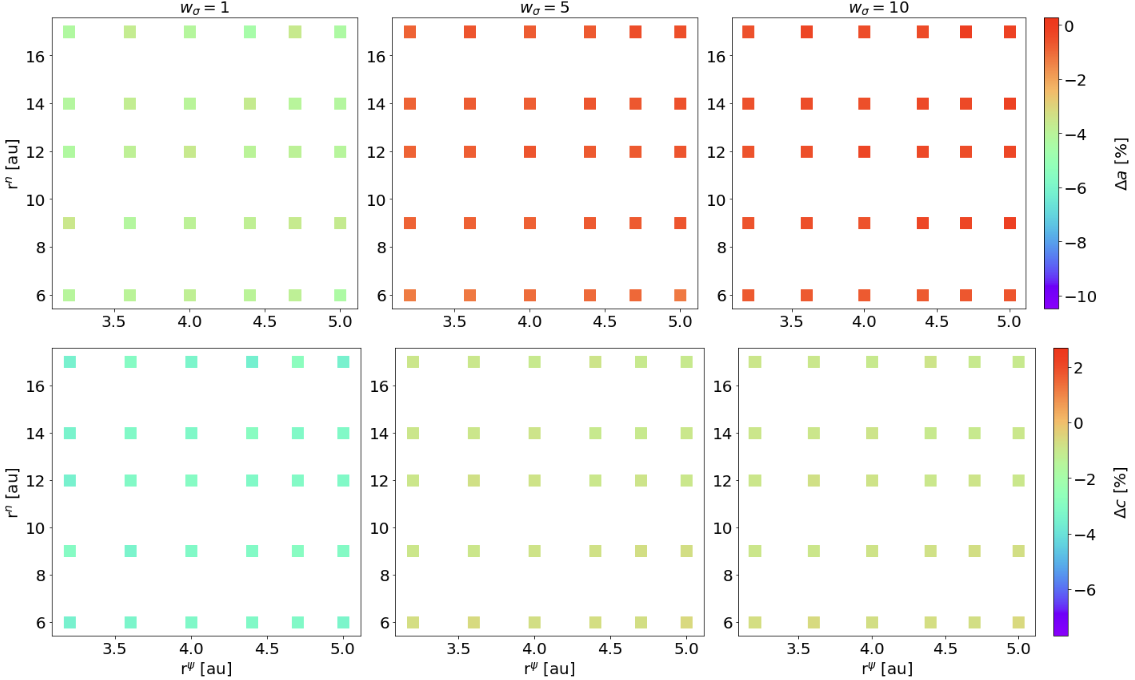}
\caption{Percentage deviation from DFT values of schreibersite lattice constant for the \{12,8,4\} set. Top panel: lattice constant $a$, bottom panel: lattice constant $c$.}
\label{fig:LATCONST_12-8-4}
\end{figure}

\begin{table}[]
\small
\caption{Lattice constants value for $w_\sigma=1$ set.}
\label{tab:RMSE_12-8-4}
\begin{tabular}{cccccccccc}
     &      & \multicolumn{4}{c}{a {[}Å{]}}  & \multicolumn{4}{c}{c {[}Å{]}}  \\ \cline{3-10} 
$R_\Psi$ & \multicolumn{1}{c}{$R_n$} & \{8,4,0\}   & \{12,8,0\}  & \{16,12,0\} & \{12,8,4\} & \{8,4,0\}  & \{12,8,0\}  & \{16,12,0\} & \{12,8,4\} \\ \hline
3.2  & 6    & 8.098 & 8.611 & 8.759 & 8.676  & 4.044 & 4.229 & 4.286 & 4.228  \\
3.2  & 9    & 8.211 & 8.641 & 8.759 & 8.724  & 4.076 & 4.234 & 4.293 & 4.249  \\
3.2  & 12   & 8.205 & 8.609 & 8.744 & 8.660  & 4.068 & 4.228 & 4.284 & 4.228  \\
3.2  & 14   & 8.203 & 8.617 & 8.749 & 8.667  & 4.067 & 4.231 & 4.286 & 4.230  \\
3.2  & 17   & 8.202 & 8.612 & 8.758 & 8.663  & 4.066 & 4.228 & 4.287 & 4.227  \\

3.6  & 6    & 8.113 & 8.623 & 8.767 & 8.687  & 4.054 & 4.236 & 4.292 & 4.235  \\
3.6  & 9    & 8.224 & 8.636 & 8.775 & 8.670  & 4.085 & 4.235 & 4.294 & 4.232  \\
3.6  & 12   & 8.221 & 8.653 & 8.774 & 8.697  & 4.081 & 4.243 & 4.292 & 4.243  \\
3.6  & 14   & 8.216 & 8.658 & 8.778 & 8.705  & 4.078 & 4.246 & 4.295 & 4.241  \\
3.6  & 17   & 8.232 & 8.629 & 8.749 & 8.709  & 4.079 & 4.241 & 4.284 & 4.248  \\

4    & 6    & 8.128 & 8.636 & 8.777 & 8.700  & 4.064 & 4.241 & 4.296 & 4.240  \\
4    & 9    & 8.228 & 8.642 & 8.776 & 8.693  & 4.091 & 4.243 & 4.297 & 4.241  \\
4    & 12   & 8.215 & 8.673 & 8.805 & 8.721  & 4.077 & 4.249 & 4.291 & 4.248  \\
4    & 14   & 8.227 & 8.642 & 8.773 & 8.684  & 4.083 & 4.245 & 4.299 & 4.240  \\
4    & 17   & 8.222 & 8.605 & 8.756 & 8.674  & 4.082 & 4.232 & 4.294 & 4.234  \\

4.4  & 6    & 8.141 & 8.620 & 8.756 & 8.689  & 4.071 & 4.243 & 4.294 & 4.242  \\
4.4  & 9    & 8.222 & 8.661 & 8.800 & 8.701  & 4.089 & 4.248 & 4.299 & 4.244  \\
4.4  & 12   & 8.227 & 8.638 & 8.766 & 8.687  & 4.087 & 4.245 & 4.289 & 4.241  \\
4.4  & 14   & 8.243 & 8.637 & 8.732 & 8.713  & 4.093 & 4.251 & 4.285 & 4.254  \\
4.4  & 17   & 8.219 & 8.604 & 8.776 & 8.636  & 4.080 & 4.230 & 4.298 & 4.224  \\

4.7  & 6    & 8.152 & 8.620 & 8.746 & 8.696  & 4.075 & 4.245 & 4.290 & 4.242  \\
4.7  & 9    & 8.226 & 8.663 & 8.785 & 8.719  & 4.096 & 4.248 & 4.296 & 4.252  \\
4.7  & 12   & 8.221 & 8.654 & 8.763 & 8.693  & 4.087 & 4.248 & 4.295 & 4.244  \\
4.7  & 14   & 8.244 & 8.584 & 8.743 & 8.680  & 4.092 & 4.232 & 4.296 & 4.242  \\
4.7  & 17   & 8.243 & 8.613 & 8.772 & 8.722  & 4.093 & 4.239 & 4.303 & 4.257  \\

5    & 6    & 8.151 & 8.568 & 8.741 & 8.651  & 4.082 & 4.233 & 4.286 & 4.235  \\
5    & 9    & 8.222 & 8.672 & 8.779 & 8.714  & 4.090 & 4.248 & 4.289 & 4.247  \\
5    & 12   & 8.246 & 8.581 & 8.736 & 8.677  & 4.095 & 4.231 & 4.294 & 4.242  \\
5    & 14   & 8.188 & 8.613 & 8.743 & 8.667  & 4.099 & 4.235 & 4.299 & 4.241  \\
5    & 17   & 8.222 & 8.590 & 8.748 & 8.653  & 4.082 & 4.226 & 4.292 & 4.228  

\end{tabular}
\end{table}

\begin{table}[]
\small
\caption{Lattice constants value for $w_\sigma=5$ set.}
\label{tab:RMSE_12-8-4}
\begin{tabular}{cccccccccc}
     &      & \multicolumn{4}{c}{a {[}Å{]}}  & \multicolumn{4}{c}{c {[}Å{]}}  \\ \cline{3-10} 
$R_\Psi$ & \multicolumn{1}{c}{$R_n$} & \{8,4,0\}   & \{12,8,0\}  & \{16,12,0\} & \{12,8,4\} & \{8,4,0\}  & \{12,8,0\}  & \{16,12,0\} & \{12,8,4\} \\ \hline
3.2  & 6    & 8.577 & 8.910 & 8.952 & 8.928  & 4.240 & 4.358 & 4.366 & 4.350  \\
3.2  & 9    & 8.625 & 8.946 & 8.971 & 8.966  & 4.265 & 4.346 & 4.356 & 4.338  \\
3.2  & 12   & 8.611 & 8.948 & 8.972 & 8.965  & 4.259 & 4.345 & 4.356 & 4.339  \\
3.2  & 14   & 8.609 & 8.950 & 8.973 & 8.966  & 4.258 & 4.346 & 4.356 & 4.339  \\
3.2  & 17   & 8.607 & 8.950 & 8.972 & 8.965  & 4.257 & 4.345 & 4.356 & 4.339  \\

3.6  & 6    & 8.586 & 8.911 & 8.952 & 8.929  & 4.249 & 4.361 & 4.368 & 4.352  \\
3.6  & 9    & 8.633 & 8.951 & 8.971 & 8.968  & 4.269 & 4.348 & 4.360 & 4.339  \\
3.6  & 12   & 8.623 & 8.956 & 8.986 & 8.974  & 4.267 & 4.348 & 4.355 & 4.345  \\
3.6  & 14   & 8.618 & 8.957 & 8.977 & 8.975  & 4.262 & 4.346 & 4.356 & 4.339  \\
3.6  & 17   & 8.620 & 8.964 & 8.984 & 8.988  & 4.262 & 4.344 & 4.355 & 4.337  \\

4    & 6    & 8.595 & 8.913 & 8.955 & 8.948  & 4.257 & 4.362 & 4.370 & 4.347  \\
4    & 9    & 8.633 & 8.950 & 8.974 & 8.967  & 4.273 & 4.349 & 4.359 & 4.342  \\
4    & 12   & 8.623 & 8.966 & 8.991 & 8.987  & 4.265 & 4.347 & 4.353 & 4.343  \\
4    & 14   & 8.621 & 8.957 & 8.980 & 8.973  & 4.267 & 4.349 & 4.358 & 4.341  \\
4    & 17   & 8.614 & 8.957 & 8.977 & 8.976  & 4.260 & 4.343 & 4.353 & 4.335  \\

4.4  & 6    & 8.605 & 8.914 & 8.953 & 8.954  & 4.265 & 4.366 & 4.373 & 4.350  \\
4.4  & 9    & 8.634 & 8.961 & 8.993 & 8.978  & 4.273 & 4.352 & 4.358 & 4.344  \\
4.4  & 12   & 8.624 & 8.959 & 8.986 & 8.981  & 4.266 & 4.345 & 4.352 & 4.337  \\
4.4  & 14   & 8.625 & 8.965 & 8.978 & 8.988  & 4.268 & 4.343 & 4.356 & 4.335  \\
4.4  & 17   & 8.614 & 8.960 & 8.982 & 8.979  & 4.259 & 4.349 & 4.355 & 4.340  \\

4.7  & 6    & 8.613 & 8.915 & 8.955 & 8.957  & 4.270 & 4.365 & 4.373 & 4.348  \\
4.7  & 9    & 8.636 & 8.960 & 8.989 & 8.979  & 4.275 & 4.351 & 4.358 & 4.347  \\
4.7  & 12   & 8.622 & 8.960 & 8.980 & 8.978  & 4.268 & 4.349 & 4.358 & 4.340  \\
4.7  & 14   & 8.626 & 8.961 & 8.985 & 8.984  & 4.270 & 4.347 & 4.356 & 4.335  \\
4.7  & 17   & 8.629 & 8.973 & 8.992 & 9.002  & 4.262 & 4.346 & 4.356 & 4.335  \\

5    & 6    & 8.610 & 8.901 & 8.960 & 8.930  & 4.268 & 4.363 & 4.369 & 4.355  \\
5    & 9    & 8.637 & 8.973 & 9.003 & 8.989  & 4.271 & 4.352 & 4.360 & 4.347  \\
5    & 12   & 8.629 & 8.963 & 8.982 & 8.987  & 4.275 & 4.347 & 4.358 & 4.336  \\
5    & 14   & 8.627 & 8.973 & 8.992 & 8.995  & 4.268 & 4.345 & 4.359 & 4.336  \\
5    & 17   & 8.620 & 8.976 & 8.992 & 9.000  & 4.261 & 4.345 & 4.352 & 4.335  

\end{tabular}
\end{table}

\begin{table}[]
\small
\caption{Lattice constants value for $w_\sigma=10$ set.}
\label{tab:RMSE_12-8-4}
\begin{tabular}{cccccccccc}
     &      & \multicolumn{4}{c}{a {[}Å{]}}  & \multicolumn{4}{c}{c {[}Å{]}}  \\ \cline{3-10} 
$R_\Psi$ & \multicolumn{1}{c}{$R_n$} & \{8,4,0\}   & \{12,8,0\}  & \{16,12,0\} & \{12,8,4\} & \{8,4,0\}  & \{12,8,0\}  & \{16,12,0\} & \{12,8,4\} \\ \hline
3.2 & 6  & 8.847 & 8.947 & 8.974 & 8.978 & 4.341 & 4.374 & 4.374 & 4.356 \\
3.2 & 9  & 8.765 & 8.978 & 8.991 & 8.990 & 4.385 & 4.361 & 4.364 & 4.350 \\
3.2 & 12 & 8.758 & 8.982 & 8.992 & 8.993 & 4.385 & 4.361 & 4.365 & 4.351 \\
3.2 & 14 & 8.757 & 8.983 & 8.993 & 8.994 & 4.386 & 4.361 & 4.365 & 4.352 \\
3.2 & 17 & 8.757 & 8.982 & 8.992 & 8.995 & 4.385 & 4.360 & 4.364 & 4.351 \\

3.6 & 6  & 8.851 & 8.946 & 8.977 & 8.978 & 4.346 & 4.376 & 4.375 & 4.357 \\
3.6 & 9  & 8.773 & 8.984 & 8.991 & 8.997 & 4.384 & 4.362 & 4.368 & 4.352 \\
3.6 & 12 & 8.765 & 8.988 & 9.004 & 8.999 & 4.385 & 4.362 & 4.363 & 4.355 \\
3.6 & 14 & 8.764 & 8.988 & 8.996 & 9.001 & 4.384 & 4.359 & 4.363 & 4.350 \\
3.6 & 17 & 8.771 & 8.997 & 9.010 & 9.019 & 4.384 & 4.357 & 4.361 & 4.345 \\

4   & 6  & 8.854 & 8.949 & 8.979 & 8.978 & 4.351 & 4.376 & 4.376 & 4.359 \\
4   & 9  & 8.773 & 8.982 & 8.993 & 8.993 & 4.388 & 4.363 & 4.367 & 4.353 \\
4   & 12 & 8.772 & 8.996 & 9.012 & 9.018 & 4.389 & 4.360 & 4.360 & 4.350 \\
4   & 14 & 8.769 & 8.989 & 8.999 & 8.999 & 4.389 & 4.362 & 4.366 & 4.353 \\
4   & 17 & 8.765 & 8.989 & 8.999 & 9.003 & 4.383 & 4.357 & 4.361 & 4.347 \\

4.4 & 6  & 8.861 & 8.972 & 8.980 & 8.981 & 4.358 & 4.374 & 4.380 & 4.362 \\
4.4 & 9  & 8.777 & 8.992 & 9.010 & 9.018 & 4.386 & 4.365 & 4.365 & 4.350 \\
4.4 & 12 & 8.770 & 8.992 & 9.006 & 9.009 & 4.382 & 4.358 & 4.360 & 4.348 \\
4.4 & 14 & 8.774 & 8.996 & 9.002 & 9.011 & 4.382 & 4.356 & 4.365 & 4.344 \\
4.4 & 17 & 8.770 & 8.993 & 9.001 & 9.008 & 4.384 & 4.363 & 4.363 & 4.351 \\

4.7 & 6  & 8.859 & 8.974 & 8.983 & 8.983 & 4.360 & 4.374 & 4.381 & 4.361 \\
4.7 & 9  & 8.777 & 8.991 & 9.008 & 9.018 & 4.387 & 4.364 & 4.366 & 4.353 \\
4.7 & 12 & 8.771 & 8.991 & 9.001 & 9.004 & 4.388 & 4.361 & 4.366 & 4.351 \\
4.7 & 14 & 8.780 & 8.994 & 9.006 & 9.010 & 4.387 & 4.360 & 4.366 & 4.346 \\
4.7 & 17 & 8.780 & 9.003 & 9.011 & 9.036 & 4.380 & 4.359 & 4.364 & 4.341 \\

5   & 6  & 8.856 & 8.941 & 8.986 & 8.988 & 4.357 & 4.381 & 4.376 & 4.361 \\
5   & 9  & 8.782 & 9.004 & 9.022 & 9.031 & 4.383 & 4.364 & 4.367 & 4.352 \\
5   & 12 & 8.781 & 8.997 & 9.004 & 9.013 & 4.386 & 4.361 & 4.367 & 4.347 \\
5   & 14 & 8.789 & 9.011 & 9.010 & 9.027 & 4.391 & 4.357 & 4.367 & 4.345 \\
5   & 17 & 8.779 & 9.012 & 9.013 & 9.025 & 4.386 & 4.357 & 4.361 & 4.345 

\end{tabular}
\end{table}

\begin{table}[]
\small
\caption{Lattice constants value for $w_\sigma=25$ set.}
\begin{tabular}{cccccccc}
     &      & \multicolumn{3}{c}{a {[}Å{]}}  & \multicolumn{3}{c}{c {[}Å{]}}  \\ \cline{3-8} 
$R_\Psi$ & \multicolumn{1}{c}{$R_n$} & \{8,4,0\}   & \{12,8,0\}  & \{16,12,0\} & \{8,4,0\}  & \{12,8,0\}  & \{16,12,0\} \\ \hline
3.2  & 6    & 8.960    & 9.004    & 9.005   & 4.472    & 4.384    & 4.389   \\
3.2  & 9    & 8.843    & 9.008    & 9.015   & 4.489    & 4.379    & 4.381   \\
3.2  & 12   & 9.020    & 9.013    & 9.016   & 4.404    & 4.378    & 4.381   \\
3.2  & 14   & 9.020    & 9.013    & 9.016   & 4.404    & 4.377    & 4.381   \\
3.2  & 17   & 9.021    & 9.013    & 9.016   & 4.401    & 4.377    & 4.380   \\

3.6  & 6    & 8.961    & 9.003    & 9.004   & 4.468    & 4.386    & 4.391   \\
3.6  & 9    & 8.852    & 9.015    & 9.014   & 4.484    & 4.378    & 4.384   \\
3.6  & 12   & 9.021    & 9.030    & 9.024   & 4.401    & 4.374    & 4.380   \\
3.6  & 14   & 9.020    & 9.018    & 9.019   & 4.399    & 4.373    & 4.379   \\
3.6  & 17   & 9.028    & 9.030    & 9.029   & 4.398    & 4.371    & 4.377   \\

4    & 6    & 8.964    & 9.004    & 9.005   & 4.462    & 4.387    & 4.392   \\
4    & 9    & 8.854    & 9.025    & 9.017   & 4.481    & 4.376    & 4.382   \\
4    & 12   & 9.028    & 9.030    & 9.030   & 4.404    & 4.373    & 4.376   \\
4    & 14   & 9.026    & 9.030    & 9.023   & 4.404    & 4.374    & 4.381   \\
4    & 17   & 9.022    & 9.021    & 9.020   & 4.399    & 4.373    & 4.378   \\

4.4  & 6    & 8.967    & 9.004    & 9.005   & 4.457    & 4.390    & 4.395   \\
4.4  & 9    & 8.859    & 9.030    & 9.025   & 4.479    & 4.377    & 4.382   \\
4.4  & 12   & 8.852    & 9.029    & 9.025   & 4.480    & 4.372    & 4.377   \\
4.4  & 14   & 9.028    & 9.026    & 9.024   & 4.396    & 4.371    & 4.381   \\
4.4  & 17   & 9.032    & 9.023    & 9.019   & 4.398    & 4.377    & 4.380   \\

4.7  & 6    & 8.963    & 9.006    & 9.007   & 4.451    & 4.389    & 4.396   \\
4.7  & 9    & 8.858    & 9.031    & 9.026   & 4.479    & 4.375    & 4.382   \\
4.7  & 12   & 9.028    & 9.032    & 9.024   & 4.403    & 4.373    & 4.381   \\
4.7  & 14   & 9.035    & 9.032    & 9.024   & 4.399    & 4.372    & 4.382   \\
4.7  & 17   & 9.035    & 9.037    & 9.026   & 4.393    & 4.370    & 4.379   \\

5    & 6    & 8.964    & 9.005    & 9.009   & 4.440    & 4.388    & 4.392   \\
5    & 9    & 9.033    & 9.041    & 9.034   & 4.400    & 4.374    & 4.383   \\
5    & 12   & 9.035    & 9.033    & 9.024   & 4.398    & 4.373    & 4.382   \\
5    & 14   & 9.046    & 9.038    & 9.027   & 4.399    & 4.371    & 4.380   \\
5    & 17   & 9.041    & 9.037    & 9.029   & 4.397    & 4.372    & 4.377   

\end{tabular}
\end{table}

\begin{table}[]
\small
\caption{Lattice constants value for $w_\sigma=50$ set.}
\begin{tabular}{cccccccc}
     &      & \multicolumn{3}{c}{a {[}Å{]}}  & \multicolumn{3}{c}{c {[}Å{]}}  \\ \cline{3-8} 
$R_\Psi$ & \multicolumn{1}{c}{$R_n$} & \{8,4,0\}   & \{12,8,0\}  & \{16,12,0\} & \{8,4,0\}  & \{12,8,0\}  & \{16,12,0\} \\ \hline
3.2 & 6  & 9.000 & 9.039 & 9.033 & 4.497 & 4.405 & 4.411 \\
3.2 & 9  & 8.995 & 9.041 & 9.036 & 4.472 & 4.398 & 4.404 \\
3.2 & 12 & 9.049 & 9.050 & 9.033 & 4.439 & 4.395 & 4.404 \\
3.2 & 14 & 9.048 & 9.050 & 9.032 & 4.440 & 4.394 & 4.404 \\
3.2 & 17 & 9.048 & 9.050 & 9.031 & 4.437 & 4.393 & 4.403 \\
3.6 & 6  & 9.002 & 9.039 & 9.033 & 4.492 & 4.406 & 4.412 \\
3.6 & 9  & 9.000 & 9.046 & 9.034 & 4.468 & 4.397 & 4.406 \\
3.6 & 12 & 9.049 & 9.053 & 9.040 & 4.434 & 4.394 & 4.402 \\
3.6 & 14 & 8.994 & 9.050 & 9.035 & 4.472 & 4.392 & 4.401 \\
3.6 & 17 & 9.055 & 9.051 & 9.043 & 4.430 & 4.392 & 4.399 \\
4   & 6  & 9.001 & 9.039 & 9.033 & 4.487 & 4.405 & 4.413 \\
4   & 9  & 9.001 & 9.043 & 9.035 & 4.467 & 4.400 & 4.406 \\
4   & 12 & 9.002 & 9.051 & 9.044 & 4.471 & 4.394 & 4.399 \\
4   & 14 & 9.054 & 9.054 & 9.036 & 4.436 & 4.394 & 4.404 \\
4   & 17 & 8.997 & 9.053 & 9.040 & 4.471 & 4.392 & 4.400 \\
4.4 & 6  & 9.002 & 9.039 & 9.032 & 4.483 & 4.408 & 4.415 \\
4.4 & 9  & 9.003 & 9.046 & 9.041 & 4.466 & 4.400 & 4.405 \\
4.4 & 12 & 9.000 & 9.052 & 9.043 & 4.467 & 4.393 & 4.399 \\
4.4 & 14 & 9.056 & 9.052 & 9.042 & 4.427 & 4.390 & 4.402 \\
4.4 & 17 & 9.060 & 9.054 & 9.036 & 4.429 & 4.395 & 4.403 \\
4.7 & 6  & 8.997 & 9.040 & 9.032 & 4.478 & 4.406 & 4.416 \\
4.7 & 9  & 9.002 & 9.047 & 9.042 & 4.464 & 4.398 & 4.405 \\
4.7 & 12 & 9.004 & 9.055 & 9.038 & 4.469 & 4.393 & 4.404 \\
4.7 & 14 & 9.060 & 9.053 & 9.042 & 4.428 & 4.391 & 4.403 \\
4.7 & 17 & 9.060 & 9.057 & 9.042 & 4.424 & 4.389 & 4.399 \\
5   & 6  & 9.001 & 9.039 & 9.032 & 4.470 & 4.407 & 4.413 \\
5   & 9  & 8.905 & 9.052 & 9.046 & 4.497 & 4.397 & 4.406 \\
5   & 12 & 9.059 & 9.053 & 9.041 & 4.427 & 4.393 & 4.403 \\
5   & 14 & 9.069 & 9.057 & 9.041 & 4.425 & 4.391 & 4.399 \\
5   & 17 & 9.066 & 9.056 & 9.045 & 4.423 & 4.392 & 4.397 
\end{tabular}
\end{table}

\newpage
\section{Bulk modulus}
\begin{figure}
\centering
\includegraphics[width=.95\linewidth]{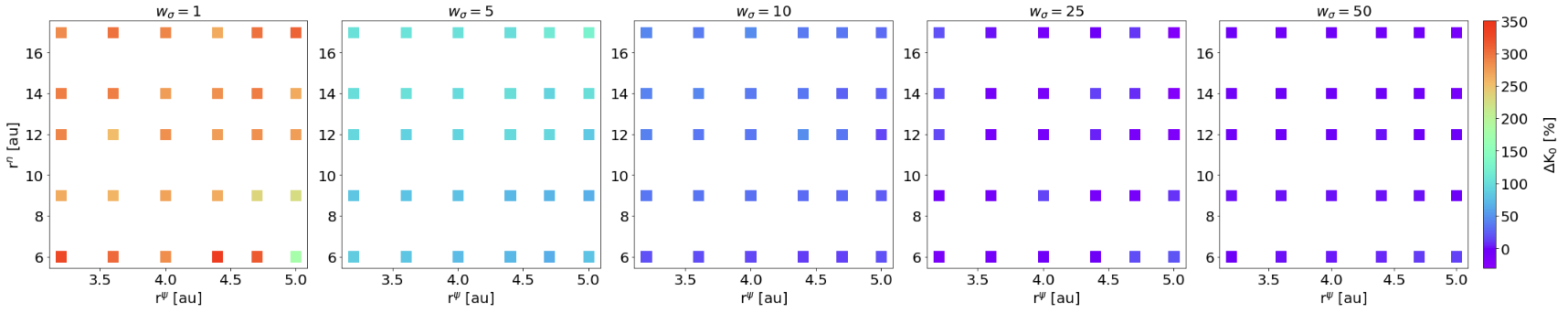}
\caption{Percentage deviation from DFT bulk modulus of schreibersite for the \{8,4,0\} set.}
\label{fig:BULKMOD_8-4-0}
\end{figure}

\begin{table}[]
\small
\caption{Bulk modulus of schreibersite for the \{8,4,0\} set as a function of $w_\sigma$.}
\begin{tabular}{ccccccc}
$R_\Psi$ & $R_n$ & $w_\sigma$=0    & $w_\sigma$=5    & $w_\sigma$=10   & $w_\sigma$=25   & $w_\sigma$=50   \\ \hline
3.2  & 6    & 675.325 & 342.759 & 243.748 & 203.069 & 168.867 \\
3.2  & 9    & 662.245 & 350.413 & 246.413 & 205.133 & 170.794 \\
3.2  & 12   & 735.415 & 321.744 & 206.077 & 148.208 & 169.259 \\
3.2  & 14   & 627.185 & 311.802 & 233.186 & 159.435 & 175.474 \\
3.2  & 17   & 668.858 & 335.509 & 241.636 & 201.894 & 167.937 \\
3.6  & 6    & 676.858 & 349.414 & 246.385 & 157.412 & 172.502 \\
3.6  & 9    & 688.528 & 353.997 & 240.218 & 175.252 & 168.672 \\
3.6  & 12   & 696.006 & 309.981 & 204.119 & 157.474 & 174.756 \\
3.6  & 14   & 616.953 & 309.119 & 230.318 & 156.93  & 174.374 \\
3.6  & 17   & 606.281 & 324.727 & 234.048 & 154.396 & 170.639 \\
4    & 6    & 678.742 & 330.986 & 219.6   & 185.257 & 167.44  \\
4    & 9    & 689.055 & 367.513 & 233.17  & 190.717 & 168.774 \\
4    & 12   & 713.52  & 284.545 & 196.738 & 203.722 & 192.399 \\
4    & 14   & 573.671 & 292.521 & 220.705 & 153.729 & 174.817 \\
4    & 17   & 656.929 & 336.166 & 234.02  & 150.631 & 170.831 \\
4.4  & 6    & 642.395 & 340.027 & 235.781 & 150.74  & 169.036 \\
4.4  & 9    & 667.769 & 349.864 & 250.896 & 153.864 & 170.42  \\
4.4  & 12   & 659.458 & 300.995 & 200.435 & 164.761 & 178.158 \\
4.4  & 14   & 637.586 & 306.621 & 228.601 & 197.872 & 175.037 \\
4.4  & 17   & 657.356 & 330.475 & 232.995 & 154.092 & 173.092 \\
4.7  & 6    & 657.437 & 347.029 & 233.364 & 194.453 & 170.411 \\
4.7  & 9    & 629.247 & 343.109 & 235.669 & 171.915 & 168.183 \\
4.7  & 12   & 756.609 & 294.303 & 192.513 & 169.591 & 183.54  \\
4.7  & 14   & 626.188 & 295.696 & 222.525 & 154.077 & 175.214 \\
4.7  & 17   & 642.74  & 339.572 & 253.669 & 180.7   & 174.726 \\
5    & 6    & 630.793 & 345.494 & 215.077 & 132.961 & 155.049 \\
5    & 9    & 704.353 & 387.903 & 224.352 & 134.715 & 154.825 \\
5    & 12   & 479.598 & 308.978 & 204.761 & 208.449 & 195.071 \\
5    & 14   & 563.202 & 287.916 & 212.805 & 188.023 & 172.169 \\
5    & 17   & 641.292 & 316.81  & 197.466 & 139.63  & 171.892 
\end{tabular}
\end{table}

\begin{figure}
\centering
\includegraphics[width=.95\linewidth]{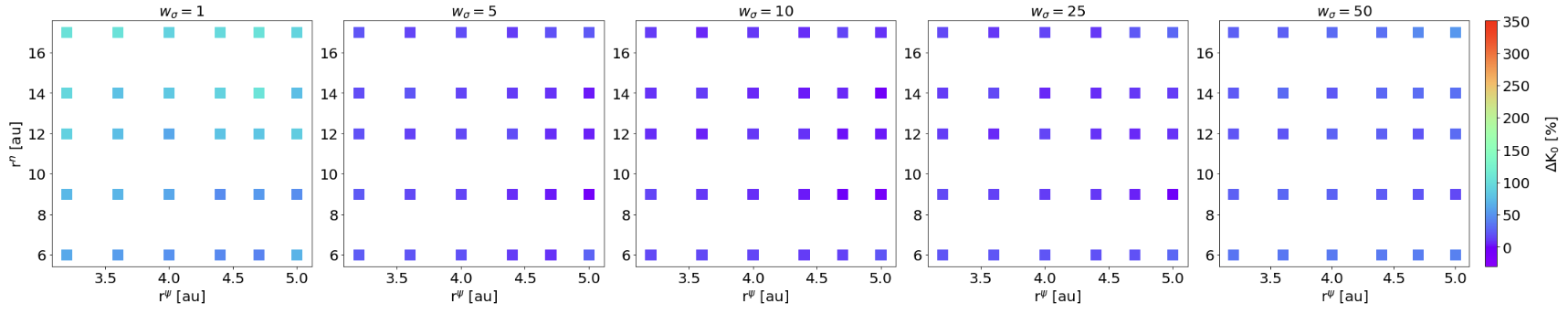}
\caption{Percentage deviation from DFT bulk modulus of schreibersite for the \{12,8,0\} set.}
\label{fig:BULKMOD_12-8-0}
\end{figure}

\begin{table}[]
\small
\caption{Bulk modulus of schreibersite for the \{12,8,0\} set as a function of $w_\sigma$.}
\begin{tabular}{ccccccc}
$R_\Psi$ & $R_n$ & $w_\sigma$=0    & $w_\sigma$=5    & $w_\sigma$=10   & $w_\sigma$=25   & $w_\sigma$=50   \\ \hline
3.2 & 6  & 337.127 & 204.39  & 189.112 & 194.282 & 212.12  \\
3.2 & 9  & 351.534 & 209.17  & 193.049 & 200.285 & 215.157 \\
3.2 & 12 & 281.093 & 212.717 & 201.052 & 207.38  & 231.889 \\
3.2 & 14 & 294.69  & 211.785 & 197.587 & 200.669 & 217.166 \\
3.2 & 17 & 326.733 & 204.756 & 186.566 & 192.311 & 209.663 \\
3.6 & 6  & 305.802 & 207.806 & 192.943 & 200.661 & 224.283 \\
3.6 & 9  & 351.242 & 198.063 & 183.487 & 190.411 & 214.892 \\
3.6 & 12 & 267.131 & 209.717 & 199.685 & 208.94  & 233.316 \\
3.6 & 14 & 292.685 & 201.877 & 189.385 & 196.483 & 220.135 \\
3.6 & 17 & 302.218 & 196.867 & 180.646 & 183.137 & 209.437 \\
4   & 6  & 353.376 & 187.784 & 182.052 & 190.908 & 225.89  \\
4   & 9  & 351.782 & 207.044 & 199.036 & 212.209 & 249.162 \\
4   & 12 & 245.048 & 187.762 & 196.091 & 206.18  & 238.49  \\
4   & 14 & 263.632 & 178.359 & 170.478 & 181.312 & 208.045 \\
4   & 17 & 310.676 & 186.559 & 175.534 & 182.324 & 209.66  \\
4.4 & 6  & 316.693 & 197.139 & 182.133 & 184.039 & 210.968 \\
4.4 & 9  & 328.699 & 202.955 & 191.813 & 197.563 & 224.648 \\
4.4 & 12 & 257.153 & 205.793 & 197.542 & 207.88  & 235.483 \\
4.4 & 14 & 276.699 & 198.571 & 186.425 & 194.256 & 218.154 \\
4.4 & 17 & 277.265 & 200.811 & 190.091 & 198.148 & 219.259 \\
4.7 & 6  & 328.308 & 194.293 & 178.705 & 186.03  & 221.904 \\
4.7 & 9  & 339.171 & 192.861 & 185.148 & 190.508 & 227.909 \\
4.7 & 12 & 250.206 & 192.34  & 193.748 & 206.294 & 237.234 \\
4.7 & 14 & 251.466 & 185.405 & 178.139 & 190.712 & 213.043 \\
4.7 & 17 & 308.688 & 204.342 & 188.687 & 189.028 & 215.767 \\
5   & 6  & 302.367 & 178.839 & 167.568 & 192.761 & 228.907 \\
5   & 9  & 330.202 & 210.378 & 187.4   & 222.454 & 263.161 \\
5   & 12 & 289.726 & 219.09  & 208.299 & 221.526 & 242.313 \\
5   & 14 & 253.627 & 160.287 & 155.322 & 170.36  & 200.064 \\
5   & 17 & 320.484 & 180.204 & 177.201 & 187.181 & 224.999 

\end{tabular}
\end{table}

\begin{figure}
\centering
\includegraphics[width=.95\linewidth]{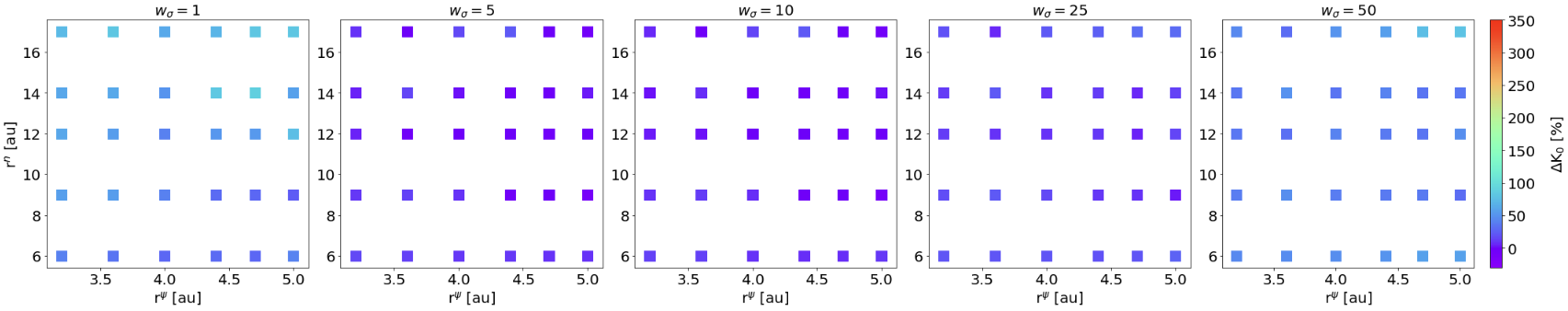}
\caption{Percentage deviation from DFT bulk modulus of schreibersite for the \{16,12,0\} set.}
\label{fig:BULKMOD_16-12-0}
\end{figure}

\begin{table}[]
\small
\caption{Bulk modulus of schreibersite for the \{16,12,0\} set as a function of $w_\sigma$.}
\begin{tabular}{ccccccc}
$R_\Psi$ & $R_n$ & $w_\sigma$=0    & $w_\sigma$=5    & $w_\sigma$=10   & $w_\sigma$=25   & $w_\sigma$=50   \\ \hline
3.2 & 6  & 278.367 & 179.738 & 176.408 & 193.048 & 235.029 \\
3.2 & 9  & 298.102 & 188.971 & 183.177 & 205.919 & 249.298 \\
3.2 & 12 & 242.182 & 199.74  & 195.941 & 208.37  & 250.515 \\
3.2 & 14 & 270.87  & 190.439 & 188.985 & 204.434 & 240.15  \\
3.2 & 17 & 278.236 & 180.764 & 178.578 & 192.903 & 234.697 \\
3.6 & 6  & 279.065 & 198.279 & 185.1   & 210.763 & 257.353 \\
3.6 & 9  & 313.904 & 172.37  & 171.34  & 182.839 & 225.092 \\
3.6 & 12 & 232.561 & 196.702 & 196.509 & 208.256 & 253.08  \\
3.6 & 14 & 265.98  & 194.459 & 192.419 & 209.99  & 254.704 \\
3.6 & 17 & 266.123 & 167.298 & 175.573 & 188.783 & 228.265 \\
4   & 6  & 322.085 & 168.912 & 168.198 & 183.106 & 232.675 \\
4   & 9  & 312.148 & 172.091 & 171.978 & 226.23  & 302.534 \\
4   & 12 & 219.97  & 187.23  & 185.52  & 209.309 & 273.473 \\
4   & 14 & 221.408 & 169.246 & 170.565 & 188.314 & 235.594 \\
4   & 17 & 265.857 & 164.756 & 165.463 & 180.331 & 231.236 \\
4.4 & 6  & 259.687 & 174.152 & 171.083 & 187.91  & 230.364 \\
4.4 & 9  & 279.183 & 201.328 & 197.177 & 211.908 & 258.259 \\
4.4 & 12 & 221.352 & 189.336 & 187.227 & 207.595 & 255.551 \\
4.4 & 14 & 241.344 & 188.903 & 185.143 & 199.574 & 239.515 \\
4.4 & 17 & 242.765 & 170.955 & 173.666 & 187.656 & 243.19  \\
4.7 & 6  & 315.605 & 171.605 & 168.718 & 194.079 & 244.603 \\
4.7 & 9  & 292.406 & 211.501 & 170.265 & 216.666 & 270.317 \\
4.7 & 12 & 222.668 & 189.622 & 186.483 & 205.991 & 259.303 \\
4.7 & 14 & 222.538 & 166.389 & 167.246 & 191.129 & 235.426 \\
4.7 & 17 & 261.451 & 171.603 & 171.824 & 191.422 & 236.177 \\
5   & 6  & 271.833 & 176.688 & 175.946 & 194.989 & 235.05  \\
5   & 9  & 312.826 & 162.262 & 161.987 & 223.132 & 308.649 \\
5   & 12 & 245.196 & 188.068 & 190.031 & 217.764 & 272.059 \\
5   & 14 & 214.466 & 155.482 & 156.69  & 178.41  & 225.836 \\
5   & 17 & 301.875 & 172.874 & 172.923 & 197.658 & 253.147 

\end{tabular}
\end{table}

\begin{figure}
\centering
\includegraphics[width=.5\linewidth]{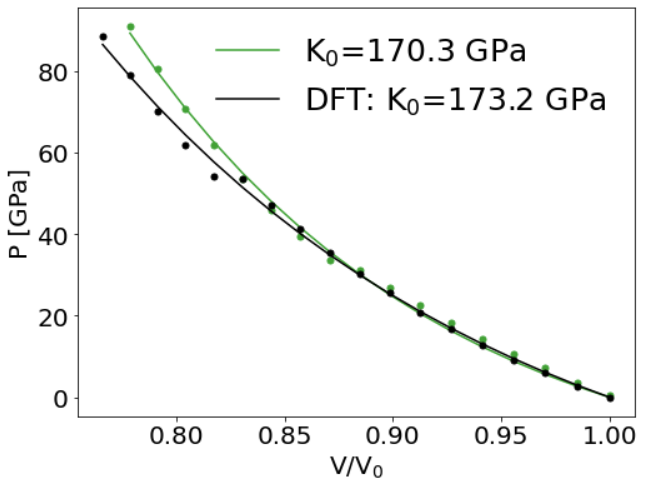}
\caption{Cold curve for the DFT and DFTB/ChIMES $R_\Psi=4.7$ au, $R_n=9.0$ au $\{16,12,0\}$/$w_{\rm \sigma} = 10$ parametrization.}
\label{fig:BULKMOD_FINAL}
\end{figure}

\newpage
\section{Surface formation energy}
\begin{figure}
\centering
\includegraphics[width=.95\linewidth]{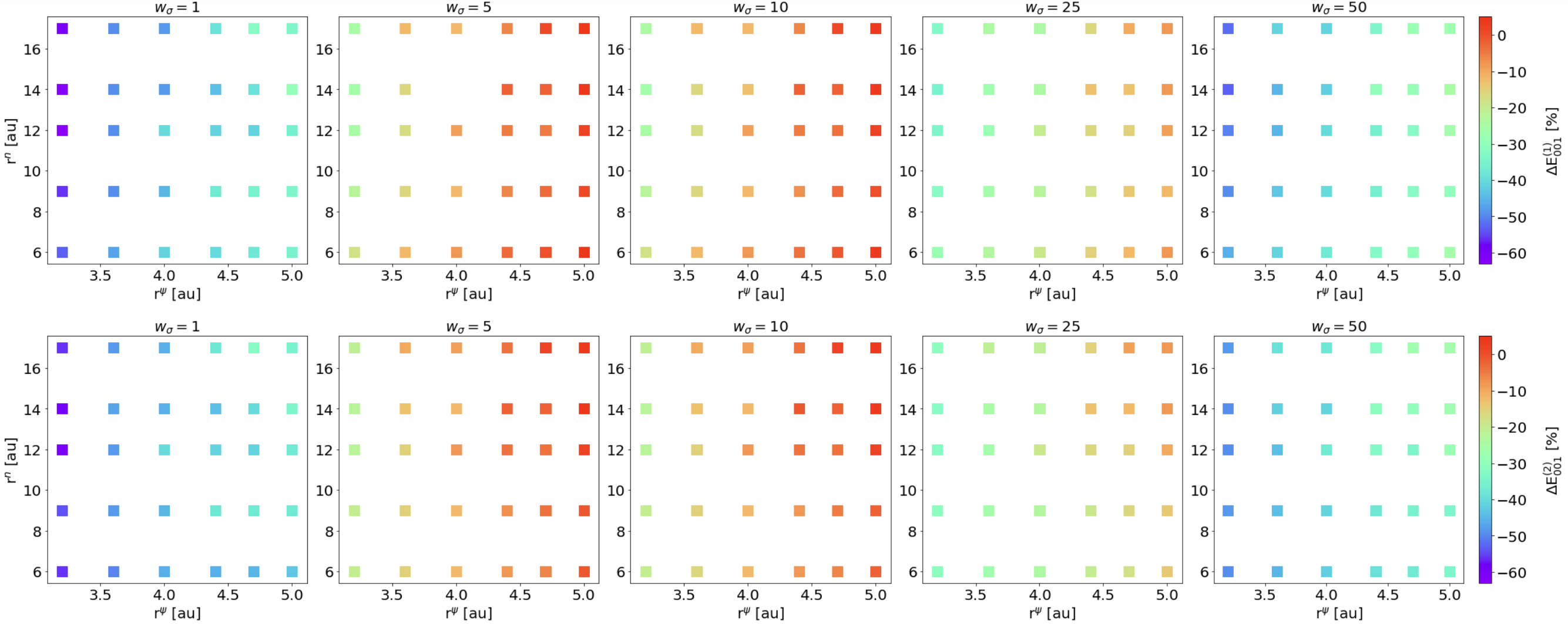}
\caption{Percentage deviation from DFT surface formation energy for (001) surface for $R_\Psi=4.7$ au, $R_n=9.0$ au $\{16,12,0\}$ parametrization, as a function of $w_\sigma$.}
\label{fig:SURFEN_001}
\end{figure}

\begin{figure}
\centering
\includegraphics[width=.95\linewidth]{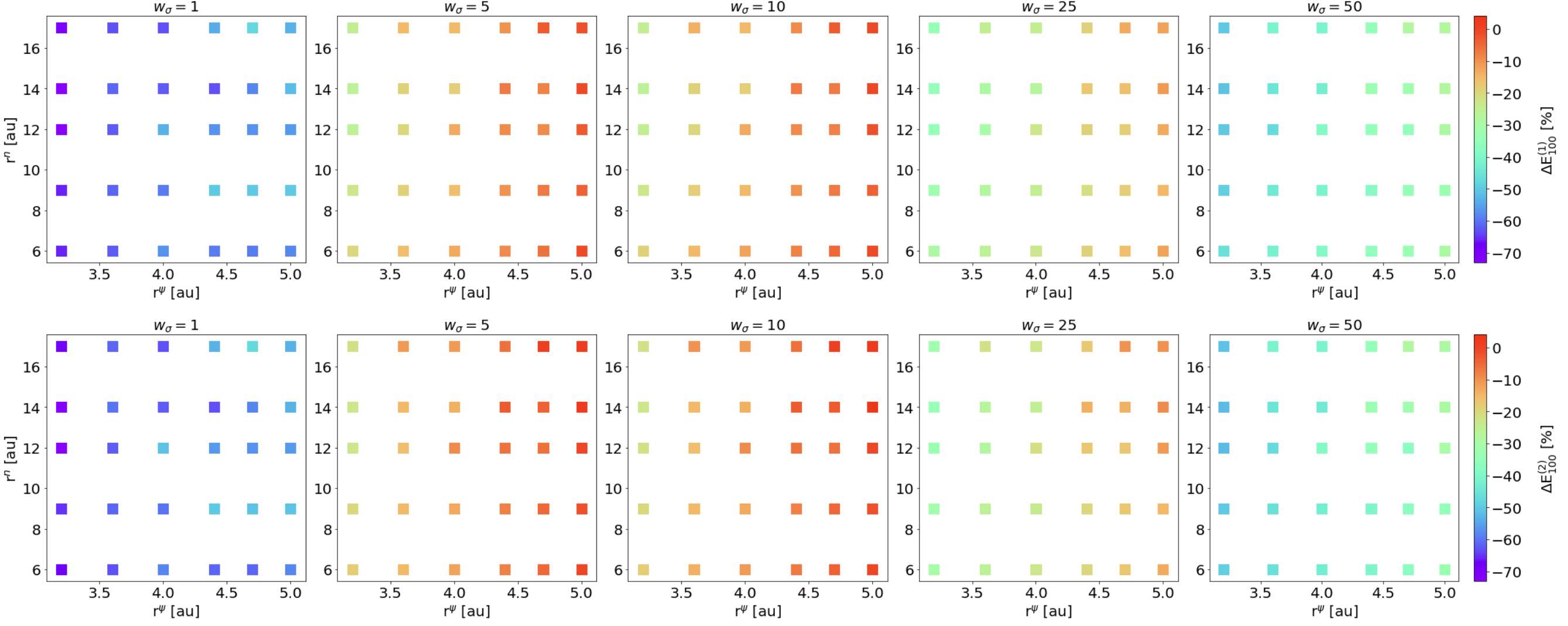}
\caption{Percentage deviation from DFT surface formation energy for (100) surface for $R_\Psi=4.7$ au, $R_n=9.0$ au $\{16,12,0\}$ parametrization, as a function of $w_\sigma$.}
\label{fig:SURFEN_100}
\end{figure}

\begin{figure}
\centering
\includegraphics[width=.95\linewidth]{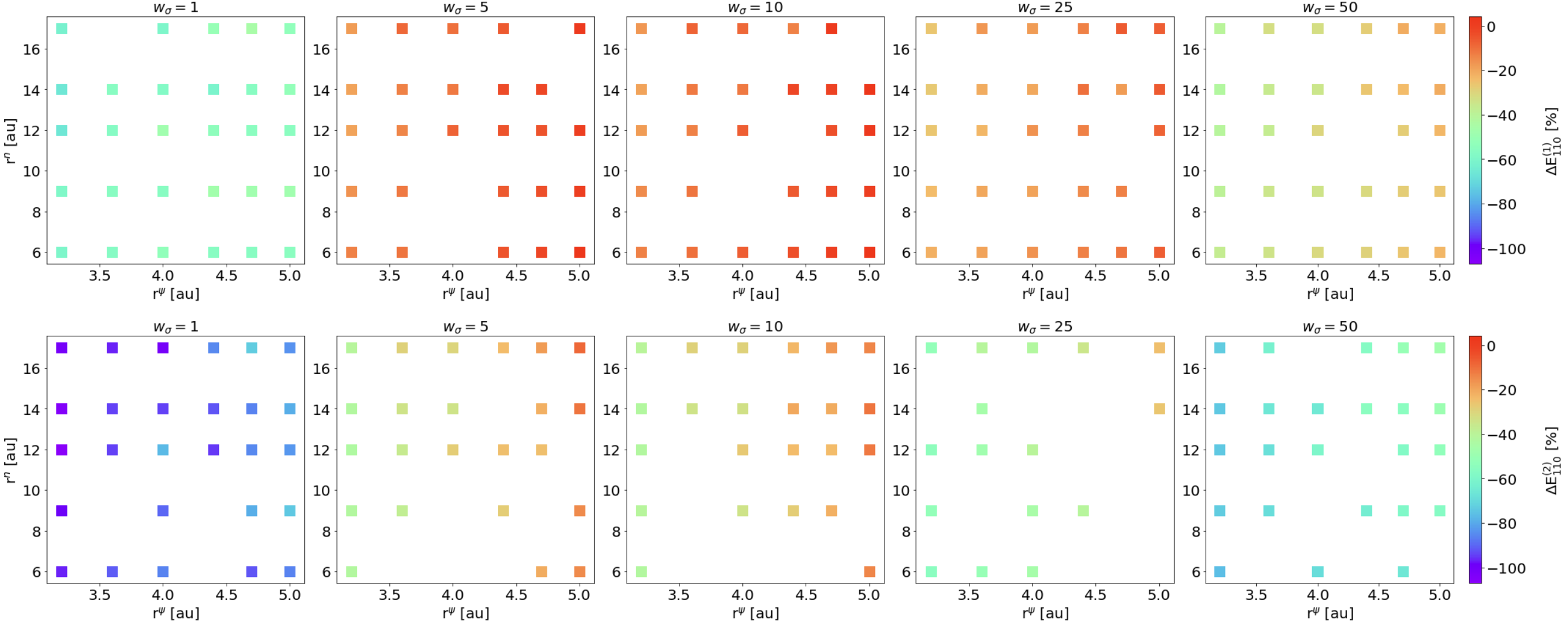}
\caption{Percentage deviation from DFT surface formation energy for (110) surface for $R_\Psi=4.7$ au, $R_n=9.0$ au $\{16,12,0\}$ parametrization, as a function of $w_\sigma$.}\label{fig:SURFEN_110}
\end{figure}

\begin{table}[]
\small
\caption{Surface formation energy $E_{\rm surface}$ for $\{16,12,0\}$/$w_{\rm \sigma} = 1$ set. Energies are reported in J/m$^2$. Zero values correspond to non-converged runs thus to missing data in the previous plots.}
\begin{tabular}{cccccccc}
$R_\Psi$ & $R_n$ & $E^{(1)}_{001}$ & $E^{(2)}_{001}$ & $E^{(1)}_{100}$ & $E^{(2)}_{100}$ & $E^{(1)}_{110}$ & $E^{(2)}_{110}$ \\ \hline
3.2  & 6    & 3.44    & 3.12    & 3.39    & 3.25    & 4.04    & 3.03    \\
3.2  & 9    & 3.49    & 3.09    & 3.39    & 3.22    & 4.02    & 3.05    \\
3.2  & 12   & 3.66    & 3.2     & 3.55    & 3.38    & 4.21    & 3.18    \\
3.2  & 14   & 3.64    & 3.18    & 3.54    & 3.35    & 4.19    & 3.16    \\
3.2  & 17   & 3.59    & 3.13    & 3.48    & 3.27    & 4.1     & 3.09    \\
3.6  & 6    & 3.29    & 3.01    & 3.31    & 3.17    & 3.97    & 2.94    \\
3.6  & 9    & 3.34    & 2.97    & 3.3     & 3.14    & 3.94    & 0       \\
3.6  & 12   & 3.34    & 2.97    & 3.32    & 3.15    & 3.98    & 2.98    \\
3.6  & 14   & 3.34    & 2.94    & 3.3     & 3.11    & 3.95    & 2.98    \\
3.6  & 17   & 3.34    & 2.96    & 3.34    & 3.14    & 0       & 3.04    \\
4    & 6    & 3.15    & 2.92    & 3.21    & 3.08    & 3.87    & 2.86    \\
4    & 9    & 3.25    & 2.91    & 3.26    & 3.11    & 3.94    & 2.92    \\
4    & 12   & 3.14    & 2.8     & 3.14    & 2.95    & 3.75    & 2.73    \\
4    & 14   & 3.32    & 2.93    & 3.31    & 3.15    & 4       & 2.98    \\
4    & 17   & 3.32    & 2.93    & 3.34    & 3.17    & 4.05    & 3.1     \\
4.4  & 6    & 3.13    & 2.92    & 3.26    & 3.13    & 3.95    & 0       \\
4.4  & 9    & 3.06    & 2.76    & 3.07    & 2.92    & 3.71    & 0       \\
4.4  & 12   & 3.15    & 2.83    & 3.21    & 3.04    & 3.9     & 3       \\
4.4  & 14   & 3.23    & 2.88    & 3.34    & 3.17    & 4.07    & 2.95    \\
4.4  & 17   & 3.11    & 2.76    & 3.15    & 2.98    & 3.82    & 2.85    \\
4.7  & 6    & 3.09    & 2.91    & 3.26    & 3.13    & 3.95    & 2.96    \\
4.7  & 9    & 3.03    & 2.75    & 3.08    & 2.94    & 3.75    & 2.75    \\
4.7  & 12   & 3.15    & 2.82    & 3.22    & 3.07    & 3.93    & 2.85    \\
4.7  & 14   & 3.11    & 2.8     & 3.23    & 3.08    & 3.95    & 2.87    \\
4.7  & 17   & 2.96    & 2.65    & 3.03    & 2.87    & 3.68    & 2.67    \\
5    & 6    & 3.02    & 2.87    & 3.23    & 3.08    & 3.92    & 2.87    \\
5    & 9    & 3.01    & 2.74    & 3.07    & 2.94    & 3.75    & 2.68    \\
5    & 12   & 3.05    & 2.75    & 3.19    & 3.05    & 3.92    & 2.82    \\
5    & 14   & 2.89    & 2.69    & 3.12    & 2.98    & 3.85    & 2.75    \\
5    & 17   & 3       & 2.71    & 3.13    & 2.98    & 3.88    & 2.84   
\end{tabular}
\end{table}

\begin{table}[]
\small
\caption{Surface formation energy $E_{\rm surface}$ for $\{16,12,0\}$/$w_{\rm \sigma} = 5$ set. Energies are reported in J/m$^2$. Zero values correspond to non-converged runs thus to missing data in the previous plots.}
\begin{tabular}{cccccccc}
$R_\Psi$ & $R_n$ & $E^{(1)}_{001}$ & $E^{(2)}_{001}$ & $E^{(1)}_{100}$ & $E^{(2)}_{100}$ & $E^{(1)}_{110}$ & $E^{(2)}_{110}$ \\ \hline
3.2 & 6  & 2.64 & 2.4  & 2.46 & 2.31 & 2.87 & 2.19 \\
3.2 & 9  & 2.74 & 2.41 & 2.52 & 2.34 & 2.93 & 2.18 \\
3.2 & 12 & 2.8  & 2.44 & 2.57 & 2.39 & 3    & 2.2  \\
3.2 & 14 & 2.82 & 2.44 & 2.58 & 2.4  & 3.01 & 2.19 \\
3.2 & 17 & 2.8  & 2.42 & 2.56 & 2.38 & 2.99 & 2.17 \\
3.6 & 6  & 2.51 & 2.31 & 2.38 & 2.24 & 2.8  & 0    \\
3.6 & 9  & 2.59 & 2.3  & 2.43 & 2.25 & 2.84 & 2.12 \\
3.6 & 12 & 2.62 & 2.3  & 2.46 & 2.27 & 2.88 & 2.09 \\
3.6 & 14 & 2.6  & 2.27 & 2.44 & 2.24 & 2.86 & 2.07 \\
3.6 & 17 & 2.52 & 2.21 & 2.36 & 2.16 & 2.75 & 1.99 \\
4   & 6  & 2.42 & 2.25 & 2.31 & 2.18 & 0    & 0    \\
4   & 9  & 2.52 & 2.24 & 2.37 & 2.21 & 0    & 0    \\
4   & 12 & 2.44 & 2.16 & 2.31 & 2.12 & 2.72 & 1.97 \\
4   & 14 & 0    & 2.25 & 2.42 & 2.23 & 2.84 & 2.06 \\
4   & 17 & 2.51 & 2.18 & 2.36 & 2.17 & 2.78 & 2    \\
4.4 & 6  & 2.32 & 2.17 & 2.23 & 2.1  & 2.65 & 0    \\
4.4 & 9  & 2.38 & 2.15 & 2.26 & 2.11 & 2.68 & 1.97 \\
4.4 & 12 & 2.35 & 2.09 & 2.24 & 2.06 & 2.65 & 1.92 \\
4.4 & 14 & 2.29 & 2.04 & 2.19 & 2.01 & 2.6  & 0    \\
4.4 & 17 & 2.37 & 2.08 & 2.25 & 2.07 & 2.67 & 1.91 \\
4.7 & 6  & 2.25 & 2.12 & 2.17 & 2.05 & 2.58 & 1.86 \\
4.7 & 9  & 2.31 & 2.09 & 2.2  & 2.05 & 2.62 & 0    \\
4.7 & 12 & 2.35 & 2.09 & 2.23 & 2.07 & 2.66 & 1.93 \\
4.7 & 14 & 2.29 & 2.04 & 2.21 & 2.02 & 2.59 & 1.88 \\
4.7 & 17 & 2.22 & 1.96 & 2.11 & 1.93 & 0    & 1.81 \\
5   & 6  & 2.14 & 2.03 & 2.07 & 1.95 & 2.45 & 1.77 \\
5   & 9  & 2.23 & 2.03 & 2.12 & 1.98 & 2.53 & 1.77 \\
5   & 12 & 2.2  & 1.97 & 2.11 & 1.95 & 2.52 & 0    \\
5   & 14 & 2.14 & 1.91 & 2.06 & 1.89 & 0    & 1.71 \\
5   & 17 & 2.14 & 1.92 & 2.08 & 1.91 & 2.49 & 1.68 
\end{tabular}
\end{table}

\begin{table}[]
\small
\caption{Surface formation energy $E_{\rm surface}$ for $\{16,12,0\}$/$w_{\rm \sigma} = 10$ set. Energies are reported in J/m$^2$. Zero values correspond to non-converged runs thus to missing data in the previous plots.}
\begin{tabular}{cccccccc}
$R_\Psi$ & $R_n$ & $E^{(1)}_{001}$ & $E^{(2)}_{001}$ & $E^{(1)}_{100}$ & $E^{(2)}_{100}$ & $E^{(1)}_{110}$ & $E^{(2)}_{110}$ \\ \hline
3.2 & 6  & 2.64 & 2.4  & 2.44 & 2.3  & 2.85 & 2.18 \\
3.2 & 9  & 2.74 & 2.41 & 2.51 & 2.34 & 2.92 & 2.17 \\
3.2 & 12 & 2.8  & 2.44 & 2.56 & 2.38 & 2.98 & 2.18 \\
3.2 & 14 & 2.82 & 2.44 & 2.57 & 2.39 & 3    & 2.18 \\
3.2 & 17 & 2.77 & 2.42 & 2.55 & 2.37 & 2.97 & 2.16 \\
3.6 & 6  & 2.51 & 2.32 & 2.36 & 2.23 & 2.78 & 0    \\
3.6 & 9  & 2.59 & 2.31 & 2.42 & 2.25 & 2.82 & 0    \\
3.6 & 12 & 2.62 & 2.31 & 2.45 & 2.27 & 2.87 & 0    \\
3.6 & 14 & 2.6  & 2.27 & 2.43 & 2.24 & 2.84 & 2.06 \\
3.6 & 17 & 2.52 & 2.2  & 2.34 & 2.15 & 2.73 & 1.99 \\
4   & 6  & 2.42 & 2.25 & 2.29 & 2.17 & 2.71 & 0    \\
4   & 9  & 2.5  & 2.25 & 2.36 & 2.2  & 0    & 2.07 \\
4   & 12 & 2.45 & 2.17 & 2.31 & 2.12 & 2.71 & 1.96 \\
4   & 14 & 2.54 & 2.25 & 2.41 & 2.22 & 2.83 & 2.05 \\
4   & 17 & 2.51 & 2.19 & 2.35 & 2.16 & 2.76 & 1.99 \\
4.4 & 6  & 2.33 & 2.17 & 2.21 & 2.09 & 2.62 & 0    \\
4.4 & 9  & 2.39 & 2.16 & 2.26 & 2.11 & 2.67 & 1.97 \\
4.4 & 12 & 2.35 & 2.09 & 2.23 & 2.06 & 0    & 1.91 \\
4.4 & 14 & 2.29 & 2.03 & 2.18 & 2    & 2.57 & 1.85 \\
4.4 & 17 & 2.37 & 2.08 & 2.24 & 2.06 & 2.85 & 1.89 \\
4.7 & 6  & 2.26 & 2.12 & 2.14 & 2.03 & 2.54 & 0    \\
4.7 & 9  & 2.32 & 2.1  & 2.2  & 2.05 & 2.61 & 1.88 \\
4.7 & 12 & 2.33 & 2.09 & 2.22 & 2.06 & 2.63 & 1.91 \\
4.7 & 14 & 2.29 & 2.04 & 2.19 & 2.01 & 2.56 & 1.86 \\
4.7 & 17 & 2.23 & 1.97 & 2.1  & 1.93 & 2.47 & 1.8  \\
5   & 6  & 2.15 & 2.04 & 2.06 & 1.94 & 2.42 & 1.76 \\
5   & 9  & 2.24 & 2.04 & 2.12 & 1.98 & 2.52 & 0    \\
5   & 12 & 2.2  & 1.97 & 2.09 & 1.94 & 2.49 & 1.72 \\
5   & 14 & 2.15 & 1.92 & 2.05 & 1.88 & 2.43 & 1.71 \\
5   & 17 & 2.16 & 1.93 & 2.07 & 1.9  & 0    & 1.76
\end{tabular}
\end{table}

\begin{table}[]
\small
\caption{Surface formation energy $E_{\rm surface}$ for $\{16,12,0\}$/$w_{\rm \sigma} = 25$ set. Energies are reported in J/m$^2$. Zero values correspond to non-converged runs thus to missing data in the previous plots.}
\begin{tabular}{cccccccc}
$R_\Psi$ & $R_n$ & $E^{(1)}_{001}$ & $E^{(2)}_{001}$ & $E^{(1)}_{100}$ & $E^{(2)}_{100}$ & $E^{(1)}_{110}$ & $E^{(2)}_{110}$ \\ \hline
3.2 & 6  & 2.88 & 2.63 & 2.65 & 2.53 & 3.08 & 2.39 \\
3.2 & 9  & 2.96 & 2.62 & 2.71 & 2.56 & 3.14 & 2.34 \\
3.2 & 12 & 3.01 & 2.65 & 2.76 & 2.6  & 3.2  & 2.37 \\
3.2 & 14 & 3.02 & 2.65 & 2.77 & 2.61 & 3.22 & 0    \\
3.2 & 17 & 2.99 & 2.62 & 2.74 & 2.58 & 3.18 & 2.34 \\
3.6 & 6  & 2.76 & 2.55 & 2.57 & 2.46 & 3.02 & 2.33 \\
3.6 & 9  & 2.82 & 2.52 & 2.62 & 2.46 & 3.04 & 0    \\
3.6 & 12 & 2.86 & 2.53 & 2.66 & 2.5  & 3.1  & 2.28 \\
3.6 & 14 & 2.85 & 2.5  & 2.64 & 2.47 & 3.07 & 2.25 \\
3.6 & 17 & 2.77 & 2.42 & 2.55 & 2.37 & 2.96 & 2.17 \\
4   & 6  & 2.67 & 2.49 & 2.51 & 2.4  & 2.94 & 2.25 \\
4   & 9  & 2.74 & 2.47 & 2.56 & 2.42 & 3    & 2.23 \\
4   & 12 & 2.7  & 2.39 & 2.52 & 2.35 & 2.93 & 2.15 \\
4   & 14 & 2.77 & 2.47 & 2.6  & 2.44 & 3.04 & 0    \\
4   & 17 & 2.77 & 2.42 & 2.56 & 2.39 & 2.99 & 2.18 \\
4.4 & 6  & 2.57 & 2.41 & 2.43 & 2.33 & 2.87 & 0    \\
4.4 & 9  & 2.64 & 2.39 & 2.47 & 2.34 & 2.9  & 2.17 \\
4.4 & 12 & 2.6  & 2.32 & 2.43 & 2.28 & 2.86 & 0    \\
4.4 & 14 & 2.54 & 2.26 & 2.39 & 2.22 & 2.79 & 0    \\
4.4 & 17 & 2.6  & 2.3  & 2.44 & 2.28 & 2.87 & 2.08 \\
4.7 & 6  & 2.51 & 2.37 & 2.37 & 2.28 & 2.8  & 0    \\
4.7 & 9  & 2.56 & 2.33 & 2.41 & 2.28 & 2.85 & 0    \\
4.7 & 12 & 2.57 & 2.31 & 2.42 & 2.28 & 0    & 0    \\
4.7 & 14 & 2.53 & 2.25 & 2.38 & 2.22 & 2.98 & 0    \\
4.7 & 17 & 2.47 & 2.19 & 2.31 & 2.15 & 2.69 & 0    \\
5   & 6  & 2.42 & 2.29 & 2.3  & 2.2  & 2.71 & 0    \\
5   & 9  & 2.5  & 2.29 & 2.36 & 2.24 & 0    & 0    \\
5   & 12 & 2.44 & 2.2  & 2.31 & 2.17 & 2.72 & 0    \\
5   & 14 & 2.42 & 2.16 & 2.27 & 2.12 & 2.67 & 1.94 \\
5   & 17 & 2.42 & 2.16 & 2.29 & 2.14 & 2.7  & 1.92
\end{tabular}
\end{table}

\begin{table}[]
\small
\caption{Surface formation energy $E_{\rm surface}$ for $\{16,12,0\}$/$w_{\rm \sigma} = 50$ set. Energies are reported in J/m$^2$. Zero values correspond to non-converged runs thus to missing data in the previous plots.}
\begin{tabular}{cccccccc}
$R_\Psi$ & $R_n$ & $E^{(1)}_{001}$ & $E^{(2)}_{001}$ & $E^{(1)}_{100}$ & $E^{(2)}_{100}$ & $E^{(1)}_{110}$ & $E^{(2)}_{110}$ \\ \hline
3.2 & 6  & 3.27 & 2.99 & 2.99 & 2.91 & 3.47 & 2.71 \\
3.2 & 9  & 3.34 & 2.97 & 3.05 & 2.93 & 3.52 & 2.66 \\
3.2 & 12 & 3.37 & 2.99 & 3.08 & 2.96 & 3.56 & 2.68 \\
3.2 & 14 & 3.43 & 2.99 & 3.09 & 2.97 & 3.58 & 2.68 \\
3.2 & 17 & 3.4  & 2.97 & 3.07 & 2.94 & 3.55 & 2.66 \\
3.6 & 6  & 3.16 & 2.91 & 2.91 & 2.84 & 3.39 & 0    \\
3.6 & 9  & 3.21 & 2.88 & 2.95 & 2.84 & 3.42 & 2.6  \\
3.6 & 12 & 3.24 & 2.88 & 3    & 2.86 & 3.47 & 2.59 \\
3.6 & 14 & 3.26 & 2.85 & 2.97 & 2.83 & 3.44 & 2.56 \\
3.6 & 17 & 3.17 & 2.78 & 2.88 & 2.74 & 3.33 & 2.49 \\
4   & 6  & 3.06 & 2.85 & 2.84 & 2.78 & 3.32 & 2.6  \\
4   & 9  & 3.13 & 2.82 & 2.89 & 2.79 & 3.38 & 0    \\
4   & 12 & 3.13 & 2.74 & 2.84 & 2.71 & 3.3  & 2.47 \\
4   & 14 & 3.19 & 2.82 & 2.92 & 2.81 & 3.42 & 2.54 \\
4   & 17 & 3.16 & 2.77 & 2.88 & 2.75 & 3.35 & 0    \\
4.4 & 6  & 2.98 & 2.79 & 2.78 & 2.72 & 3.26 & 0    \\
4.4 & 9  & 3.05 & 2.77 & 2.82 & 2.73 & 3.31 & 2.52 \\
4.4 & 12 & 3.05 & 2.67 & 2.76 & 2.64 & 0    & 0    \\
4.4 & 14 & 2.92 & 2.63 & 2.72 & 2.6  & 3.18 & 2.38 \\
4.4 & 17 & 3    & 2.65 & 2.77 & 2.64 & 3.23 & 2.41 \\
4.7 & 6  & 2.91 & 2.74 & 2.72 & 2.67 & 3.19 & 2.56 \\
4.7 & 9  & 2.97 & 2.71 & 2.76 & 2.68 & 3.25 & 2.45 \\
4.7 & 12 & 2.97 & 2.68 & 2.77 & 2.67 & 3.25 & 2.44 \\
4.7 & 14 & 2.89 & 2.61 & 2.71 & 2.58 & 3.14 & 2.38 \\
4.7 & 17 & 2.88 & 2.54 & 2.63 & 2.5  & 3.06 & 2.32 \\
5   & 6  & 2.83 & 2.68 & 2.65 & 2.59 & 3.11 & 0    \\
5   & 9  & 2.92 & 2.68 & 2.72 & 2.64 & 3.2  & 2.41 \\
5   & 12 & 2.83 & 2.57 & 2.64 & 2.54 & 3.1  & 2.35 \\
5   & 14 & 2.81 & 2.54 & 2.62 & 2.51 & 3.07 & 2.3  \\
5   & 17 & 2.85 & 2.53 & 2.62 & 2.51 & 3.08 & 2.27
\end{tabular}
\end{table}

\newpage
\section{Bulk interstitial water absorption}
\begin{table}[]
\small
\caption{Bulk water absorption energies $\{16,12,0\}$ set obtained for both interstitial sites and for $w_\sigma = 5$ and $w_\sigma = 10$. Zero values correspond to non-converged runs thus to missing data in the plots in the main text.}
\begin{tabular}{cccccc}
     &      & \multicolumn{2}{c}{$w_\sigma$=5} & \multicolumn{2}{c}{$w_\sigma$=10} \\ \cline{3-6} 
$R_\Psi$ & $R_n$ & Site 1      & Site 2     & Site 1      & Site 2      \\ \hline
3.2  & 6    & -2.17       & -1.88      & -2.15       & -1.84       \\
3.2  & 9    & -2.28       & -1.99      & -2.25       & -1.94       \\
3.2  & 12   & -2.35       & -2.06      & -2.33       & -2.01       \\
3.2  & 14   & -1.07       & -0.97      & -1.05       & -0.92       \\
3.2  & 17   & -1.86       & -1.56      & -1.86       & -1.52       \\
3.6  & 6    & -2.54       & -2.16      & -2.49       & -2.09       \\
3.6  & 9    & -2.63       & -2.22      & -2.61       & -2.18       \\
3.6  & 12   & -2.69       & -2.3       & -2.68       & -2.23       \\
3.6  & 14   & -1.57       & -1.59      & -1.55       & -1.54       \\
3.6  & 17   & -2.2        & -1.87      & -2.19       & -1.82       \\
4    & 6    & -2.94       & -2.47      & -2.93       & -2.4        \\
4    & 9    & -3.02       & -2.57      & -3.01       & -2.53       \\
4    & 12   & -3.09       & -2.62      & -3.11       & -2.59       \\
4    & 14   & -2.08       & -2.23      & -2.06       & -2.17       \\
4    & 17   & -2.62       & -2.22      & -2.59       & -2.17       \\
4.4  & 6    & -3.38       & -2.85      & -3.37       & -2.79       \\
4.4  & 9    & -3.5        & -2.99      & -3.48       & -2.92       \\
4.4  & 12   & -3.5        & -3         & -3.52       & -2.98       \\
4.4  & 14   & -2.55       & -2.79      & -2.53       & -2.74       \\
4.4  & 17   & -3.03       & -2.61      & -2.99       & -2.54       \\
4.7  & 6    & -3.66       & -3.16      & -3.64       & -3.1        \\
4.7  & 9    & -3.79       & -3.27      & -3.78       & -3.22       \\
4.7  & 12   & -3.87       & -3.36      & -3.86       & -3.31       \\
4.7  & 14   & -2.86       & -3.18      & -2.84       & -3.12       \\
4.7  & 17   & -3.34       & -2.88      & -3.29       & -2.81       \\
5    & 6    & -4          & -3.45      & -3.99       & -3.39       \\
5    & 9    & -4.07       & -3.55      & 0           & 0           \\
5    & 12   & -4.21       & -3.65      & 0           & 0           \\
5    & 14   & -3.15       & -3.53      & 0           & 0           \\
5    & 17   & -3.62       & -3.16      & 0           & 0          
\end{tabular}
\end{table}